\documentclass[longauth]{aa} 

\usepackage{graphicx}
\usepackage{txfonts}
\usepackage{color}
\usepackage{natbib}
\bibpunct{(}{)}{;}{a}{}{,} 
\def\kms{\,{\rm km\, s}^{-1}}

\begin{document}

\title{The progeny of a Cosmic Titan: a massive multi-component
  proto-supercluster in formation at z=2.45 in VUDS\thanks{Based on
    data obtained with the European Southern Observatory Very Large
    Telescope, Paranal, Chile, under Large Program 185.A-0791.  }}

\titlerunning{Hyperion: a proto-supercluster at $z\sim2.45$ in VUDS}  

\author{O.~Cucciati\inst{1}
        \and B.~C.~Lemaux\inst{2,3}
        \and G.~Zamorani\inst{1}
        \and O. Le F\`evre\inst{3}
        \and L.~A.~M.~Tasca\inst{3}
        \and N.~P.~Hathi\inst{4}
        \and K-G.~Lee\inst{5,6}
        \and S. Bardelli\inst{1}
        \and P.~Cassata\inst{7} 
        \and B.~Garilli\inst{8} 
        \and V.~Le Brun\inst{3}
        \and D.~Maccagni\inst{8}  
        \and L.~Pentericci\inst{9}  
        \and R.~Thomas\inst{10} 
        \and E.~Vanzella\inst{1}
        \and E.~Zucca\inst{1}
        \and L.~M.~Lubin\inst{2}
        \and R.~Amorin\inst{11,12}
        \and L.~P.~Cassar\`a\inst{8}
        \and A.~Cimatti\inst{13,1}
        \and M.~Talia\inst{13}
        \and D.~Vergani\inst{1}
        \and A.~Koekemoer\inst{4}
        \and J.~Pforr\inst{14}
        \and M.~Salvato\inst{15}
}

\institute{
           INAF - Osservatorio di Astrofisica e Scienza dello Spazio di Bologna, via Gobetti 93/3 - 40129 Bologna - Italy\\
           \email{olga.cucciati@oabo.inaf.it}
           \and 
           Department of Physics, University of California, Davis, One Shields Ave., Davis, CA 95616, USA
           \and 
           Aix Marseille Universit\'e, CNRS, LAM (Laboratoire d'Astrophysique de Marseille) UMR 7326, 13388, Marseille, France
          \and 
          Space Telescope Science Institute, 3700 San Martin Drive,  Baltimore, MD 21218, USA
           \and 
           Kavli IPMU (WPI), UTIAS, The University of Tokyo, Kashiwa, Chiba 277-8583, Japan
           \and 
           Lawrence Berkeley National Laboratory, 1 Cyclotron Road, Berkeley, CA 94720, USA
          \and 
          University of Padova, Department of Physics and Astronomy, Vicolo Osservatorio 3, 35122, Padova - Italy
          \and 
          INAF--IASF Milano, via Bassini 15, I--20133, Milano, Italy
          \and 
          INAF--Osservatorio Astronomico di Roma, via di Frascati 33, I-00040, Monte Porzio Catone, Italy
          \and 
          European Southern Observatory, Avenida Alonso de C\'ordova 3107, Vitacura, 19001 Casilla, Santiago de Chile, Chile
          \and 
          Kavli Institute for Cosmology, University of Cambridge, Madingley Road, Cambridge CB3 0HA, UK
          \and 
          Cavendish Laboratory, University of Cambridge, 19 J.J. Thomson Avenue, Cambridge CB3 0HE, UK
           \and 
           University of Bologna, Department of Physics and Astronomy (DIFA), via Gobetti 93/2 - 40129, Bologna - Italy
           \and 
           ESA/ESTEC SCI-S, Keplerlaan 1, 2201 AZ, Noordwijk, The Netherlands
           \and 
           Max-Planck-Institut f\"ur Extraterrestrische Physik, Postfach 1312, D-85741, Garching bei M\"unchen, Germany
}

\date{Received - ; accepted -}

\abstract{We unveil the complex shape of a proto-supercluster at
  $z\sim2.45$ in the COSMOS field exploiting the synergy of both
  spectroscopic and photometric redshifts. Thanks to the spectroscopic
  redshifts of the VIMOS Ultra-Deep Survey (VUDS), complemented by the
  zCOSMOS-Deep spectroscopic sample and high-quality photometric
  redshifts, we compute the three-dimensional (3D) overdensity field in a volume of
  $\sim100\times100\times250$ comoving Mpc$^3$ in the central region
  of the COSMOS field, centred at $z\sim2.45$ along the line of sight.
  The method relies on a two-dimensional (2D) Voronoi tessellation in overlapping
  redshift slices that is converted into a 3D density field, where the
  galaxy distribution in each slice is constructed using a statistical
  treatment of both spectroscopic and photometric redshifts.  In this
  volume, we identify a proto-supercluster, dubbed ``Hyperion" for its
  immense size and mass, which extends over a volume of
  $\sim60\times60\times150$ comoving Mpc$^3$ and has an estimated
  total mass of $\sim 4.8\times 10^{15}{\rm M}_{\odot}$. This
  immensely complex structure contains at least seven density peaks within
  $2.4\lesssim z \lesssim 2.5$ connected by filaments that exceed the
  average density of the volume. We estimate the total mass of the
  individual peaks, $M_{\rm tot}$, based on their inferred average
  matter density, and find a range of masses from
  $\sim 0.1\times 10^{14}{\rm M}_{\odot}$ to
  $\sim 2.7\times 10^{14}{\rm M}_{\odot}$. By using spectroscopic
  members of each peak, we obtain the velocity dispersion of the
  galaxies in the peaks, and then their virial mass $M_{\rm vir}$
  (under the strong assumption that they are virialised). The
  agreement between $M_{\rm vir}$ and $M_{\rm tot}$ is surprisingly
  good, at less than $1-2\sigma$, considering that (almost all) the
  peaks are probably not yet virialised. According to the spherical
  collapse model, these peaks have already started or are about to
  start collapsing, and they are all predicted to be virialised by
  redshift $z\sim0.8-1.6$. We finally perform a careful comparison
  with the literature, given that smaller components of this
  proto-supercluster had previously been identified using either
  heterogeneous galaxy samples (Ly$\alpha$ emitters, sub-mm
  starbursting galaxies, CO emitting galaxies) or 3D Ly$\alpha$
    forest tomography on a smaller area.  With VUDS, we obtain, 
    for the first time across the central $\sim1$ deg$^2$ of the COSMOS field, a panoramic
  view of this large structure, that encompasses, connects, and
  considerably expands in a homogeneous way on all previous detections
  of the various sub-components. The characteristics of this
  exceptional proto-supercluster, its redshift, its richness over a
  large volume, the clear detection of its sub-components, together
  with the extensive multi-wavelength imaging and spectroscopy granted
  by the COSMOS field, provide us the unique possibility to study a
  rich supercluster in formation.}



   \maketitle
%


\section{Introduction}\label{intro}

Proto-clusters are crucial sites for studying how environment
  affects galaxy evolution in the early universe, both in observations
  (see e.g.
  \citealp{steidel05,peter07,miley08,tanaka10,strazzullo13}) and
  simulations (e.g. \citealp{chiang17,muldrew18}).  Moreover, since
  proto-clusters mark the early stages of structure formation, they
  have the potential to provide additional constraints on the already
  well established probes on standard and non-standard cosmology based
  on galaxy clusters at low and intermediate redshift (see e.g.
  \citealp{allen11,heneka18,schmidt09,roncarelli15}, and references
  therein).  

Although the sample of confirmed or candidate proto-clusters is
increasing in both number (see e.g. the systematic searches in
\citealp{diener2013_list,chiang2014_list,franck16_CCPC,lee2016_colossus,toshikawa18_goldrush})
and maximum redshift (e.g. \citealp{higuchi18_silverrush}), our
knowledge of high-redshift ($z>2$) structures is still limited, as it
is broadly based on heterogeneous data sets. These structures span
from relaxed to unrelaxed systems, and are detected by using
different, and sometimes apparently contradicting, selection
criteria. As a non-exhaustive list of examples, high-redshift clusters
and proto-clusters have been identified as excesses of either
star-forming galaxies (e.g.
\citealp{steidel00,ouchi05,lemaux09,capak11}) or red galaxies
(e.g. \citealp{kodama07,spitler12}),  as excesses of infrared(IR)-luminous galaxies \citep{gobat11}, or via SZ signatures \citep{foley11_SZ} or diffuse X-ray emission \citep{fassbender11}. Other
detection methods include the search for photometric redshift
overdensities in deep multi-band surveys
\citep{salimbeni09,scoville13_env} or around active galactic nuclei (AGNs) and radio galaxies
\citep{pentericci00,galametz12},  the identification of large
  intergalactic medium reservoirs via Ly$\alpha$ forest absorption
  \citep{cai16_method, lee2016_colossus, cai17_z23}, and the exploration
  of narrow redshift slices via narrow band imaging
  \citep{venemans02,lee14_NB}.

The identification and study of proto-structures can be boosted by two
factors: 1) the use of spectroscopic redshifts, and 2) the use of
unbiased tracers with respect to the underlying galaxy population.  On
the one hand, the use of spectroscopic redshifts is crucial for a
robust identification of the overdensities themselves, for the study
of the velocity field, especially in terms of the galaxy velocity
dispersion which can be used as a proxy for the total mass, and
finally for the identification of possible
sub-structures. On the other hand, if such proto-structures are found and
mapped by tracers that are representative of the dominant galaxy
population at the epoch of interest, we can recover an unbiased view
of such environments.

In this context, we used the VUDS (VIMOS Ultra Deep Survey)
spectroscopic survey \citep{lefevre2015_vuds} to systematically search
for proto-structures. VUDS targeted approximately
$10000$ objects presumed to be at high redshift for spectroscopic observations, confirming over
$5000$ galaxies at $z>2$. These galaxies generally have stellar
masses $\gtrsim 10^{9} {\rm M}_{\odot}$, and are broadly
representative in stellar mass, absolute magnitude, and rest-frame
colour of all star-forming galaxies (and thus, the vast majority of
galaxies) at $2 \lesssim z \lesssim 4.5$ for $i\leq25$.  We identified
a preliminary sample of $\sim50$ candidate proto-structures (Lemaux et
al, in prep.) over $2<z<4.6$ in the COSMOS, CFHTLS-D1 and ECDFS fields
(1 deg$^2$ in total).  With this `blind' search in the COSMOS field
we identified the complex and rich proto-structure at $z\sim2.5$
presented in this paper.

This proto-structure, extended over a volume of
$\sim60\times60\times150$ comoving Mpc$^3$, has a very complex shape,
and includes several density peaks within $2.42<z<2.51$, possibly
connected by filaments, that are more dense than the average volume
density. Smaller components of this proto-structure have already
been identified in the literature from heterogeneous galaxy
samples, like for example Ly$\alpha$ emitters (LAEs), three-dimensional (3D) Ly$\alpha$-forest
tomography, sub-millimetre starbursting galaxies, and CO-emitting galaxies (see
\citealp{diener2015_z245, chiang2015_z244, casey2015_z247,
  lee2016_colossus,wang2016_z250}). Despite the sparseness of
previous identifications of sub-clumps, a part of this structure was already
dubbed “Colossus” for its extension \citep{lee2016_colossus}.

With VUDS, we obtain a more complete and unbiased panoramic view of
this large structure, placing the previous sub-structure detections
reported in the literature in the broader context of this extended
large-scale structure. The characteristics of this proto-structure,
its redshift, its richness over a large volume, the clear detection
of its sub-components, the extensive imaging and spectroscopy coverage
granted by the COSMOS field, provide us the unique possibility to
study a rich supercluster in its formation.

From now on we refer to this huge structure as a
`proto-supercluster'. On the one hand, throughout the paper we show that it is as extended and as massive as known superclusters at
lower redshift. Moreover, it presents a very complex shape, which
includes several density peaks embedded in the same large-scale
structure, similarly to other lower-redshift structures defined
superclusters. In particular, one of the peaks has already been
identified in the literature \citep{wang2016_z250} as a possibly
virialised structure. On the other hand, we also show that the
evolutionary status of some of these peaks is compatible with that
of overdensity fluctuations which are collapsing and are foreseen to
virialise in a few gigayears. For all these reasons, we consider this
structure a proto-supercluster.

In this work, we aim to characterise the 3D shape of the
proto-supercluster, and in particular to study the properties of its
sub-components, for example their average density, volume, total mass,
velocity dispersion, and shape. We also perform a thorough comparison of
our findings with the previous density peaks detected in the
literature on this volume, so as to put them in the broader context of a
large-scale structure.

The paper is organised as follows.  In Sect.~\ref{data} we present our
data set and how we reconstruct the overdensity field. The discovery
of the proto-supercluster, and its total volume and mass, are discussed
in Sect.~\ref{supercluster}. In Sect.~\ref{3D_peaks} we describe the
properties of the highest density peaks embedded in the
proto-supercluster (their individual mass, velocity dispersion, etc.)
and we compare our findings with the literature. In
Sect.~\ref{discussion} we discuss how the peaks would evolve according
to the spherical collapse model, and how we can compare the
proto-supercluster to similar structures at lower redshifts. Finally,
in Sect.~\ref{summary} we summarise our results.

Except where explicitly stated otherwise, we assume a flat $\Lambda$CDM
cosmology with $\Omega_m=0.25$,
$\Omega_{\Lambda}=0.75$, $H_0=70\kms {\rm Mpc}^{-1}$ and
$h=H_0/100$. Magnitudes are expressed in the AB system
\citep{oke74,fukugita96}. Comoving and physical Mpc(/kpc) are
expressed as cMpc(/ckpc) and pMpc(/pkpc), respectively.


\section{The data sample and the density field}\label{data}

VUDS is a spectroscopic survey performed with VIMOS on the ESO-VLT
\citep{lefevre2003}, targeting approximately $10000$ objects in the three
fields COSMOS, ECDFS, and VVDS-2h to study galaxy evolution at
$2 \lesssim z \lesssim 6$. Full details are given in
\cite{lefevre2015_vuds}; here we give only a brief review.

VUDS spectroscopic targets have been pre-selected using four different
criteria. The main criterion is a photometric redshift ($z_p$) cut
($z_p+1\sigma \geq 2.4$, with $z_p$ being either the $1^{st}$ or
$2^{nd}$ peak of the $z_p$ probability distribution function) coupled
with the flux limit $i\leq 25$. This main criterion provided 87.7\% of
the primary sample. Photometric redshifts were derived as described in
\cite{ilbert2013} with the code {\it Le
  Phare\footnote{http://www.cfht.hawaii.edu/$\sim$arnouts/LEPHARE/lephare.html}}
\citep{arnouts99,ilbert2006_pz}. The remaining targets include
galaxies with colours compatible with Lyman-break galaxies, if not
already selected by the $z_p$ criterion, as well as drop-out
galaxies for which a strong break compatible with $z>2$ was identified
in the $ugrizYJHK$ photometry. In addition to this primary sample, a
purely flux-limited sample with $23 \leq i \leq 25$ has been
targeted to fill-up the masks of the multi-slit observations.

VUDS spectra have an extended wavelength coverage from 3600 to
$9350\AA$, because targets have been observed with both the LRBLUE and
LRRED grisms (both with R$\sim230$), with 14h integration each.  With
this integration time it is possible to reach S/N $\sim 5$ on the
continuum at $\lambda\sim8500$\AA~ (for $i = 25$), and for an emission
line with flux $F = 1.5\times10^{-18}$erg s$^{-1}$ cm$^{2}$. The
redshift accuracy is $\sigma_{zs}= 0.0005 (1 + z)$, corresponding to
$\sim150\kms$ (see also \citealp{lefevre2013a}).

We refer the reader to \cite{lefevre2015_vuds} for a detailed
description of data reduction and redshift measurement. Concerning the reliability of the measured redshifts, here it is
important to stress that each measured redshift is given a reliability
flag equal to X1, X2, X3, X4, or X9\footnote{$X=0$ is for galaxies,
  $X=1$ for broad line AGNs, and $X=2$ for secondary objects falling
  serendipitously in the slits and spatially separable from the main
  target. The case $X=3$ is as $X=2$ but for objects not separable
  spatially from the main target.}, which correspond to a probability
of being correct of 50-75\%, 75-85\%, 95-100\%, 100\%, and $\sim80$\%
respectively. In the COSMOS field, the VUDS sample comprises 4303
spectra of unique objects, out of which 2045 have secure spectroscopic
redshift (flags X2, X3, X4, or X9) and $z\ge 2.$

Together with the VUDS data, we used the zCOSMOS-Bright
\citep{lilly2007,lilly2009} and zCOSMOS-Deep (Lilly et al, in prep.,
\citealp{diener2013_list}) spectroscopic samples. The flag system for
the robustness of the redshift measurement is basically the same as in
the VUDS sample, with very similar flag probabilities (although they
have never been fully assessed for zCOSMOS-Deep). In the zCOSMOS
samples, the spectroscopic flags have also been given a decimal digit
to represent the level of agreement of the spectroscopic redshift
($z_s$) with the photometric redshift ($z_p$). A given $z_p$ is
defined to be in agreement with its corresponding $z_s$ when
$|z_s-z_p|<0.08(1+z_s)$, and in these cases the decimal digit of the
spectroscopic flag is `5'. For the zCOSMOS samples, we define secure
$z_s$ those with a quality flag X2.5, X3, X4, or X9, which means that
for flag X2 we used only the $z_s$ in agreement with their respective
$z_p$, while for higher flags we trust the $z_s$ irrespectively of the
agreement with their $z_p$. With these flag limits, we are left with
more than 19000 secure $z_s$, of which 1848 are at $z\geq 2$.  We
merged the VUDS and zCOSMOS samples, removing the duplicates between
the two surveys as follows. For each duplicate, that is, objects
observed in both VUDS and zCOSMOS, we retained the redshift with the
most secure quality flag, which in the vast majority of cases was
the one from VUDS. In case of equal flags, we retained the VUDS
spectroscopic redshift.  Our final VUDS$+$zCOSMOS spectroscopic
catalogue consists of 3822 unique secure $z_s$ at $z\geq 2$.

We note that we did not use spectroscopic redshifts from
any other survey, although other spectroscopic samples in this area
are already publicly available in the literature (see e.g.
\citealp{casey2015_z247, chiang2015_z244, diener2015_z245,
  wang2016_z250}). These samples are often follow-up of small regions
around dense regions, and we did not want to be biased in the
identification of already known density peaks. Unless specified otherwise, our
spectroscopic sample always refers only to the good quality flags
in VUDS and zCOSMOS discussed above.  We also did not include public
$z_s$ from more extensive campaigns, like for example the COSMOS AGN spectroscopic survey \citep{trump09}, the MOSDEF
survey \citep{kriek15}, or the DEIMOS 10K spectroscopic survey
\citep{hasinger18}.

We matched our spectroscopic catalogue with the photometric COSMOS2015
catalogue \citep{laigle2016}. The matching was done by selecting the closest source within a matching radius of $0.55^{\prime \prime}$. Objects in the COSMOS2015 have been
detected via an ultra-deep $\chi^2$ sum of the $YJHK_s$ and $z^{++}$
images. $YJHK_s$ photometry was obtained by the VIRCAM instrument on
the VISTA telescope (UltraVISTA-DR2
survey\footnote{https://www.eso.org/sci/observing/phase3/data\_releases/uvista\_dr2.pdf
}, \citealp{mcCracken12}), and the $z^{++}$ data, taken using the Subaru Suprime-Cam, are a
(deeper) replacement of the previous $z-$band COSMOS data
\citep{taniguchi2007,taniguchi2015}. With this match with the
COSMOS2015 catalogue we obtained a uniform target coverage of the COSMOS
field down to a given flux limit (see Sect.\ref{method}), using
spectroscopic redshifts for the objects in our original spectroscopic
sample or photometric redshifts for the remaining sources. The
photometric redshifts in COSMOS2015 are derived using
$3^{\prime \prime}$ aperture fluxes in the 30 photometric bands of
COSMOS2015. According to Table 5 of \cite{laigle2016}, a direct
comparison of their photometric redshifts with the spectroscopic
redshifts of the entire VUDS survey in the COSMOS field (median
redshift $z_{\rm med}=2.70$ and median $i^+-$band
$i^+_{\rm med}=24.6$) gives a photometric redshift accuracy of
$\Delta z = 0.028(1+z)$. The same comparison with the zCOSMOS-Deep
sample (median redshift $z_{\rm med}=2.11$ and median $i^+-$band
$i^+_{\rm med}=23.8$) gives $\Delta z = 0.032(1+z)$.

\begin{figure*} \centering
\includegraphics[width=6.cm]{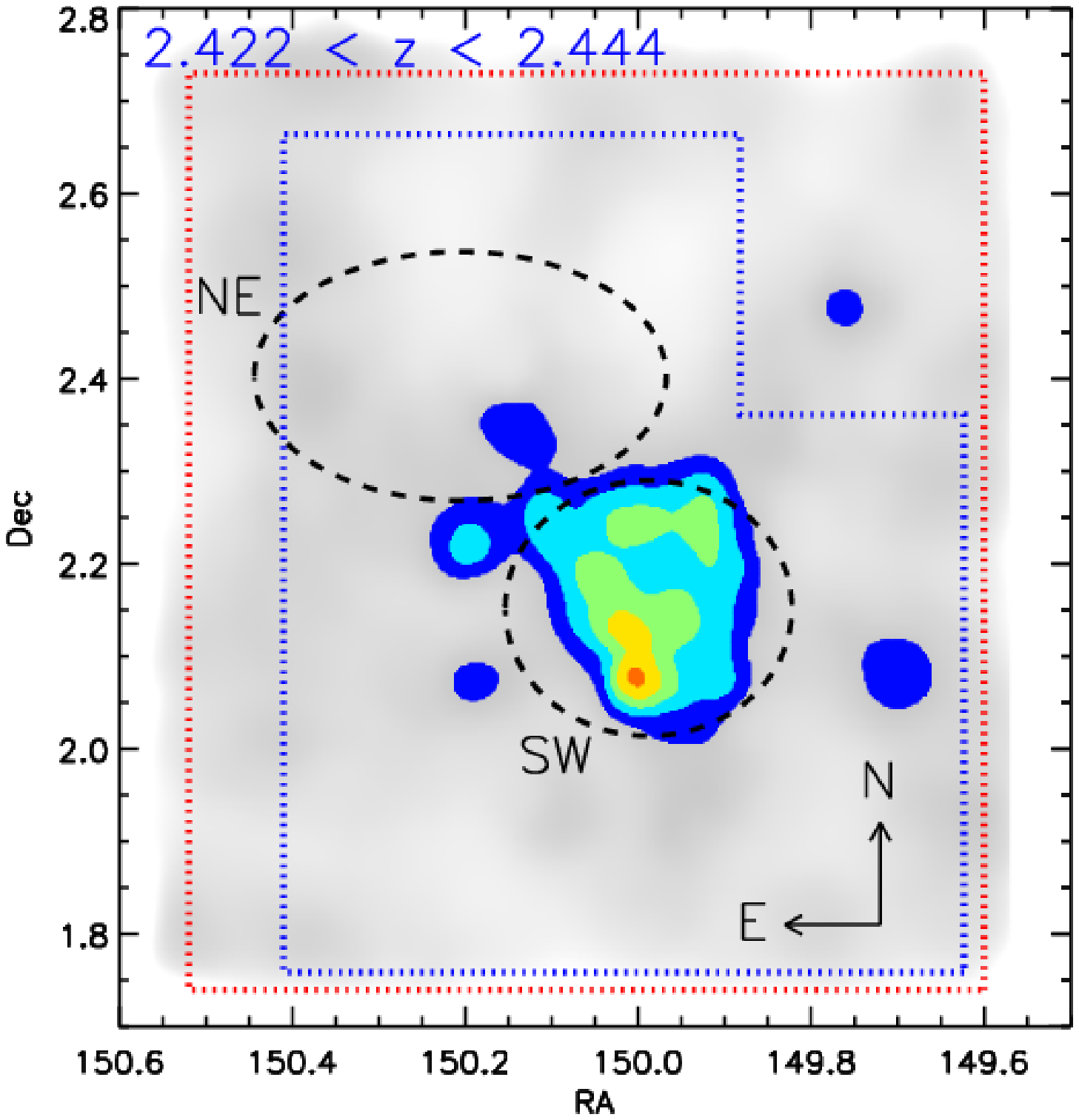}
\includegraphics[width=6.cm]{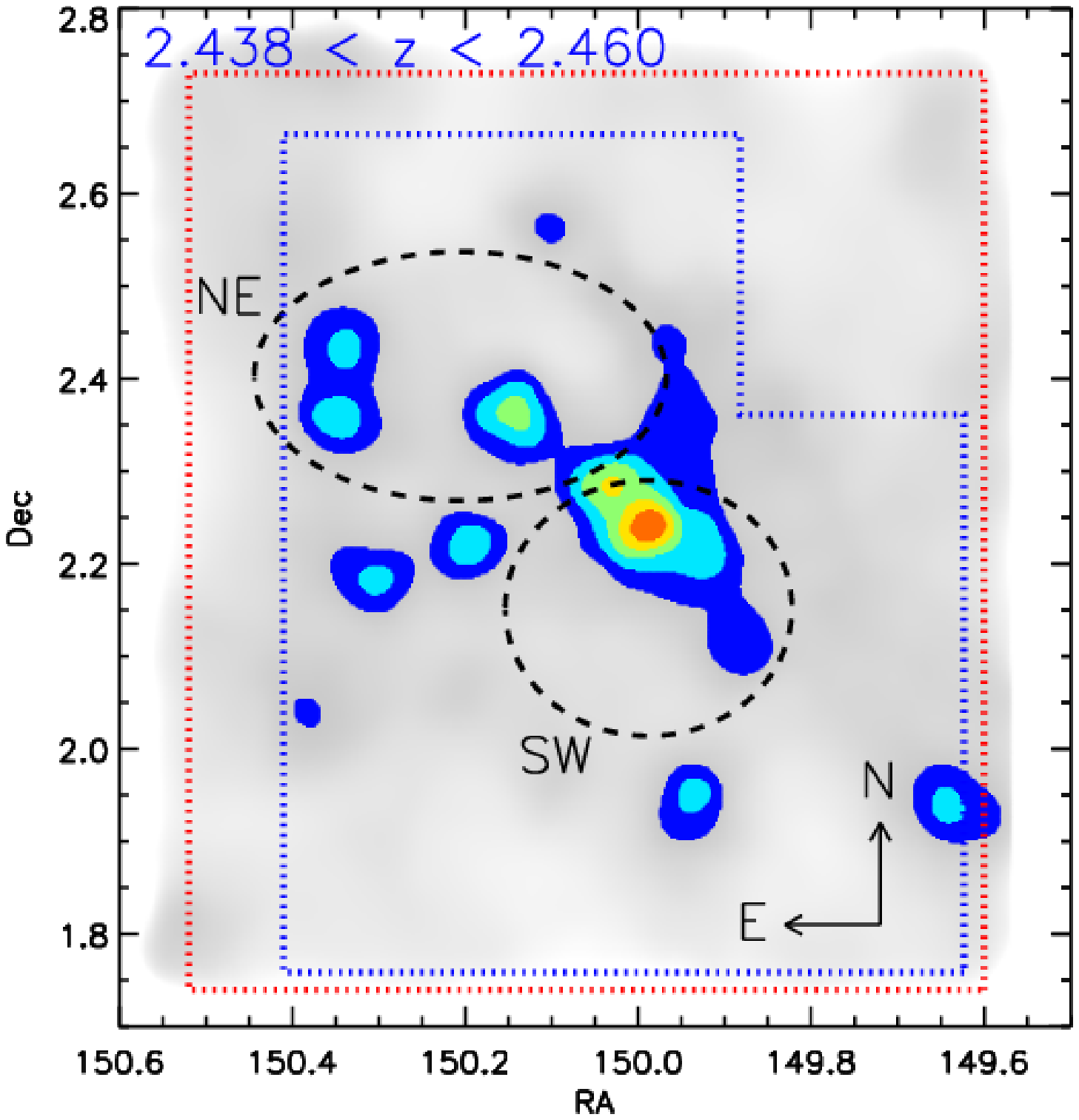}
\includegraphics[width=6.cm]{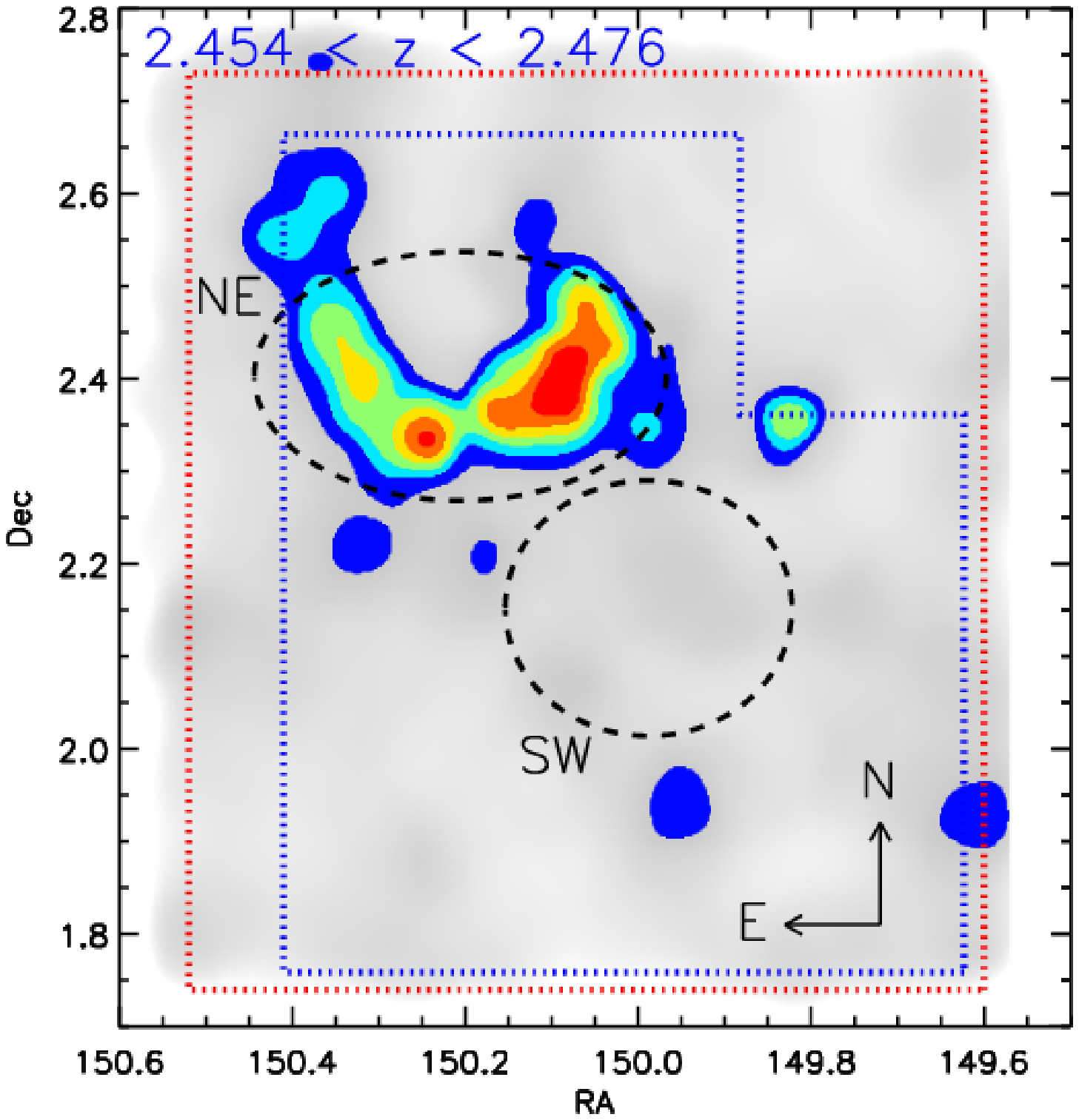}
\caption{RA- Dec overdensity maps, in three redshift slices
  as indicated in the labels. The background grey-scale indicates the
  overdensity ${\rm log}(1+\delta_{\rm gal})$ value (darker grey is
  for higher values). Regions with ${\rm log}(1+\delta_{\rm gal})$
  above 2, 3, 4, 5, 6, and 7 $\sigma_{\rm \delta}$ above the mean are
  indicated with blue, cyan, green, yellow, orange, and red colours.
  respectively. The red dotted line encloses the region retained for
  the analysis, and the blue dotted line is the region covered by the
  VUDS survey. The two black dashed ellipses (repeated in all panels
  for reference) show the rough positions of the two main components
  of the proto-supercluster identified in this work, dubbed ``NE''
  (rightmost panel) and ``SW'' (leftmost panel). The field dimensions
  in RA and Dec correspond roughly to $\sim120\times130$ cMpc in
  the redshift range spanned by the three redshift slices. }
\label{2D_maps} 
\end{figure*}

The method to compute the density field and identify the density peaks
is the same as described in \cite{lemaux2018_z45}; we describe it here briefly. The method is based on the Voronoi Tessellation,
which has already been successfully used at different redshifts to
characterise the local environment around galaxies and identify the
highest density peaks, including the search for groups and clusters
(see e.g. \citealp{marinoni02, coooper05, cucciati10, gerke12,
  scoville13_env, darvish15,smolcic17}). Its main advantage is that
the local density is measured both on an adaptive scale and with an
adaptive filter shape, allowing us to follow the natural distribution
of tracers.

In our case, we worked in two dimensions in overlapping redshift slices. We used
as tracers the spectroscopic sample complemented by a photometric
sample which provides us with the photometric redshifts of all the galaxies
for which we did not have any $z_s$ information.

For each redshift slice, we generated a set of Monte Carlo (MC)
realisations. Galaxies (with $z_s$ or $z_p$) to be used in each
realisation were selected observing the following steps, in this order:

\begin{itemize}

\item[1)] irrespectively of their redshift, galaxies with a $z_s$ were
  retained in a percentage of realisations equal to the probability
  associated to the reliability flag; namely, in each realisation,
  before the selection in redshift, for each galaxy we drew a number
  from a uniform distribution from 0 to 100 and retained that galaxy
  only if the drawn number was equal to or less than the galaxy
  redshift reliability;

\item[2)] galaxies with only $z_p$ were first selected to complement
  the retained spectroscopic sample (i.e. the photometric sample
  comprises all the galaxies without a $z_s$ or for which we threw
  away their $z_s$ for a given iteration), then they were assigned a
  new photometric redshift $z_{\rm p,new}$ randomly drawn from an
  asymmetrical Gaussian distribution centred on their nominal $z_p$
  value and with negative and positive sigmas equal to
  the lower and upper uncertainties in the $z_p$ measurement, respectively; with
  this approach we do not try to correct for catastrophic redshift
  errors, but only for the shape of the PDF of each $z_p$;

\item[3)] among the samples selected at steps 1 and 2, we retained all
  the galaxies with $z_s$ (from step 1) or $z_{\rm p,new}$ (from step
  2) falling in the considered redshift slice.

\end{itemize}

We performed a 2D Voronoi tessellation for each $i^{th}$ MC
realisation, and assigned to each Voronoi polygon a surface density
$\Sigma_{VMC,i}$ equal to the inverse of the area (expressed in
Mpc$^2$) of the given polygon. Finally, we created a regular grid of
$75\times75$ pkpc cells, and assigned to each grid point the
$\Sigma_{VMC,i}$ of the polygon enclosing the central point of the
cell. For each redshift slice, the final density field $\Sigma_{VMC}$
is computed on the same grid, as the median of the density fields
among the realisations, cell by cell. As a final step, from the median
density map we computed the local over-density at each grid point as
$\delta_{\rm gal} = \Sigma_{VMC}/ \tilde{\Sigma}_{VMC} -1 $, where
$\tilde{\Sigma}_{VMC}$ is the mean $\Sigma_{VMC}$ for all grid points.
In our analysis we are more interested in $\delta_{\rm gal}$ 
than in $\Sigma_{VMC}$ because we want to identify the regions that
are overdense with respect to the mean density at each redshift, a
density which can change not only for astrophysical reasons but also
due to characteristics of the imaging/spectroscopic survey. Moreover,
as we see in the following sections, the computation of
$\delta_{\rm gal}$ is useful to estimate the total mass of our
proto-cluster candidates and their possible evolution.

Proto-cluster candidates were identified by searching for extended
regions of contiguous grid cells with a $\delta_{\rm gal}$ value above
a given threshold.  The initial systematic search for proto-clusters
in the COSMOS field (which will be presented in Lemaux et al., in prep.)
was run with the following set of parameters: redshift slices of 7.5
pMpc shifting in steps of 3.75 pMpc (so as to have redshift slices
overlapping by half of their depth); 25 Monte Carlo realisations per
slice; and spectroscopic and photometric catalogues with $[3.6]\leq 25.3$
(IRAC Channel 1). With this `blind' search we re-identified two
proto-clusters at $z\sim3$ serendipitously discovered at the beginning
of VUDS observations \citep{lemaux2014_z33,cucciati2014_z29}, together
with other outstanding proto-structures presented separately in
companion papers (\citealp{lemaux2018_z45}, Lemaux et al. in prep.).


\section{Discovery of a rich extended
  proto-supercluster}\label{supercluster}

The preliminary overdensity maps showed two extended overdensities at
$z\sim2.46$, in a region of $0.4\times0.25$ deg$^2$. Intriguingly,
there were several other smaller overdensities very close in right
ascension (RA), declination (Dec), and redshift. We therefore explored
in more detail the COSMOS field by focusing our attention on the
volume around these overdensities. This focused analysis revealed the
presence of a rich extended structure, consisting of density peaks
linked by slightly less dense regions.

\subsection{The method}\label{method}

We re-ran the computation of the density field and the search for
overdense regions with a fine-tuned parameter set (see below), in the
range $2.35 \lesssim z \lesssim 2.55$, which we studied by considering
several overlapping redshift slices. Concerning the angular
extension of our search, we computed the density field in the central
$\sim1\times1$ deg$^2$ of the COSMOS field, but then used only the
slightly smaller 0.91~deg$^2$ region at $149.6\leq RA \leq 150.52$ and
$1.74 \leq Dec \leq2.73$ to perform any further analysis (computation
of the mean density etc.). This choice was made to avoid the regions
close to the field boundaries, where the Voronoi tessellation is
affected by border effects. In this smaller area, considering a flux
limit at $i=25$, about 24\% of the objects with a redshift ($z_s$ or
$z_p$) falling in the above-mentioned redshift range have a
spectroscopic redshift.  If we reduce the area to the region covered
by VUDS observations, which is slightly smaller, this percentage
increases to about 28\%.

We also verified the robustness of our choices for what concerns the
following issues:

{\bf Number of Monte Carlo realisations}. With respect to Lemaux et
al. (in prep.), we increased the number of Monte Carlo realisations
from the initial 25 to 100 to obtain a more reliable median value
(similarly to, e.g. \citealp{lemaux2018_z45}). We verified that our
results did not significantly depend on the number of realisations
$n_{\rm MC}$ as long as $n_{\rm MC}\geq 100$, and, therefore, all analyses
presented in this paper are done on maps which used $n_{\rm
  MC}=100$.
This high number of realisations allowed us to produce not only the
median density field for each redshift slice, but also its associated
error maps, as follows. For each grid cell, we considered the
distribution of the 100 $\Sigma_{VMC}$ values, and took the $16^{th}$
and $84^{th}$ percentiles of this distribution as lower and upper
limits for $\Sigma_{VMC}$. We produced density maps with these lower
and upper limits, in the same way as for the median $\Sigma_{VMC}$,
and then computed the corresponding overdensities that we call
$\delta_{\rm gal,16}$ and $\delta_{\rm gal,84}$.

{\bf Spectroscopic sample.} As in \cite{lemaux2018_z45}, we assigned a
probability to each spectroscopic galaxy to be used in a given
realisation equal to the reliability of its $z_s$ measurement, as
given by its quality flag. Namely, we used the quality flags X2 (X2.5
for zCOSMOS), X3, X4, and X9 with a reliability of 80\%, 97.5\%, 100\%
and 80\% respectively (see Sect.~\ref{data}; here we adopt the mean
probability for the flags X2 and X3, for which
\citealp{lefevre2015_vuds} give a range of probabilities). These
values were computed for the VUDS survey, but we applied them also to
the zCOSMOS spectroscopic galaxies in our sample, as discussed in
Sect.~\ref{data}. We verified that our results do not qualitatively
change if we choose slightly different reliability percentages or if
we used the entire spectroscopic sample (flag=X2/X2.5, X3, X4, X9) in
all realisations instead of assigning a probability to each
spectroscopic galaxy. The agreement between these results is due to
the very high flag reliabilities, and to the dominance of objects with
only $z_p$. With the cut in redshift at $2.35 \leq z \leq 2.55$, the
above-mentioned quality flag selection, and the magnitude limit at
$i\leq 25$ (see below), we are left with 271 spectroscopic
  redshifts from VUDS and 309 from zCOSMOS, for
  a total of 580 spectroscopic redshifts used in our analysis. This
  provides us with a spectroscopic sampling rate of $\sim24\%$,
  considering the above mentioned redshift range and magnitude cut.
We remind the reader that we use only VUDS and zCOSMOS spectroscopic
redshift, and do not include in our sample any other $z_s$ found in
the literature.

{\bf Mean density.} To compute the mean density $\tilde{\Sigma}_{VMC}$
we proceeded as follows. Given that $\Sigma_{VMC}$ has a log-normal
distribution \citep{coles91}, in each redshift slice we fitted the
distribution of ${\rm log}(\Sigma_{VMC})$ of all pixels with a
$3\sigma$-clipped Gaussian. The mean $\mu$ and standard deviation
$\sigma$ of this Gaussian are related to the average density
$\langle \Sigma_{VMC} \rangle$ by the equation
$\langle \Sigma_{VMC}\rangle = 10^{\mu}e^{2.652\sigma^{2}}$. We used
this $\langle \Sigma_{VMC} \rangle$ as the average density
$\tilde{\Sigma}_{VMC}$ to compute the density contrast
$\delta_{gal}$. $\tilde{\Sigma}_{VMC}$ was computed in this way in
each redshift slice.

{\bf Overdensity threshold.} In each redshift slice, we fitted the distribution of ${\rm log}(1+\delta_{\rm gal})$ with a
Gaussian,
obtaining its $\mu$ and $\sigma$. We call these parameters
$\mu_{\rm \delta}$ and $\sigma_{\rm \delta}$, for simplicity, although
they refer to the Gaussian fit of the ${\rm log}(1+\delta_{\rm gal})$
distribution and not of the $\delta_{\rm gal}$ distribution. We then fitted
$\mu_{\rm \delta}$ and $\sigma_{\rm \delta}$ as a function of redshift
with a second-order polynomial, obtaining $\mu_{\rm \delta,fit}$ and
$\sigma_{\rm \delta,fit}$ at each redshift. Our detection thresholds
were then set as a certain number of $\sigma_{\rm \delta,fit}$ above
the mean overdensity $\mu_{\rm \delta,fit}$, that is, as
$ {\rm log}(1+\delta_{\rm gal}) \geq \mu_{\rm \delta,fit}(z_{\rm
  slice}) + n_{\sigma}\sigma_{\rm \delta,fit}(z_{\rm slice})$,
where $z_{\rm slice}$ is the central redshift of each slice, and
$n_{\sigma}$ is chosen as described in Sects. \ref{3D} and
\ref{3D_peaks}. From now, when referring to setting  a
`$n_{\sigma}\sigma_{\rm \delta}$ threshold'  we mean that we
consider the volume of space with
$ {\rm log}(1+\delta_{\rm gal}) \geq \mu_{\rm \delta,fit}(z_{\rm
  slice}) + n_{\sigma}\sigma_{\rm \delta,fit}(z_{\rm slice})$.

{\bf Slice depth and overlap.} We used overlapping redshift slices with a full depth of
7.5 pMpc, which corresponds to $\delta z \sim 0.02$ at $z\sim2.45$,
running in steps of $\delta z \sim 0.002$. We also tried with thinner
slices (5 pMpc), but we adopted a depth of 7.5 pMpc as a
compromise between i) reducing the line of sight (l.o.s.) elongation of the density
peaks (see Sect.~\ref{3D}) and ii) keeping a low noise in the density
reconstruction. We define `noise' as the difference between
$\delta_{\rm gal}$ and its lower and upper uncertainties
$\delta_{\rm gal,16}$ and $\delta_{\rm gal,84}$\footnote{In this work
  we neglect the correlations in the noise between the cells in the
  same slice and those in different slices.}. The choice of small
steps of $\delta z \sim 0.002$ is due to the fact that we do not want
to miss the redshift where each structure is more prominent.

{\bf Tracers selection} We fine-tuned our search method (including the
$\delta_{\rm gal}$ thresholds etc...) for a sample of galaxies limited
at $i=25$. We verified the robustness of our findings by using also a
sample selected with $K_S\leq 24$ and one selected with
${\rm[3.6]}\leq24$ (IRAC Channel 1). With these two latter cuts, in
the redshift range $2.3\leq z \leq 2.6$ we have a number of galaxies
with spectroscopic redshift corresponding to $\sim87\%$ and $\sim94\%$
of the number of spectroscopic galaxies with $i\leq25$, respectively,
but not necessarily the same galaxies, while roughly 65\% and 85\%
more objects, respectively, with photometric redshifts entered in our
maps than did with $i\leq25$. Although the $K_S\leq 24$ and
${\rm[3.6]}\leq24$ samples might be distributed in a different way in
the considered volume because of the different clustering properties
of different galaxy populations, with these samples we recovered the
overdensity peaks in the same locations as with $i\leq25$. Clearly,
the $\delta_{\rm gal}$ distribution is slightly different, so the
overdensity threshold that we used to define the overdensity peaks
(see Sect.~\ref{3D_peaks}) encloses regions with slightly different
shape with respect to those recovered with a sample flux-limited at
$i\leq25$. We defer a more precise analysis of the kind of galaxy
populations which inhabit the different density peaks to future work.

Figure \ref{2D_maps} shows three 2D overdensity ($\delta_{\rm gal}$)
maps obtained as described above, in the redshift slices
$2.422<z<2.444$, $2.438<z<2.460$, and $2.454<z<2.476$. We can
distinguish two extended and very dense components at two different
redshifts and different RA-Dec positions: one at $z\sim 2.43$, in
the left-most panel, that we call the ``South-West'' (SW) component, and
the other at $z\sim 2.46$, at higher RA and Dec, that we call here the
``North-East'' (NE) component (right-most panel). The NE and
SW components seem to be connected by a region of relatively high
density, shown in the middle panel of the figure. This sort of
filament is particularly evident when we fix a threshold around
$2\sigma_{\rm \delta}$, as shown in the figure.  For this reason, we
retained the $2\sigma_{\rm \delta}$ threshold as the threshold used to
identify the volume of space occupied by this huge overdensity.  As a
reference, a $2\sigma_{\rm \delta}$ threshold corresponds to
$\delta\sim0.65$, while 3, 4, and 5$\sigma_{\rm \delta}$ thresholds correspond to
$\delta\sim1.1$, $\sim1.7$, and $\sim2.55$, respectively

To better understand the complex shape of the structure, we performed
an analysis in three dimensions, as described in the following sub-section.

\subsection{The 3D matter distribution}\label{3D}

We built a 3D overdensity cube in the following way. First, we
considered each redshift slice to be placed at $z_{\rm slice}$ along
the line of sight, where $z_{\rm slice}$ is the central redshift of
the slice. All the 2D maps were interpolated at the positions of the
nodes in the 2D grid of the lowest redshift ($z=2.35$). This way we
have a 3D data cube with RA-Dec pixel size corresponding to
$\sim75\times75$ pkpc at $z=2.38$, and a l.o.s. pixel size equal to
$\delta z \sim 0.002$ (see Sect.~\ref{method}). From now on we
use `pixels' and `grid cells' with the same meaning, referring to
the smallest components of our data cube. We smoothed our data cube in
RA and Dec with a Gaussian filter with sigma equal to 5 pixels. Along
the l.o.s., we used instead a boxcar filter with a depth of 3
pixels. The shape and dimension of the smoothing in RA-Dec was chosen
as a compromise between the two aims of i) smoothing the shapes of the
Voronoi polygons and ii) not washing away the highest density
peaks. The smoothing along the l.o.s. was done to link each redshift
slice with the previous and following slice. Different choices on the
smoothing filters do not significantly affect the 2D maps in terms of
the shapes of the over-dense regions, and have only a minor effect on
the values of $\delta_{\rm gal}$, even if the highest-density peaks
risk to be washed away in case of excessive smoothing.  We produced
data cubes for the lower and upper limits of $\delta_{\rm gal}$
($\delta_{\rm gal,16}$ and $\delta_{\rm gal,84}$) in the same
way. These two latter cubes are used for the treatment of
uncertainties in our following analysis.

\begin{figure} \centering
\includegraphics[width=9.0cm]{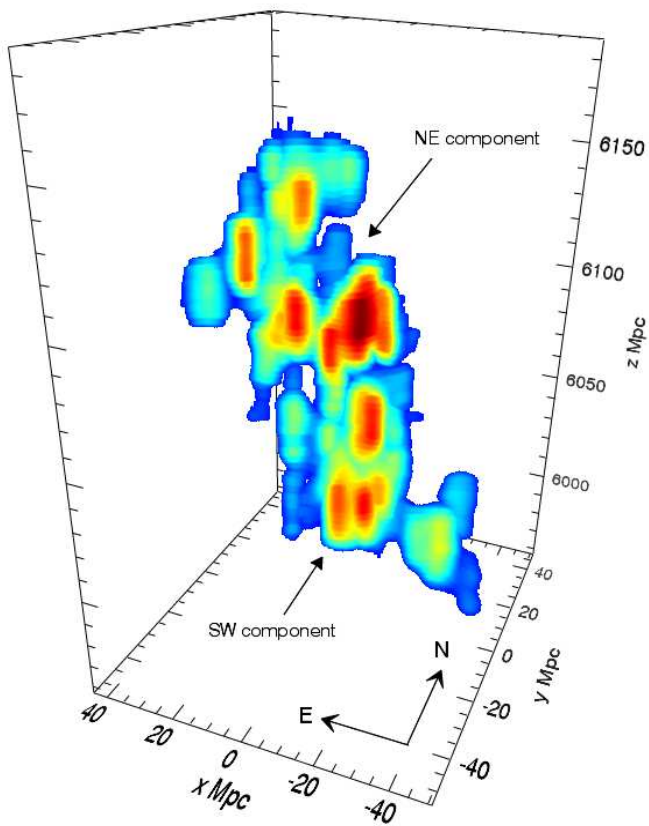}
\caption{3D overdensity map of the Hyperion
  proto-supercluster, in comoving megaparsecs. Colours scale with
  $\rm log(\sigma_{\rm \delta})$, exactly as in Fig.~\ref{2D_maps},
  from blue ($2\sigma_{\rm \delta}$) to the darkest red
  ($\sim8.3\sigma_{\rm \delta}$, the highest measured value in our 3D
  cube).  The $x-$, $y-$ and $z-$axes span the ranges
  $149.6 \leq RA \leq 150.52$, $1.74 \leq Dec \leq2.73$ and
  $2.35 \leq z \leq 2.55$. The NE and SW components are indicated. We
  highlight the fact that this figure shows only the proto-supercluster, and omits
  other less extended and less dense density peaks which fall in the
  plotted volume (see discussion in Sect.~\ref{3D}.)}
\label{3D_cube} 
\end{figure}

Figure \ref{2D_maps} shows that around the main components of the
proto-supercluster there are less extended density peaks. Since we
wanted to focus our attention on the proto-supercluster, we excluded
from our analysis all the density peaks not directly connected to the
main structure. To do this, we proceeded as follows: we started from
the pixels of the 3D grid which are enclosed in the
$2\sigma_{\rm \delta}$ contour of the ``NE'' region in the redshift
slice $2.454<z<2.476$ (right panel of Fig.~\ref{2D_maps}). Starting
from this pixel set, we iteratively searched in the 3D cube for all the
pixels, contiguous to the previous pixels set, with a
${\rm log}(1+\delta_{\rm gal})$ higher than $2\sigma_{\rm \delta}$
above the mean, and we added those pixels to our pixel set. We stopped
the search when there were no more contiguous pixels satisfying the
threshold on ${\rm log}(1+\delta_{\rm gal})$.  In this way we define a
single volume of space enclosed in a $2\sigma_{\rm \delta}$ surface,
and we define our proto-supercluster as the volume of space comprised
within this surface. The final 3D overdensity map of the
proto-supercluster is shown in Fig.~\ref{3D_cube}, with the three axes
in comoving megaparsecs.

The 3D shape of the proto-supercluster is very irregular. The NE
and SW components are clearly at different average redshifts, and
have very different 3D shapes. Figure \ref{3D_cube} also shows that both
components contain some density peaks (visible as the reddest regions
within the $2\sigma_{\rm \delta}$ surface) with a very high average
$\delta_{\rm gal}$.  We discuss the properties of
these peaks in detail in Sect.~\ref{3D_peaks}.

The volume occupied by the proto-supercluster shown in
Fig.~\ref{3D_cube} is about $9.5\times 10^4$ cMpc$^3$ (obtained by
adding up the volume of all the contiguous pixels bounded by the
2$\sigma_{\rm \delta}$ surface), and the average overdensity is
$\langle \delta_{\rm gal} \rangle \sim 1.24$. We can give a rough
estimate of the total mass $M_{\rm tot}$ of the proto-supercluster by
using the formula (see \citealp{steidel98}):

\begin{equation} \displaystyle
M_{\rm tot}=\rho_{\rm m} V (1+\delta_{\rm m}),
\label{eq_mass} 
\end{equation}

\noindent where $\rho_m$ is the comoving matter density, $V$ the
volume\footnote{In \cite{cucciati2014_z29} we corrected the volume of
  the proto-cluster under analysis by a factor which took into account
  the Kaiser effect, which causes the observed volume to be smaller
  than the real one, due to the coherent motions of galaxies towards
  density peaks on large scales. Here we show that we are
  concerned rather by an opposite effect, i.e. our volumes might be
  artificially elongated along the l.o.s..} that encloses the
proto-cluster and $\delta_{\rm m}$ the matter overdensity in our
proto-cluster. We computed $\delta_{\rm m}$ by using the relation
$\delta_{\rm m}=\langle \delta_{\rm gal} \rangle/b$, where $b$ is the
bias factor. Assuming $b=2.55$, as derived in \cite{durkalec15b} at
$z\sim2.5$ with roughly the same VUDS galaxy sample we use here, we
obtain $M_{\rm tot} \sim 4.8\times10^{15}{\rm M}_{\odot}$.  There are
at least two possible sources of uncertainty in this
computation\footnote{Excluding the possible uncertainty on the bias
  factor $b$, which does not depend on our reconstruction of the
  overdensity field. For instance, if we assume $b=2.59$, as derived
  in \cite{bielby13} at $z\sim3$, we obtain a total mass $<1\%$
  smaller.}. The first is the chosen $\sigma_{\rm \delta}$
threshold. If we changed our threshold by $\pm0.2\sigma_{\rm \delta}$
around our adopted value of $2\sigma_{\rm \delta}$,
$\langle \delta_{\rm gal} \rangle$ would vary by $\sim \pm 10\%$ and
the volume would vary by $\sim \pm 17\%$, for a variation of the estimated mass
of $\sim \pm 15\%$ (a higher threshold means a higher
$\langle \delta_{\rm gal} \rangle$ and a smaller volume, with a net
effect of a smaller mass; the opposite holds when we use a lower
threshold). Another source of uncertainty is related to the
uncertainty in the measurement of $\delta_{\rm gal}$ in the 2D
maps. If we had used the 3D cube based on
$\delta_{\rm gal,16}$(/$\delta_{\rm gal,84}$), we would have obtained
$\langle \delta_{\rm gal} \rangle \sim 1.23(/1.26)$ and a volume of
1.06(/0.75)$\times 10^5$ cMpc$^3$, for an overall total mass
$\sim10$\% larger (/ $\sim20$\% smaller). If we sum quadratically the
two uncertainties, the very liberal global statistical error on the
mass measurement is of about $+18\%/-25\%$. Irrespectively of the
errors, it is clear that this structure has assembled an immense mass
($> 2\times10^{15}{\rm M}_{\odot}$) at very early times. This
structure is referred to hereafter as the "Hyperion
proto-supercluster"\footnote{Hyperion, one of the Titans according to
  Greek mythology, is the father of the sun god Helios, to whom the
  Colossus of Rhodes was dedicated.} or simply "Hyperion" (officially
PSC J1001$+$0218) due to its immense size and mass and because one of its
subcomponents (peak [3], see Sect. \ref{peak3}) is broadly
coincident with the Colossus proto-cluster discovered by
\cite{lee2016_colossus}.

We remark that the volume computed in our data cube is most probably
an overestimate, at the very least because it is artificially
elongated along the l.o.s. This elongation is mainly due to 1) the
photometric redshift error ($\Delta z\sim0.1$ for
$\sigma_{zp}=0.03(1+z)$ at $z=2.45$), 2) the depth of the redshift
slices ($\Delta z\sim0.02$) used to produce the density field, and 3) the
velocity dispersion of the member galaxies, which might create the
feature known as the Fingers of God ($\Delta z\sim0.006$ for a
velocity dispersion of $500\kms$).  Although the velocity dispersion
should be important only for virialised sub-structures, these three
factors should all work to surreptitiously increase the dimension of
the structure along the l.o.s. and at the same time decrease the local
overdensity $\delta_{\rm gal}$. In this transformation there is no
mass loss (or, equivalently, the total galaxy counts remain the same,
with galaxies simply spread on a larger volume). Therefore, the total
mass of our structure, computed with Eq.~\ref{eq_mass}, would not
change if we used the real (smaller) volume and the real (higher)
density instead of the elongated volume and its associated lower
overdensity.

We also ran a simple simulation to verify the effects of the depth of
the redshift slices on the elongation.  We built a simple mock galaxy
catalogue at $z=2.5$ following a method similar to that described in
\cite{tomczak17}, a method which is based on injecting a mock galaxy
cluster and galaxy groups onto a sample of mock galaxies that are
intended to mimic the coeval field. As in \cite{tomczak17}, the three
dimensional positions of mock field galaxies are randomly distributed
over the simulated transverse spatial and redshift ranges, with the
number of mock field galaxies set to the number of photometric objects
within an identical volume in COSMOS at $z\sim2.5$ that is devoid of
known proto-structures. Galaxy brightnesses were assigned by sampling
the $K-$band luminosity function of \cite{cirasuolo10}, with cluster
and group galaxies perturbed to slightly brighter luminosities (0.5
and 0.25 mag, respectively). Member galaxies of the mock cluster and
groups were assigned spatial locations based on Gaussian sampling with
$\sigma$ equal to 0.5 and 0.33 h$_{70}^{-1}$ pMpc, respectively, and
were scattered along the l.o.s. by imposing Gaussian velocity
dispersions of 1000 and 500 km s$^{-1}$, respectively. We then applied
a magnitude cut to the mock catalogue similar to that used in our
actual reconstructions, applied a spectroscopic sampling rate of 20\%,
and, for the remainder of the mock galaxies, assigned photometric
redshifts with precision and accuracy identical to those in our
photometric catalogue at the redshift of interest. We then ran the
exact same density field reconstruction and method to identify peaks
as was run on our real data, each time varying the depth of the
redshift slices used. Following this exercise, we observed a smaller
elongation for decreasing slice depth, with a $\sim40\%$ smaller
elongation observed when dropping the slice size from 7.5 to 2.5
pMpc. This result confirmed that we need to correct for the elongation
if we want to give a better estimate of the volume and/or the density
of the structures in our 3D cube. We will apply a correction for the
elongation to the highest density peaks found in the Hyperion
proto-supercluster, as discussed in Sect.~\ref{3D_peaks}.

\begin{figure} \centering
\includegraphics[width=8.0cm]{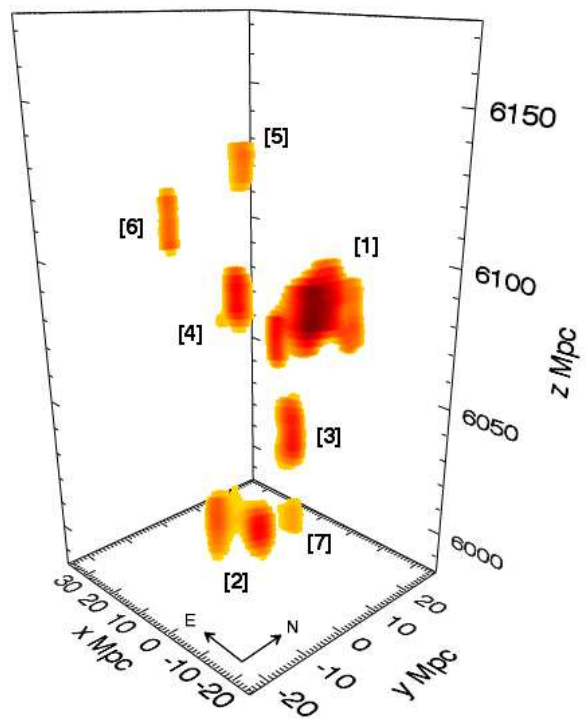}
\caption{Zoom-in of Fig.~\ref{3D_cube}. The angle of view is slightly
  rotated with respect to Fig.~\ref{3D_cube} so as to distinguish all the
  peaks. The colour scale is the same as in Fig.~\ref{3D_cube}, but
  here only the highest density peaks are shown, that is, the 3D volumes
  where ${\rm log}(1+\delta_{\rm gal})$ is above the
  $5\sigma_{\rm \delta}$ threshold discussed in
  Sect.~\ref{3D_peaks}. Peaks are numbered as in
  Fig.~\ref{3D_sph_peaks} and Table~\ref{peaks_tab}. }
\label{3D_cube_peaks} 
\end{figure}

\section{The highest density peaks}\label{3D_peaks}

We identified the highest density peaks in the 3D cube by considering
only the regions of space with ${\rm log}(1+\delta_{\rm gal})$ above
$5\sigma_{\rm \delta}$ from the mean density. In our work, this
threshold corresponds to $\delta_{\rm gal} \sim 2.6$, which corresponds to 
$\delta_m\sim 1 $ when using the bias factor $b=2.55$ found by
\cite{durkalec15b}. We also verified, {\it a posteriori}, that with
this choice we select density peaks which are about to begin or have
just begun to collapse, after the initial phase of
expansion (see Sect.~\ref{discussion}). This is very important if we
want to consider these peaks as proto-clusters.

With the overdensity threshold defined above, we identified seven
separated high-density sub-structures. We show their 3D position and
shape in Fig.~\ref{3D_cube_peaks}. We computed the barycenter of each
peak by weighting the $(x,y,z)$ position of each pixel belonging to
the peak by its $\delta_{\rm gal}$. For each peak, we computed its
volume, its $\langle \delta_{\rm gal} \rangle$, and derived its
$M_{\rm tot}$  using Eq.~\ref{eq_mass} (the bias factor is always
$b=2.55$, found by \citealp{durkalec15b} and discussed in
Sect.~\ref{3D}).  Table \ref{peaks_tab} lists barycenter,
$\langle \delta_{\rm gal} \rangle$, volume, and $M_{\rm tot}$ of the
seven peaks, numbered in order of decreasing $M_{\rm tot}$. We applied
the same peak-finding procedure on the data cubes with
$\delta_{\rm gal,16}$ and $\delta_{\rm gal,84}$, and computed the
total masses of their peaks in the same way. We used these values as
lower and upper uncertainties for the $M_{\rm tot}$ values quoted in the
table.

From Table~\ref{peaks_tab} we see that the overall range of masses
spans a factor of $\sim30$, from $\sim0.09$ to $\sim2.6$ times
$10^{14}$M$_\odot$. The total mass enclosed within the peaks
($\sim5.0 \times 10^{14}$M$_\odot$) is about 10\% of the total mass in the Hyperion 
proto-supercluster, while the volume enclosing all the peaks is a
lower fraction of the volume of the entire proto-supercluster
($\sim6.5\%$), as expected given the higher average overdensity within
the peaks. The most massive peak (peak [1]) is included in the NE
structure, together with peak [4] which has one fifth the total mass
of peak [1]. Peak [2], which corresponds to the SW structure, has
a $M_{\rm tot}$ comparable to peak [4], and it is located at lower
redshift. Peak [3], with a $M_{\rm tot}$ similar to peaks [2] and [4],
is placed in the sort of filament shown in the middle panel of
Fig.~\ref{2D_maps}. At smaller $M_{\rm tot}$ there is peak [5], with
the highest redshift ($z=2.507$), and peak [6], at slightly lower
redshift. They both have
$M_{\rm tot} \sim0.2 \times 10^{14}$M$_\odot$. Finally, peak [7] is
the least massive, and is very close in RA-Dec to peak [2], and at
approximately the same redshift. In Appendix \ref{app_Mtot_sigma} we
show that the computation of $M_{\rm tot}$ is relatively stable if we
slightly change the overdensity threshold used to define the peaks,
with the exception of the least massive peak (peak [7]).

Figure \ref{3D_cube_peaks} shows that the peaks have very different
shapes, from irregular to more compact. We verified that their shape
and position are not possibly driven by spectral sampling issues, by
checking that the peaks persist through the $2^{\prime}$ gaps between
the VIMOS quadrants from VUDS. This also implies that we are not
missing high-density peaks that might fall in the gaps. We remind the
reader that the zCOSMOS-Deep spectroscopic sample, which we use
together with the VUDS sample, has a more uniform distribution in
RA-Dec, and does not present gaps.

Concerning the shape of the peaks, we tried to take into
account the artificial elongation along the l.o.s.. As mentioned at
the end of Sect.~\ref{3D}, this elongation is probably due to the
combined effect of the velocity dispersion of the member galaxies, the
depth of the redshift slices, and the photometric redshift error 
  (although we refer the reader to e.g. \citealp{lovell18} for an
  analysis of the shapes of proto-clusters in simulations). We used a
simple approach to give an approximate statistical estimate of this
elongation, starting from the assumption that on average our peaks
should have roughly the same dimension in the $x$, $y,$ and $z$
dimensions\footnote{This assumption is more suited for a virialised
  object than for a structure in formation. Nevertheless, our approach
  does not intend to be exhaustive, and we just want to compute a
  rough correction. }, and any measured systematic deviation from this
assumption is artificial. In each of the three dimensions we measured
a sort of effective radius $R_e$ defined as
$R_{e,x}=\sqrt{ \sum _{i}w_i(x_{i}-x_{peak})^2 / \sum_i(w_i) }$ (and
similarly for $R_{e,y}$ and $R_{e,z}$), where the sum is over all the
pixels belonging to the given peak, the weight $w_i$ is the value of
$\delta_{\rm gal}$, $x_{i}$ the position in cMpc along the $x-$axis
and $x_{peak}$ is the barycenter of the peak along the $x-$axis, as
listed in Table~\ref{peaks_tab}. We defined the elongation
$E_{\rm z/xy}$ for each peak as the ratio between $R_{e,z}$ and
$R_{e,xy}$, where $R_{e,xy}$ is the mean between $R_{e,x}$ and
$R_{e,y}$. The effective radii and the elongations are reported in
Table~\ref{peaks_elongation_tab}. If the measured volume $V_{\rm meas}$
of our peaks is affected by this artificial elongation, the real
corrected volume is $V_{\rm corr} = V_{\rm meas} / E_{\rm z/xy}$.
Moreover, given that the elongation has the opposite and compensating
effects of increasing the volume and decreasing $\delta_{\rm gal}$, as
discussed at the end of Sect.~\ref{3D}, $M_{\rm tot}$ remains the
same. For this reason, inverting Eq.~\ref{eq_mass} it is possible to
derive the corrected (higher) average overdensity
$\langle \delta_{\rm gal, corr} \rangle$ for each peak, by using
$V_{\rm corr}$ and the mass in Table~\ref{peaks_tab}. $V_{\rm corr}$
and $\langle \delta_{\rm gal, corr} \rangle$ are listed in
Table~\ref{peaks_elongation_tab}. We note that by definition $R_e$ is
smaller than the total radial extent of an overdensity peak, because
it is computed by weighting for the local $\delta_{\rm gal}$, which is
higher for regions closer to the centre of the peak. For this reason,
the $V_{\rm corr}$ values are much larger than the volumes that one
would naively obtain by using $R_{e,xy}$ as intrinsic total radius of
our peaks. We use $\langle \delta_{\rm gal, corr} \rangle$ in
Sect.~\ref{discussion} to discuss the evolution of the peaks. We refer
the reader to \ref{app_elongation} for a discussion on the robustness
of the computation of $E_{\rm z/xy}$ and its empirical dependence on
$R_{e,xy}$.

\begin{figure} \centering
\includegraphics[width=8.0cm]{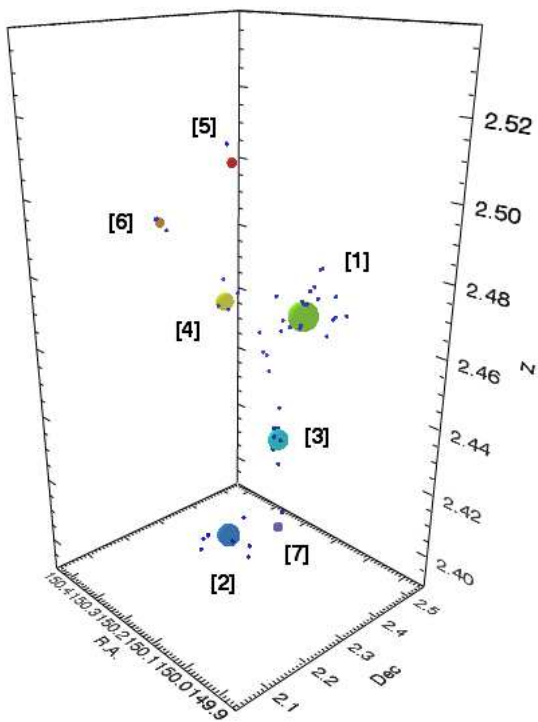}
\caption{Same volume of space as Fig.~\ref{3D_cube_peaks}, but in
  RA-Dec-$z$ coordinates.  Each sphere represents one of the
  overdensity peaks, and is placed at its barycenter (see
  Table~\ref{peaks_tab}). The colour of the spheres scales with
  redshift (blue = low $z$, dark red = high $z$), and the dimension
  scales with the logarithm of $M_{\rm tot}$ quoted in
  Table~\ref{peaks_tab}. Small blue dots are the spectroscopic galaxies
  which are members of each overdensity peak, as described in
  Sect.\ref{3D_peaks}.}
\label{3D_sph_peaks} 
\end{figure}

We also assigned member galaxies to each peak. We defined a
spectroscopic galaxy to be a member of a given density peak if the
given galaxy falls in one of the $\geq 5 \sigma_{\delta}$ pixels that
comprise the peak.  The 3D distribution of the spectroscopic members
is shown in Fig.~\ref{3D_sph_peaks}, where each peak is schematically
represented by a sphere placed in a $(x,y,z)$ position corresponding
to its barycenter. It is evident that the 3D distribution of the
member galaxies mirrors the shape of the peaks (see
Fig.~\ref{3D_cube_peaks}). The number of spectroscopic members
$n_{\rm zs}$ is quoted in Table~\ref{peaks_tab}. The most extended and
massive peak, peak [1], has 24 spectroscopic members. All the other
peaks have a much smaller number of members (from 7 down to even only
one member). We remind the reader that these numbers depend on the
chosen overdensity threshold used to define the peaks, because the
threshold defines the volume occupied by the peaks. Moreover, here we
are counting only spectroscopic galaxies with good quality flags (see
Sect.~\ref{data}) from VUDS and zCOSMOS, excluding other spectroscopic
galaxies identified in the literature (but see Sect.~\ref{vel_disp}
for the inclusion of other samples to compute the velocity
dispersion).

\subsection{Velocity dispersion and virial mass}\label{vel_disp}

We computed the l.o.s. velocity dispersion $\sigma_{\rm v}$ of the
galaxies belonging to each peak. For this computation we used a more
relaxed definition of membership with respect to the one described
above, so as to include also the galaxies residing in the tails of the
velocity distribution of each peak. Basically, we used all the
available good-quality spectroscopic galaxies within $\pm2500\kms$
from $z_{\rm peak}$ comprised in the RA-Dec region corresponding to
the largest extension of the given peak on the plane of the sky.
Moreover, we did not impose any cut in $i-$band magnitude, because, in
principle, all galaxies can serve as reliable tracers of the
underlying velocity field. We also included in this computation the
spectroscopic galaxies with lower quality flag (flag = X1 for VUDS,
all flags with X1.5$\leq$flag$<$2.5 for zCOSMOS), but only if they
could be defined members of the given peak, with membership defined as
at the end of the previous section. This less restrictive choice
allows us to use more galaxies per peak than the pure spectroscopic
members, although we still have only $\leq 4$ galaxies for three of the
peaks. We quote these larger numbers of members in Table~\ref{peaks_tab_veldisp}.

With these galaxies, we computed $\sigma_{\rm v}$ for each peak  by applying the
biweight method (for peak [1]) or the gapper method (for all the other
peaks), and report the results of these computations in
Table~\ref{peaks_tab_veldisp}.  The choice of these methods followed
the discussion in \cite{beers90}, where they show that for the
computation of the scale of a distribution the gapper method is more
robust for a sample of $\lesssim 20$ objects (all our peaks but peak
[1]), while it is better to use the biweight method for $\gtrsim 20$
objects (our peak [1]).  We computed the error on $\sigma_{\rm v}$
with the bootstrap method, which was taken as the reference method in
\cite{beers90}. In the case of peak [7], with only three spectroscopic
galaxies available to compute $\sigma_{\rm v}$, we had to use the
jack-knife method to evaluate the uncertainty on $\sigma_{\rm v}$; see
also Sect.~\ref{app_vdisp_sigma} for more details on $\sigma_{\rm v}$
of peak [7]. 

We found a range of $\sigma_{\rm v}$ between $320\kms$ and
$731\kms$. The most massive peak, peak [1], has the largest velocity
dispersion, but for the other peaks the ranking in $M_{\rm tot}$ is
not the same as in $\sigma_{\rm v}$. The uncertainty on
$\sigma_{\rm v}$ is mainly driven by the number of galaxies used to
compute $\sigma_{\rm v}$ itself, and it ranges from $\sim12$\% for
peak [1] to $\sim65$\% for peak [7], for which we used only three
galaxies to compute $\sigma_{\rm v}$. As we see below, other
identifications in the literature of high-density peaks at the same
redshift cover broadly the same $\sigma_{\rm v}$ range.

As we already mentioned, there are some works in the literature that
identified/followed up some overdensity peaks in the COSMOS field at
$z\sim2.45$, such as for example \cite{casey2015_z247}, \cite{diener2015_z245},
\cite{chiang2015_z244}, and \cite{wang2016_z250}.  Moreover, the
COSMOS field has also been surveyed with spectroscopy by other
campaigns, such as for example the COSMOS AGN spectroscopic survey \citep{trump09}, the MOSDEF survey \citep{kriek15}, and the DEIMOS
10K spectroscopic survey \citep{hasinger18}. We collected the
spectroscopic redshifts of these other samples (including in this
search also much smaller samples, like e.g. the one by
\citealp{perna15}), removed the possible duplicates with our sample
and between samples, and assigned these new objects to our peaks, by
applying the same membership criterion as applied to our
VUDS$+$zCOSMOS sample. We re-computed the velocity dispersion using
our previous sample plus the new members found in the literature. We
note that many objects in the COSMOS field have been observed
spectroscopically multiple times, and in most of the cases the new
redshifts were concordant with previous observations. This is a
further proof of the robustness of the $z_s$ we use here.

In the literature we only find new members for the peaks [1], [3],
[4], and [5]. For each of these peaks, Table \ref{peaks_tab_veldisp}
reports the number $n_{\rm lit}$ of spectroscopic redshifts added to
our original sample, together with the new estimates of
$\sigma_{\rm v}$ and $M_{\rm vir}$. The new $\sigma_{\rm v}$ is always
in very good agreement (below $1\sigma$) with our previous
computation, but it has a smaller uncertainty. We will see that this
translates into new $M_{\rm vir}$ values which are in very good
agreement with those based on the original $\sigma_{\rm v}$.

As a by-product of the use of the
spectroscopic member galaxies, we also computed a second estimate of the
redshift of each peak (after the barycenter, see
above). \cite{beers90} show that the biweight method is the most
robust to compute the central location of a distribution of objects
(in our case, the average redshift) also in the case of
relatively few objects ($5-50$). This central redshift, $z_{\rm BI}$,
is reported in Table~\ref{peaks_tab_veldisp}, and is in excellent
agreement with $z_{\rm peak}$, that is, the barycenter along the
l.o.s. quoted in Table~\ref{peaks_tab}.

The use of the gapper and/or biweight methods is to be favoured when
estimating the scale of a distribution also because they apply
when the distribution is not necessarily a Gaussian, and certainly the shape of
the galaxy velocity distribution in a proto-cluster may not follow a
Gaussian distribution. In addition, it is questionable to
assume that proto-clusters are virialised systems. Nevertheless, a
crude way to estimate the mass of the peaks is to assume the validity of the
virial theorem. In this way we can estimate the virial mass
$M_{\rm vir}$ by using the measured velocity dispersion and some known
scaling relations. We follow the same procedure as \cite{lemaux12},
where $M_{\rm vir}$ is defined as:

\begin{equation} \displaystyle
M_{\rm vir}=\frac{3 \sqrt{3} \sigma_{\rm v}^{3}}{\alpha~ 10~ G~ H(z)}.
\label{mvir} 
\end{equation}

In Eq.~\ref{mvir}, $\sigma_{\rm v}$ is the line of sight velocity
dispersion, $G$ is the gravitational constant, and $H(z)$ is the
Hubble parameter at a given redshift. Equation~\ref{mvir} is derived from
i) the definition of the virial mass,

\begin{equation} \displaystyle
M_{\rm vir}=\frac{3}{G}\sigma_{\rm v}^{2}~R_{\rm v} ,
\label{vir_theo} 
\end{equation}

\noindent where $R_{\rm v}$ is the virial radius; ii) the relation between
$R_{\rm 200}$ and $R_{\rm v}$,

\begin{equation} \displaystyle
R_{\rm 200}=\alpha ~ R_{\rm v},
\label{rvir_r200} 
\end{equation}

\noindent where $R_{\rm 200}$ is the radius within which the density
is 200 times the critical density, and iii) the relation between
$R_{\rm 200}$ and $\sigma_{\rm v}$,

\begin{equation} \displaystyle
R_{\rm 200}=\frac{\sqrt{3}~\sigma_{\rm v}}{10~H(z)}.
\label{r200_sigma} 
\end{equation}

Equations \ref{vir_theo} and \ref{r200_sigma} are from
\cite{carlberg97}. Differently from \cite{lemaux12}, we use
$\alpha\simeq0.93$, which is derived comparing the radii where a NFW
profile with concentration parameter $c=3$ encloses a density 200
times ($R_{\rm 200}$) and 173 times ($R_{\rm v}$) the critical density
at $z\simeq2.45$. Here we consider a structure to be virialised when
its average overdensity is $\Delta_{\rm v} \simeq 173$, which
corresponds, in a $\Lambda$CDM Universe at $z\simeq2.45$, to the more
commonly used value $\Delta_{\rm v} \simeq 178$, constant at all
redshifts in an Einstein-de Sitter Universe (see the discussion in
Sect.~\ref{collapse}).

\begin{figure} \centering
\includegraphics[width=9.0cm]{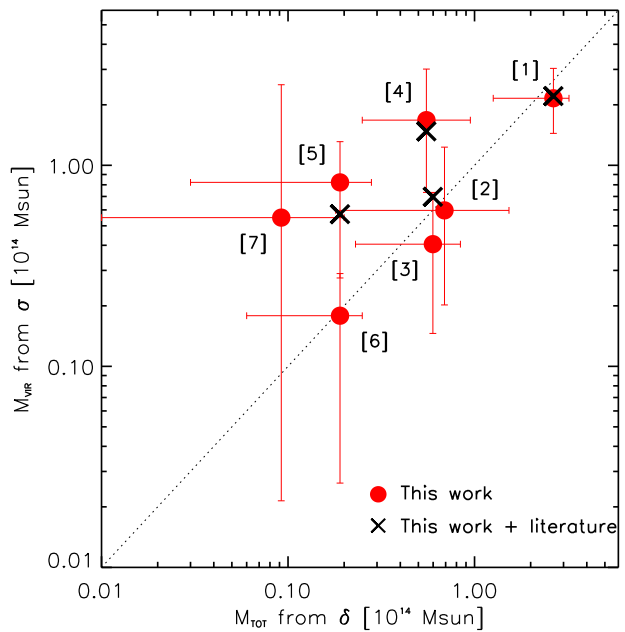}
\caption{Virial mass $M_{\rm vir}$ of the seven identified peaks, as
  in Table~\ref{peaks_tab_veldisp}, vs. the total mass $M_{\rm tot}$
  as in Table~\ref{peaks_tab}. We show both the virial mass computed
  only with our spectroscopic sample (red dots, column 6 of
  Table~\ref{peaks_tab_veldisp}) and how it would change if we add to
  our sample other spectroscopic sources found in the literature
  (black crosses, column 9 of Table~\ref{peaks_tab_veldisp}). Only
  peaks [1], [3], [4], and [5] have this second estimate of
  $M_{\rm vir}$. The dotted line is the bisector, as a reference.}
\label{virial_mass} 
\end{figure}

The virial masses of our density peaks, computed with Eq.~\ref{mvir},
are listed in Table~\ref{peaks_tab_veldisp}, together with the virial
masses obtained from the $\sigma_{\rm v}$ computed by using also other
spectroscopic galaxies in the literature. Figure~\ref{virial_mass}
shows how our $M_{\rm vir}$ compared with the total masses $M_{\rm tot}$ obtained
with Eq.~\ref{eq_mass}. For four of the seven peaks, the two mass
estimates basically lie on the 1:1 relation. In the three other cases,
the virial mass is higher than the mass estimated with the overdensity
value: namely, for peaks [4] and [5] the agreement is at $<2\sigma$,
while for peak [7] the agreement is at less than $1\sigma$
given the very large uncertainty on $M_{\rm vir}$. 

The overall agreement between the two sets of masses is surprisingly
good, considering that $M_{\rm vir}$ is computed under the strong (and
probably incorrect) assumption that the peaks are virialised, and that
$M_{\rm tot}$ is computed above a reasonable but still arbitrary
density threshold. Indeed, although the adopted density threshold
corresponds to selecting peaks which are about to begin or have just
begun to collapse (see Sect.~\ref{3D_peaks}), the evolution of a
density fluctuation from the beginning of collapse to virialisation
can take a few gigayears (see Sect.~\ref{discussion}). Moreover, the
galaxies used to compute $\sigma_v$ and hence $M_{\rm vir}$ are drawn
from slightly larger volumes than the volumes used to compute $M_{\rm
  tot}$, because we included galaxies in the tails of the velocity
distribution along the l.o.s., outside the peaks' volumes. We also
find that $M_{\rm tot}$ continuously varies by changing the
overdensity threshold to define the peaks (see Appendix
\ref{app_Mtot_sigma}), while the computation of the velocity
dispersion in our peaks is very stable if we change this same
threshold (see Appendix \ref{app_vdisp_sigma}). As a consequence, we
do not expect the estimated $M_{\rm vir}$ to change either. In addition to these caveats, peaks [1], [2] and [3] show an irregular 3D shape
(see Appendix \ref{app_peaks}), and they might be multi-component
structures. In these cases, the limited physical meaning of $M_{\rm
  vir}$ is evident.

We also note that peak [5] has already been identified in the
literature as a virialised structure (see \citealp{wang2016_z250} and
our discussion in Sect.~\ref{peak5}), meaning that its $M_{\rm vir}$ is possibly
the most robust among the peaks, but in our reconstruction it is the
most distant from the 1:1 relation between $M_{\rm vir}$ and
$M_{\rm tot}$. This might suggest that our $M_{\rm tot}$ is
underestimated, at least for this peak.

We also remark that there is not a unique scaling relation between
$\sigma_{\rm v}$ and $M_{\rm vir}$. For instance, \cite{munari13}
study the relation between the masses of groups and clusters and their
1D velocity dispersion $\sigma_{\rm 1D}$. Clusters are extracted from $\Lambda$CDM
cosmological N-body and hydrodynamic simulations, and the authors recover the
velocity dispersion by using three different tracers, that is,  dark-matter
particles, sub-halos, and member galaxies. They find a relation in the form:

\begin{equation} \displaystyle
\sigma_{\rm 1D}=A_{1D}\left[ \frac{h(z)~M_{\rm 200}}{10^{15}M_{\odot}} \right]^{\alpha} ,
\label{m200_munari} 
\end{equation}

\noindent where $A_{1D} \simeq 1180 \kms$ and $\alpha \simeq 0.38$, as
from their Fig.~3 for $z=2$ (the highest redshift they consider) and
by using galaxies as tracers for $\sigma_{\rm 1D}$. \cite{evrard08}
find a relation based on the same principle as Eq.~\ref{m200_munari},
but they use DM particles to trace $\sigma_{\rm 1D}$. On the
observational side, \cite{sereno15} find a relation in perfect
agreement with \cite{munari13} by using observed data, with cluster
masses derived via weak lensing. We also used Eq.~\ref{m200_munari} to
compute $M_{\rm vir}$\footnote{First we computed $M_{\rm 200}$ as in
  Eq.~\ref{m200_munari}, then converted $M_{\rm 200}$ into
  $M_{\rm vir}$ based on the same assumptions as for the conversion
  between $R_{\rm 200}$ and $R_{\rm v}$. This gives
  $M_{\rm vir} =1.06~M_{\rm 200}$.}. We found that the $M_{\rm vir}$
computed via Eq.~\ref{m200_munari} are systematically smaller (by
20-40\%) than the previous ones computed with Eq.~\ref{mvir}. This
change would not appreciably affect the high degree of concordance
between $M_{\rm vir}$ and $M_{\rm tot}$ for our peaks.

In summary, the comparison between $M_{\rm vir}$ and $M_{\rm tot}$ is
meaningful only if we fully understand the evolutionary status of our
overdensities and know their intrinsic shapes (and we remind the
reader that in this work the shape of the peaks depends at the very
least on the chosen threshold, and it is not supposed to be their
intrinsic shape).  On the other hand, it would be very interesting to
understand whether it is possible to use this comparison to infer the
level of virialisation of a density peak, provided that its shape is
known. This might be studied with simulations, and we defer this
analysis to a future work.

\subsection{The many components of the proto-supercluster}\label{peak_list}

The COSMOS field is one of the richest fields in terms of data
availability and quality. It was noticed early on that it contains
extended structures at several redshifts (see
e.g. \citealp{scoville07_env,guzzo07_z07,cassata07_z07,kovac10_density,delatorre10_clustering,scoville13_env,iovino16_wall}). Besides
using galaxies as direct tracers, as in the above-mentioned works, the
large-scale structure of the COSMOS field has been revealed with other
methods like weak lensing analysis (e.g. \citealp{massey07}) and
Ly$\alpha$-forest tomography \citep{lee2016_colossus,lee2017_clamato}.
Systematic searches for galaxy groups and clusters have also been
performed up to $z\sim1$ (for instance \citealp{knobel09_groups} and
\citealp{knobel12_groups20k}), and in other works we find compilations
of candidate proto-groups \citep{diener2013_list} and candidate
proto-clusters \citep{chiang2014_list,franck16_CCPC,lee2016_colossus}
at $z\gtrsim1.6$. In some cases, the search for (proto-)clusters was
focused around a given class of objects, like radio galaxies (see
e.g. \citealp{castignani14}).

In particular, it has been found that the volume of space in the
redshift range $2.4 \lesssim z \lesssim 2.5 $ hosts a variety of
high-density peaks, which have been identified by means of different
techniques/galaxy samples, and in some cases as part of dedicated follow-ups of
interesting density peaks found in the previous compilations. Some
examples are the studies by \cite{diener2015_z245}, \cite{chiang2015_z244},
  \cite{casey2015_z247}, \cite{lee2016_colossus}, and \cite{wang2016_z250}. In this paper, we generally refer to the findings in the literature as density
peaks when referring to the ensemble of the previous works; we use the definition adopted in each single paper (e.g. `proto-groups',
`proto-cluster candidates', etc.) when we mention a specific study.

We note that in the vast majority of these previous works there
  was no attempt to put the analysed density peaks in the broader
  context of a large-scale structure. The only exceptions are the
  works by \cite{lee2016_colossus} and \cite{lee2017_clamato}, based
  on the Ly$\alpha$-forest tomography. \cite{lee2016_colossus} explore
  an area of $\sim14\times16$ h$^{-1}$ cMpc, which is roughly one
  ninth of the area covered by Hyperion, while \cite{lee2017_clamato}
  extended the tomographic map up to an area roughly corresponding to
  one third of the area spanned by Hyperion. Both these works do
  mention the complexity and the extension of the overdense region at
  $z\sim 2.45$, and the fact that it embeds three previously
  identified overdensity peaks \citep{diener2015_z245, casey2015_z247,
    wang2016_z250}. Nevertheless, they did not expand on the
  characteristics of this extended region, and were unable to identify the
  much larger extension of Hyperion, because of the smaller explored
  area.

In this section we describe the characteristics of our seven peaks,
and compare our findings with the literature.  The aim of
this comparison is to show that some of the pieces of the Hyperion
proto-supercluster have already been sparsely observed in the
literature, and with our analysis we are able to add new pieces and
put them all together into a comprehensive scenario of a very large
structure in formation.  We also try to give a detailed description of
the characteristics (such as volume, mass, etc.) of the structures
already found in the literature, with the aim to show that different
selection methods are able to find the same very dense structures, but
these methods in some cases are different enough to give disparate
estimates of the peaks' properties. For this comparison, we refer
to Fig.~\ref{3D_map_lit} and Table~\ref{literature_tab}, as detailed
below. Moreover, in Appendix \ref{app_peaks} we show more details on
our four most massive peaks, which we dub ``Theia'', ``Eos'',
``Helios'', and ``Selene''\footnote{According to
  Greek mythology, Theia is a Titaness, sister and spouse of Hyperion. Eos,
Helios, and Selene are their offspring.}. Among the previous findings, we
discuss only those falling in the volume where our peaks are
contained. We remind that we did not make use of the samples used in
these previous works. The only exception is that the zCOSMOS-Deep
sample, included in our data set, was also used by
\cite{diener2013_list}.

\subsubsection{Peak [1] - ``Theia''}\label{peak1}

Peak [1] is by far the most massive of the peaks we detected.  Figure
\ref{3D_cube_peaks} shows that its shape is quite complex. The
peak is composed of two substructures that indeed become two separated
peaks if we increase the threshold for the peak detection
from $5\sigma_{\delta}$ to $6.6\sigma_{\delta}$. In
Fig.~\ref{peak1_fig} of Appendix \ref{app_peaks} we show two 2D
projections of peak [1], which indicate the complexity of the 3D
structure of this peak. 

\begin{figure} \centering
\includegraphics[width=9.0cm]{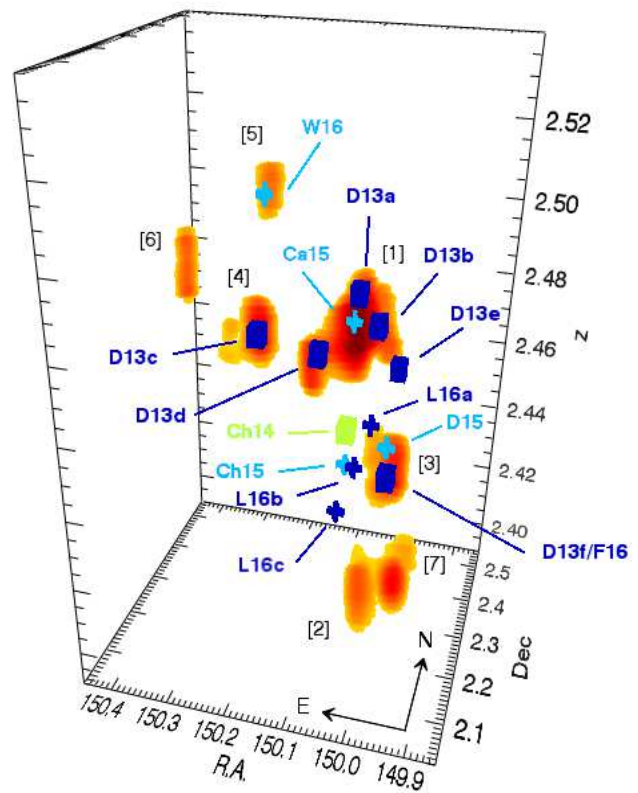}
\caption{Same as Fig.~\ref{3D_cube_peaks}, but in RA-Dec-z
  coordinates. Moreover, we overplot the location of the overdensity
  peaks/proto-clusters/proto-groups detected in other works in the
  literature (blue and green cubes, and blue and cyan
  crosses). Different colours and shapes are used for the symbols for clarity purposes only.  Labels correspond to the IDs in
  Table~\ref{literature_tab}. The dimensions of the symbols are
  arbitrary and do not refer to the extension of the overdensity peaks
  found in the literature. }
\label{3D_map_lit} 
\end{figure}

Figure \ref{3D_map_lit} is the same as Fig.~\ref{3D_cube_peaks}, but
we also added the position of the overdensity peaks found in the
literature. We verified that our peak [1] includes three of the
proto-groups in the compilation by \cite{diener2013_list}, called
D13a, D13b, and D13d in our figure. Proto-goups D13a and D13b are very close to
each other ($\sim 3$ arcmin on the RA-Dec plane) and together they are
part of the main component of our peak [1]. D13d corresponds to the
secondary component of peak [1], which detaches from the main
component when we increase the overdensity threshold to
$6.6\sigma_{\delta}$. Another proto-group (D13e) found by
\cite{diener2013_list} falls just outside the westernmost and northernmost
border of peak[1]. It is not unexpected that our peaks (see also peaks
[3] and [4]) have a good match with the proto-groups found by
\cite{diener2013_list}, given that their density peaks have been
detected using the zCOSMOS-Deep sample, which is also included in our
total sample\footnote{In our case the zCOSMOS-Deep sample, used
  together with the VUDS sample, is cut at $I=25$. Moreover we do not
  use the zCOSMOS-Deep quality flag 1.5. \cite{diener2013_list} used
  also flag=1.5 and did not apply any magnitude cut.}. In our peak [1]
we find 24 spectroscopic members (see Table~\ref{peaks_tab}), 14 of
which come from the VUDS survey and 10 from the zCOSMOS-Deep sample.

The shape of peak[1] (a sort of `L', or triangle) is mirrored by the
shape of the proto-cluster found by \cite{casey2015_z247}, as shown in
their Fig.~2. In our Fig.~\ref{3D_map_lit} their proto-cluster is
marked as Ca15, and we placed it roughly at the coordinates of the
crossing of the two arms of the `L' in their figure, where they found
an X-ray detected source. In their figure, the S-N arm extends to the
north and has a length of $\sim14$ arcmin, and the E-W arm extends
towards east and its length is about 10 arcmin. They also show that
their proto-cluster encloses the three proto-groups D13a, D13b, and
D13d.

Although we found a correspondence between the position/extension of
our peak [1] and the position/extension of some overdensities in the
literature, it is harder to compare the properties of peak [1] and
such overdensities. This difficulty is given mainly by the different
detection techniques. We attempted this comparison and show the
results in Table~\ref{literature_tab}. In this table, for each
overdensity in the literature we show its redshift,
$\delta_{\rm gal}$, velocity dispersion, and total mass, when available
in the respective papers. We also computed its total volume, based on
the information in its respective paper, and computed its
$\delta_{\rm gal}$ and total mass (using Eq.~\ref{eq_mass}) in that
same volume in our 3D cube. In the case of a 1:1 match with our
peak (like in the case of Ca15 and our peak [1]), we also reported the
properties of our matched peak.

In the case of the proto-groups D13a, D13b, D13d and D13e, we found in
the literature only their $\sigma_{\rm v}$, which we cannot compare
directly with our peak [1] given that there is not a 1:1 match. The
$\delta_{\rm gal}$ recovered in our 3D cube in the volumes
corresponding to the four proto-groups are broadly consistent with the
typical $\delta_{\rm gal}$ of our peaks, with the exception of D13e
which in fact falls outside our peak [1]. These proto-groups have all
relatively small volumes and masses compared to our peaks. At most,
the largest one (D13a) is comparable in volume and mass with our
smallest peaks ([5],[6], and [7]). The average difference in volume
between our peaks and the proto-groups found in \cite{diener2013_list}
might be due to the fact that they identified groups with a
Friend-of-Friend algorithm with a linking length of 500 pkpc,
i.e. $\sim1.7$ cMpc at $z=2.45$, which is smaller than the effective
radius of our largest peaks (although their linking lengths and our
effective radii do not have the same physical meaning).

The properties of Ca15 were computed in a volume almost three times as large as our peak
[1]. Nevertheless, its $\delta_{\rm gal}$ is much higher, probably
because of the different tracers (they use dusty star forming
galaxies, `DSFGs'). Despite our lower density in the Ca15 volume, we
find a higher total mass
($M_{\rm tot} = 4.82 \times 10^{14} M_{\odot}$ instead of their total
mass of $>0.8 \times 10^{14} M_{\odot}$). This is probably due to the
different methods used to compute $M_{\rm tot}$: we use
Eq.~\ref{eq_mass}, while \cite{casey2015_z247}  use abundance
matching techniques to assign a halo mass to each galaxy, and then sum
the estimated halo masses for each galaxy in the structure. Moreover,
they state that their mass estimate is a lower limit.

\subsubsection{Peak [2] - ``Eos''}\label{peak2}

As peak [1], this peak seems to be composed by two sub-structures, as
shown in details in Fig.~\ref{peak2_fig}. The two substructures detach
from each other when we increase the overdensity threshold to
$5.3\sigma_{\delta}$. On the contrary, by decreasing the overdensity
threshold to $4.5\sigma{\delta}$ we notice that this peak merges with the current peak [7].

We did not find any direct match of peak [2] with previous detections
of proto-structures in the literature. We note that this part of the
COSMOS field is only partially covered by the tomographic search
performed by \cite{lee2016_colossus} and \cite{lee2017_clamato}. This
could be the reason why they do not find any prominent density peak
there.

\subsubsection{Peak [3] - ``Helios''}\label{peak3}

The detailed shape of peak [3] is shown in Fig.~\ref{peak3_fig}. From
our density field, it is hard to say whether its shape is due to the
presence of two sub-structures. Even by increasing the overdensity
threshold, the peak does not split into two sub-components.

Peak [3] is basically coincident with the group D13f from
\cite{diener2013_list}, and its follow-up by \cite{diener2015_z245},
which we call D15 in our Fig.~\ref{3D_map_lit}. The barycenter of our
peak [3] is closer to the position of D13f than to the position of
D15, on both the RA-Dec plane ($<8^{\prime\prime}$ to D13f,
$\sim50^{\prime\prime}$ on the Dec axis to D15) and the redshift
direction ($\Delta z \sim 0.004$ with D13f, and $\Delta z \sim 0.05$
with D15). This very good match is possibly due also to the fact that
our sample includes the zCOSMOS-Deep data (see comment in
Sect.~\ref{peak1}).  Indeed, out of the seven spectroscopic members that
we identified in peak [3], five come from the zCOSMOS-Deep sample and two
from VUDS.  We note that the list of candidate proto-clusters by
\cite{franck16_CCPC} includes a candidate that corresponds, as stated
by the authors, to D13f. Interestingly, \cite{diener2015_z245}
  mention that D15 might be linked to the radio galaxy COSMOS-FRI 03
  \citep{chiaberge09}, around which \cite{castignani14} found an
  overdensity of photometric redshifts. Although the overdensity of
  photometric redshifts surrounding the radio galaxy is formally at
  slightly lower redshift than D15 (see also \citealp{chiaberge10}),
  it is possibly identifiable with D15, given the photometric redshift
  uncertainty.

Table \ref{literature_tab} shows that the velocity dispersion found by
\cite{diener2015_z245} for D15 is very similar to the one we find for
our peak [3], although the density that they recover is much larger
($\delta_{\rm gal}= 10$ vs $\delta_{\rm gal} \sim 3.$). We note that
D15 is defined over a volume which is almost twice as large as peak
[3]. The velocity dispersion of F16 is instead almost double the one
we recover for peak [3]. Their search volume is huge ($\sim10000$ cMpc$^3$) compared to the volume of peak [3]. Considering
that they also find quite high $\delta_{\rm gal}$, they compute a
total mass of $\sim15\times10^{14}$ M$_\odot$, which is approximately three times
larger than the one we find in our data in their same volume
($4.89\times10^{14}$ M$_\odot$), but about a factor of 30 larger
than the mass of our peak [3].

Very close to peak [3] there are the three components of the extended
proto-cluster dubbed `Colossus' in
\cite{lee2016_colossus}\footnote{\cite{lee2016_colossus} mention that
  from their unsmoothed tomographic map this huge overdensity is
  composed of several lobes (see e.g. their Figs. 4 and 13), but it
  is more continuous after applying a smoothing with a $4 h^{-1}$Mpc
  Gaussian filter.}. Here we call the three sub-structures L16a, L16b
and L16c, in order of decreasing redshift. This proto-cluster was
detected by IGM tomography (see also \citealp{lee2017_clamato})
performed by analysing the spectra of galaxies in the background of
the proto-cluster. The three peaks form a sort of chain from
$z\sim2.435$ to $z\sim2.45$, which extends over $\sim2^{\prime}$ in RA
and $\sim6^{\prime}$ in Dec. We derived the positions of the first and
third peaks from Fig.~12 of \cite{lee2016_colossus}, and assumed that
the intermediate peak was roughly in between (see their Figs. 4 and
13). Neither L16a, L16b, or L16c coincide precisely with one of our
peaks, but they fall roughly 3 arcmin eastwards of the barycenter of
our peak [3]. The declination and redshift of the intermediate
component correspond to those of our peak [3]. Given the extension of
the three peaks in RA-Dec (they have a radius from $\sim2$ to $\sim4$
arcmin) and the extension of our peak [3] ($\sim2$ arcmin radius), the
`Colossus' overlaps with, and it might be identified with, our peak
[3].

\cite{lee2016_colossus} compute the total mass of their overdensity, and
find that it is $1.6\pm0.9\times10^{14}$ M$_\odot$. Computing the
overall mass in the volumes of the three components L16a, L16b, and
L16c in our data cube, we find a smaller mass ($0.83\times10^{14}$
M$_\odot$), which is still consistent with the value found by
\cite{lee2016_colossus}.

We additionally compared our results with those by
\cite{lee2016_colossus} by directly using the smoothed IGM
overdensity, $\delta_F^{\rm sm} $, estimated from the latest
tomographic map \citep{lee2017_clamato}. We measured their average
$\delta_F^{\rm sm}$ in the volume enclosing our peak [3] and found
that this volume of space corresponds to an overdense region with
respect to the mean intergalactic medium (IGM) density at these
redshifts. Specifically, using the definition in
\cite{lee2016_colossus}, for which negative values of
$\delta_F^{\rm sm} $ signify overdense regions, we found that our peak
has $\langle \delta_F^{\rm sm} \rangle \sim -2.4\sigma_{\rm sm}$, with
$\sigma_{\rm sm}$ denoting the effective sigma of the
$\delta_F^{\rm sm} $ distribution. We repeated the same analysis in
the volumes enclosed by our other peaks (with the exception of peak
[2], which lies almost entirely outside the tomographic map), and we
found that their $\langle \delta_F^{\rm sm} \rangle$ fall in the range
from $-1.9\sigma_{\rm sm}$ to $-1\sigma_{\rm sm}$ meaning that all of
our peaks appear overdense with respect to the mean IGM density at
these redshifts. This persistent overdensity measured across the six
peaks that we are able to measure in the tomographic map strongly hint
at a coherent overdensity also present in the IGM maps. Further, all
peaks have measured $\langle \delta_F^{\rm sm} \rangle$ values
consistent with the expected IGM absorption signal due to the presence
of at least some fraction of simulated massive
($M_{\rm tot, z=0}>10^{14} M_{\odot}$) proto-clusters (see section 4
of \citealp{lee2016_colossus}). We note, however, that none of our
peaks have $\langle \delta_F^{\rm sm} \rangle < -3\sigma_{\rm sm}$,
which is the threshold suggested by \cite{lee2016_colossus} to safely
identify proto-clusters (see their Fig. 6) with IGM
tomography. Additionally, the level of the galaxy overdensity or
$M_{\rm tot}$ from our galaxy density reconstruction does not
necessarily correlate well with the $\langle \delta_F^{\rm sm} \rangle$
measured for the ensemble of proto-supercluster peaks likely due to a
variety of astrophysical reasons as well as reasons drawing from the
slight differences in the samples employed and reconstruction
method. Regardless, this comparison demonstrates the complementarity
of our method and IGM tomography to identify proto-clusters. This
comparison will be expanded in future work to investigate differences
in the signals in the two types of maps according to physical
properties (like gas temperature, etc.) of individual proto-clusters.

\cite{lee2016_colossus} identify their proto-cluster with
one of the candidate proto-clusters found by
\cite{chiang2014_list} (proto-cluster referred to here as Ch14). These latter authors
systematically searched for proto-clusters using photometric
redshifts and \cite{chiang2015_z244}
presented a follow-up of Ch14, presenting a proto-cluster that we refer to here  as Ch15. From
\cite{chiang2015_z244}, it is not easy to derive an official RA-Dec
position of Ch15, so we assume it is at the same RA-Dec coordinates as
Ch14. The redshifts of Ch14 and Ch15 are slightly different ($z=2.45$
and $z=2.445$, respectively). Our peak [3] is $\lesssim5$ arcmin away
on the plane of the sky from Ch14 and Ch15, and this is in agreement
with the distance that \cite{chiang2015_z244} mention from their
proto-cluster to the proto-group D15, which matches with our peak
[3]. Moreover, \cite{chiang2015_z244} associate a size of
$\sim10\times7$ arcmin$^2$ to Ch15, which makes Ch15 overlap with peak
[3]. According to \cite{chiang2015_z244}, Ch15 has an overdensity of  LAEs of
$\sim4$ , computed over a volume of $\sim12000$ cMpc$^3$. Over
this volume, the overdensity in our data cube is very low
($\delta_{\rm gal} = 0.53$), because it encompasses also regions well
outside the highest peaks and even outside the
proto-supercluster. Despite the low density, the volume is so huge
that the mass of Ch15 that we compute in our data cube exceeds
$5\times10^{14}$ M$_\odot$. \cite{chiang2015_z244} do not mention 
any mass estimate for Ch15.

\subsubsection{Peak [4] - ``Selene''}\label{peak4}

Peak [4] seems to be composed of a main component, which includes most
of the mass/volume, and a tail on the RA-Dec plane, which is as long as about
twice the length of the main component. This is shown in
Fig.~\ref{peak4_fig}. We did not find spectroscopic members in the
tail.

The barycenter of peak [4], centred on its main component, is
coincident with the position of the proto-group D13c from
\cite{diener2013_list}. Their distance on the plane of the sky is
$\lesssim 30^{\prime\prime}$ arcsec, and they have the same
redshift. Also in this case, this perfect agreement might be due to
our use of the zCOSMOS-Deep sample (see Sect.~\ref{peak1}), although
only half (2 out of 4) of the spectroscopic members of peak [4] come
from the zCOSMOS-Deep survey.

\cite{diener2013_list} compute a velocity dispersion of 239 km
s$^{-1}$ for D13c, while we measured $\sigma_{\rm v}=672$ km s$^{-1}$
for peak [4]. This discrepancy, which holds even if we consider our
uncertainty of $\sim 150$ km s$^{-1}$, might be due to the larger number
of galaxies that we use to compute $\sigma_{\rm v}$ (9 vs. their
3 members). Moreover, the volume over which their proto-group is
defined is much smaller (one seventh) than the volume covered by peak
[4].

\subsubsection{Peak [5]}\label{peak5}

Peak [5] has a regular roundish shape on the RA-Dec plane, so we do
not show any detailed plot in Appendix \ref{app_peaks}; it corresponds
to the cluster found by \cite{wang2016_z250}, which we call W16 in
this work. We remark that \cite{wang2016_z250} find an extended X-ray
emission associated to this cluster, and indeed they define W16 as a
`cluster' and not a `proto-cluster' because they claim that there is
evidence that it is already virialised. We refer to their paper for a
more detailed discussion. The RA-Dec coordinates of W16 are offset by
$\sim30^{\prime\prime}$ on the RA axis and $\sim5^{\prime\prime}$ on
the Dec axis from peak [5]. The redshift of our peak [5] is
$\Delta z=0.001$ higher than the redshift of W16.

The velocity dispersion of our peak [5] is in remarkably good agreement
with the one computed by \cite{wang2016_z250} (533 and 530
km~s$^{-1}$, respectively), and, as a consequence, there is a very
good agreement between the two virial masses. We note that peak [5]
is one of the cases in our work where the total mass computed from
$\delta_{\rm gal}$ is much smaller than the virial mass computed from
the $\sigma_{\rm v}$. What is interesting in W16 is that it is
extremely compact: the extended X-ray detection has a radius of about
$24^{\prime \prime}$, and the majority of its member galaxies are also
concentrated on the same area. Should we consider this small radius,
its volume would be five times smaller than the one of our peak
[5]. Instead, in Table \ref{literature_tab} we used a larger volume
for the comparison (429 cMpc$^3$), derived from the maximum RA-Dec
extension of the member galaxies quoted in \cite{wang2016_z250}.

\subsubsection{Peak [6]}\label{peak6}

Peak [6] has a regular shape on the plane of the sky. We did not find
any other overdensity peak or proto-cluster detected in the literature
matching its position.

\subsubsection{Peak [7]}\label{peak7}

Peak [7] has also a roughly round shape on the RA-Dec plane. It merges
with peak [2] if we decrease the overdensity threshold to
$4.5\sigma{\delta}$. We could not match it with any previous detection
of proto-structures in the literature.

\begin{table*} 
  \caption{Properties of the density peaks identified in
    Fig.~\ref{3D_cube_peaks}, ranked by decreasing $M_{\rm tot}$: (1) ID of
    the peak as in Fig.~\ref{3D_cube_peaks}); (2), (3), (4) are the RA,
    Dec and redshift of the barycenter of the peak; (5) is the number of
    spectroscopic members; (6), (7), and (8) are the average
    $\delta_{\rm gal}$, the total volume, and the total mass $M_{\rm tot}$ of the given
    peak, respectively; $M_{\rm tot}$ is computed by using
    Eq.~\ref{eq_mass}, and its uncertainties are discussed in the text. We
    remind the reader that the properties listed in this table are computed using only pixels and galaxies contained within
    the $5\sigma_{\rm \delta}$ contours.  See Sect.~\ref{3D_peaks} for
    more details.}
\label{peaks_tab} 
\centering 
\begin{tabular}{c c c c c c c c} 
  \hline
  \hline   
  ID & RA$_{\rm peak}$ & Dec$_{\rm peak}$ & $z_{\rm peak}$ & n$_{\rm zs}$ & $\langle \delta_{\rm gal} \rangle $ & Volume  &  $M_{\rm tot}$ \\
  (Fig.~\ref{3D_cube_peaks}) & [deg] & [deg] & & & &[cMpc$^3$]  &   [$10^{14}$ M$_\odot$]    \\
 (1) &  (2) &  (3) &  (4) &  (5) &  (6) &  (7)  & (8) \\
  \hline   
   1 &   150.0937  &    2.4049 &   2.468  & 24 & 3.79 & 3134 &  2.648$_{-1.39}^{+0.56}$  \\      
   2 &   149.9765  &    2.1124 &   2.426  &  7 & 2.89 &  951 &  0.690$_{-0.51}^{+0.84}$  \\       
   3 &   149.9996  &    2.2537 &   2.444  &  7 & 3.03 &  805 &  0.598$_{-0.37}^{+0.24}$  \\       
   4 &   150.2556  &    2.3423 &   2.469  &  4 & 3.20 &  720 &  0.552$_{-0.30}^{+0.40}$  \\       
   5 &   150.2293  &    2.3381 &   2.507  &  1 & 3.11 &  252 &  0.190$_{-0.16}^{+0.09}$  \\       
   6 &   150.3316  &    2.2427 &   2.492  &  4 & 3.12 &  251 &  0.190$_{-0.13}^{+0.06}$  \\       
   7 &   149.9581  &    2.2187 &   2.423  &  1 & 2.58 &  134 &  0.092$_{-0.09}^{+0.11}$  \\       
  \hline
  \hline 
\end{tabular} 
\end{table*}

\begin{table*} 
  \caption{Properties of the density peaks identified in
    Fig.~\ref{3D_cube_peaks}, ranked as in Table~\ref{peaks_tab}: (1) and
    (2) are the ID and redshift of the peaks as in columns (1) and (4) of
    Table~\ref{peaks_tab}; (3), (4), and (5) are the effective radii on the $x-$, $y-$, and $z-$axis, respectively; (6) ratio of the effective radius along the line of sight over the average size in RA-Dec; (7)
    and (8) are the average $\delta_{\rm gal}$ and total volume derived by
    correcting columns (6) and (7) of Table~\ref{peaks_tab} by the
    elongation in column (6) of this table.  See
    Sect.~\ref{3D_peaks} for more details.}
\label{peaks_elongation_tab} 
\centering 
\begin{tabular}{c c c c c c c c } 
  \hline
  \hline   
  ID &  $z_{\rm peak}$ &  R$_{e,x}$ & R$_{e,y}$ & R$_{e,z}$ & $E_{\rm z/xy}$ & $\langle \delta_{\rm gal, corr} \rangle $ & V$_{\rm corr}$  \\
  (Fig.~\ref{3D_cube_peaks}) & & cMpc & cMpc & cMpc & & & [cMpc$^3$]   \\
 (1) &  (2) &  (3) &  (4) &  (5) &  (6) &  (7) &  (8)  \\
  \hline   
   1 &   2.468  & 3.37 & 4.07 & 7.76 &  2.09 & 10.84 &  1500 \\ 
   2 &   2.426  & 2.31 & 3.25 & 5.18 &  1.87 &  7.74 &   509 \\ 
   3 &   2.444  & 1.94 & 1.82 & 6.15 &  3.26 & 15.92 &   247 \\  
   4 &   2.469  & 2.77 & 2.12 & 6.00 &  2.45 & 11.73 &   294 \\  
   5 &   2.507  & 1.05 & 1.27 & 4.07 &  3.52 & 17.70 &    72 \\  
   6 &   2.492  & 0.88 & 1.05 & 5.83 &  6.03 & 32.29 &    42 \\     
   7 &   2.423  & 1.22 & 0.90 & 2.71 &  2.55 & 10.73 &    53 \\ 
  \hline
  \hline 
\end{tabular} 
\end{table*}

\begin{table*} 
  \caption{Properties of the density peaks identified in
    Fig.~\ref{3D_cube_peaks}: (1) and (2) are the ID and redshift as in Table~\ref{peaks_tab}; (3) number of spectroscopic
    galaxies used to compute the velocity dispersion (see Sect.~\ref{vel_disp} for details); (4) redshift
    computed with the biweight method; (5) velocity dispersion computed
    with the biweight method (for peak [1]) and gapper method (all other
    peaks), with their uncertainties estimated with the bootstrap method;
    (6) virial mass computed as described in
    Sect. \ref{3D_peaks}; (7) number of spectroscopic galaxies found in the
    literature and different from the galaxies listed in column (3); (8)
    and (9) are as column (5) and (6) but computed by using the ensemble
    of galaxies of columns (3) and (7); (10) references where the
    literature spectroscopic redshifts are taken from: 1-
    \cite{casey2015_z247}; 2-\cite{kriek15}; 3- \cite{trump09}; 4- \cite{diener2015_z245}; 5-
    \cite{chiang2015_z244}; 6-\cite{perna15}; 7- \cite{wang2016_z250}.  Values in columns
    (4), (5), and (6) are computed using the number of galaxies
    mentioned in column (3). See Sect.~\ref{3D_peaks} for more
    details. $^*$For the velocity dispersion of peak [7] we refer the
    reader to the discussion in Appendix \ref{app_vdisp_sigma}. }
\label{peaks_tab_veldisp} 
\centering 
\begin{tabular}{c c c c c c | c c c c} 
\multicolumn{6}{c}{This work} & \multicolumn{4}{c}{This work + literature}\\
  \hline
  \hline   
  ID & $z_{\rm peak}$ & n$_{zs,\sigma}$ & $z_{\rm BI}$  & $\sigma_{\rm v}$ &  $M_{\rm vir}$ & n$_{\rm lit}$ &  $\sigma_{\rm v}$ &  $M_{\rm vir}$ & Ref.\\
  (Fig.~\ref{3D_cube_peaks}) & & & & [km s$^{-1}$]  & [$10^{14}$ M$_\odot$]  & & [km s$^{-1}$]  & [$10^{14}$ M$_\odot$] & \\
 (1) &  (2) &  (3) &  (4) &  (5) &  (6) &  (7) &  (8) &  (9) &  (10)  \\
  \hline   
   1 &       2.468  & 29 & 2.467 & 731$_{-92}^{+88}$    &   2.16$_{-0.71}^{+0.88}$   & 11  & 737$_{-86}^{+85}$ & 2.21$_{-0.69}^{+0.85}$  & 1,2,3 \\
   2 &       2.426  & 8 & 2.426 & 474$_{-144}^{+129}$    & 0.60$_{-0.40}^{+0.63}$     & - & - &  -   & - \\
   3 &       2.444  & 7 & 2.445 & 417$_{-121}^{+91}$    &  0.41$_{-0.26}^{+0.33}$     & 7 & 500$_{-87}^{+79}$& 0.70$_{-0.30}^{+0.39}$   & 4,5,6 \\
   4 &       2.469  & 9 & 2.467 & 672$_{-162}^{+145}$    & 1.68$_{-0.94}^{+1.33}$     & 1 & 644$_{-158}^{+142}$&  1.47$_{-0.84}^{+1.21}$   & 1 \\
   5 &       2.507  & 4 & 2.508 & 533$_{-163}^{+87}$    &  0.82$_{-0.55}^{+0.49}$     & 13 & 472$_{-80}^{+86}$&  0.57$_{-0.24}^{+0.37}$  & 7 \\
   6 &       2.492  & 4 & 2.490 & 320$_{-151}^{+56}$    &  0.18$_{-0.15}^{+0.11}$     & - & - & -   & - \\
   7$^*$ &   2.423  & 3 & 2.428 & 461$_{-304}^{+304}$    & 0.55$_{-0.53}^{+1.97}$      & - & - & -   & - \\
 \hline
  \hline 
\end{tabular} 
\end{table*}

\begin{sidewaystable*}
  \caption{List of density peaks/proto-clusters/proto-groups already
    found in the literature.  (1) ID of the proto-structures, as the
    labels in Fig.~\ref{3D_map_lit}. (2) References: 1-
    \cite{diener2015_z245}; 2- \cite{casey2015_z247}; 3-
    \cite{chiang2015_z244}; 4- \cite{lee2016_colossus}; 5-
    \cite{wang2016_z250}; 6- \cite{diener2013_list}; 7-
    \cite{chiang2014_list}; 8-\cite{franck16_CCPC}. (3), (4), (5), and
    (6) are the redshift, the overdensity value, the velocity
    dispersion and the total halo mass, taken from the corresponding
    paper, when available; in some cases, the redshift is the central
    redshift of the used redshift slice. When necessary, total masses
    from the literature are converted so as to correspond $h=0.70$.  Column
    (7) is the volume of the overdensity peaks as described in their
    respective papers, while (8) and (9) are the average
    $\delta_{\rm gal}$ and total mass (computed with
    Eq.~\ref{eq_mass}) as computed in our 3D data cube in the volume
    quoted in column (7). (10) matching peak of this work (see also
    the discussions in Sect.~\ref{peak_list}); the asterisks mark the
    cases when the match is not one-to-one, or there is a slight
    mis-match between the centres, and in these cases we quote our
    closest peak, as discussed in Sect.~\ref{peak_list}.  Columns from
    (11) to (15) are average overdensity, volume, $M_{\rm tot}$,
    $\sigma_{\rm v}$ and $M_{\rm vir}$ of the matching peak in this
    work (see Tables~\ref{peaks_tab} and \ref{peaks_tab_veldisp}) in
    the cases of a clear match. Notes: $^{a}$ The overdensity is
    computed using LAE galaxies. $^{b}$ The three subcomponents L16a,
    L16b and L16c are treated together as one single proto-cluster by
    \cite{lee2016_colossus} when they compute the total mass, so the
    quoted number is the overall mass comprising the three components,
    for both the values in their paper (column 6) and as recovered in
    this work (columns 7, 8 and 9). $^{c}$ The overdensity is computed
    using DSFG galaxies. $^{d}$ The first mass is the overdensity
    mass, the second the virial mass. $^{e}$ For the sake of clarity,
    we omit the uncertainties, which are already reported in the
    previous tables. }
\label{literature_tab} 
\centering 
\begin{tabular}{c c c c c c | c c c c c c c c c} 
  \hline
\multicolumn{6}{c|}{Literature} & \multicolumn{9}{c}{From this work}\\
  \hline   
  ID & Ref. & z &  $\delta_{\rm gal}$ & $\sigma_{\rm v}$ & M$_{\rm tot}$  &  Volume &  $\langle \delta_{\rm gal} \rangle$  &   M$_{\rm tot}$ &   Match with & $\langle \delta_{\rm gal} \rangle$ & Volume & M$_{\rm tot}$$^{e}$ & $\sigma_{\rm v}$$^{e}$ & M$_{\rm vir}$$^{e}$ \\
  (Fig.~\ref{3D_map_lit}) & & & & [km s$^{-1}$] & [$10^{14}$ M$_\odot$] &                 cMpc$^3$                     &   &   [$10^{14}$ M$_\odot$]&   this work &  & cMpc$^3$ &  [$10^{14}$ M$_\odot$]  &         [km s$^{-1}$] & [$10^{14}$ M$_\odot$] \\

(1)  & (2) &   (3)    &  (4)                &  (5)         &  (6)               & (7)        &  (8)     &  (9)       &  (10) &  (11) &  (12) &  (13) &  (14) &  (15)  \\
\hline
L16a &  4  &   2.450  &    -                &  -           & 1.6$\pm$0.9$^{b}$   & 1568$^{b}$   & 1.50$^{b}$ & 0.83$^{b}$ & [3]*   & -     &   -    &  -     & -   &    -  \\     
L16b &  4  &   2.443  &    -                &  -           & 1.6$\pm$0.9$^{b}$   & 1568$^{b}$   & 1.50$^{b}$ & 0.83$^{b}$ & [3]*  & -     &   -    &  -     & -   &    -  \\    
L16c &  4  &   2.435  &    -                &  -           & 1.6$\pm$0.9$^{b}$   & 1568$^{b}$   & 1.50$^{b}$ & 0.83$^{b}$ & [3]*  & -     &   -    &  -     & -   &   -   \\     
W16  &  5  &   2.506  &    -                & 530$\pm120$  & 0.79$^{+0.46}_{-0.29}$ & 429         & 2.46      & 0.29       & [5]   & 3.11  & 252   & 0.190   & 533 & 0.82  \\    
F16  &  8  &   2.442  & 9.27$\pm$4.93       & 770          & 15.5/14.1$^{d}$      & $\sim10000$ & 1.04      &  4.89 & [3]   & 3.03  & 805   & 0.598  &  417 &  0.41 \\  
D15  &  1  &   2.450  & 10                  & 426          &   -                 & 1513        & 1.99       &  0.92     & [3]   & 3.03  & 805   &  0.598  & 417 & 0.41   \\ 
Ca15 &  2  &   2.472  & 11$^{c}$             &  -           & $>0.8\pm0.3$         & 8839       & 1.55       &  4.82       & [1]   & 3.79  & 3134  & 2.648   & 731 & 2.16  \\   
Ch15 &  3  &   2.440  &  4$^{a}$             &  -           &   -                 & $\sim12000$ &  0.53      & $\sim5.6$ & [3]*  & -     &   -   &   -     & -    &   -   \\  
Ch14 &  7  &   2.450  & 1.34$^{+0.49}_{-0.40}$ & -             & -                   & $\sim23000$ &  0.37    & $\sim9.1$  & [3]*  &  -    &  -    &   -     & -    & -     \\  
D13a &  6  &   2.476  &  -                  & 264          & -                   & 87          &  3.12      & 0.07      & [1]*  & -     &  -    &   -     & -    & -     \\
D13b &  6  &   2.469  &  -                  & 488          & -                   & 253         &  3.73      & 0.21      & [1]*  & -     &  -    &   -     & -    &  -    \\  
D13c &  6  &   2.469  &  -                  & 239          & -                   & 108         &  4.26      & 0.10      & [4]   & 3.20  & 720   & 0.552   & 672  &  1.68  \\  
D13d &  6  &   2.463  &  -                  &  30          & -                   & 26          &  4.08      & 0.02      & [1]*  & -     &  -    &    -    & -    & -     \\  
D13e &  6  &   2.452  &  -                  & 476          & -                   & 38          &  0.89      & 0.02      & [1]*  & -     &  -    &    -    & -    & -     \\  
D13f &  6  &   2.440  &  -                  & 526          & -                   & 425         &  2.87      & 0.31      & [3]   & 3.03  & 805   & 0.598   &  417 & 0.41  \\

  \hline
  \hline 
\end{tabular} 
\end{sidewaystable*}


\section{Discussion}\label{discussion}

The detection of such a huge, massive structure, caught during its
formation, poses challenging questions. On the one hand, one would
like to know whether we can predict the evolution of its
components. On the other, it would be interesting to understand
whether at least some of these components are going to
interact with one another, or at the very least, how much they are
going to interact with the surrounding large-scale structure as a
whole. Moreover, the existence of superclusters at lower redshifts
begs the question of whether this proto-structure will evolve to
become similar to one of these closer superclusters. We address
these issues below in a qualitative way, and defer any further
analysis to a future work.

\subsection{The evolution of the individual density peaks.}\label{collapse}

Assuming the framework of the spherical collapse model, we computed
the evolution of our overdensity peaks as if they were isolated
spherical overdensities. This is clearly a significant assumption (see
e.g. \citealp{despali13} for the evolution of ellipsoidal halos), but
it can help us in roughly understanding the evolutionary status of
these peaks, and how peaks with similar overdensities would evolve
with time.

According to the spherical collapse model, any spherical overdensity
will evolve like a sub-universe, with a matter-energy density higher
than the critical overdensity at any given epoch.  In our case, we
reasonably assume that the average matter overdensity
$\langle \delta_{\rm m} \rangle $ in our peaks corresponds to a
non-linear regime, because it is already well above 1. We report
$\langle \delta_{\rm m} \rangle $ in Table~\ref{peaks_evol_tab} as
$\langle \delta_{\rm m,corr} \rangle $, given that we define it as
$\langle \delta_{\rm m,corr} \rangle = \langle \delta_{\rm gal,corr}
\rangle / b$,
with $\langle \delta_{\rm gal,corr} \rangle $ as reported in
Table~\ref{peaks_elongation_tab} and $b$ the bias measured by
\cite{durkalec15b} as in Sect.~\ref{3D}.

Given that it is much easier to compute the
evolution of an overdensity in linear regime than in non-linear
regime, we transform \citep{padmanabhan} our
$\langle \delta_{\rm NL} \rangle $ into their corresponding values in
linear regime, $\langle \delta_{\rm L} \rangle $, and make them evolve
according to the spherical linear collapse model.

In particular, the overdense sphere passes through three specific
evolutionary steps. The first one is the point of turn-around, when
the overdense sphere stops expanding and starts collapsing, becoming a
gravitationally bound structure. This happens when the overdensity in
linear regime is $\delta_{\rm L,ta}\simeq 1.062$ (in non-linear regime
it would be $\delta_{\rm NL,ta}\simeq 4.55$). After the turn-around,
when the radius of the sphere becomes half of the radius at
turn-around, the overdense sphere reaches the virialisation. In this
moment, we have $\delta_{\rm L,vir}\simeq 1.58$ and
$\delta_{\rm NL,vir}\simeq 146$. The sphere then continues the
collapse process, till the moment of maximum collapse which
theoretically happens when its radius becomes zero with an infinite
density. In the real universe the collapse stops before the density
becomes infinite, and at that time the system, which still
satisfies the virial theorem, reaches $\delta_{\rm L,c}\simeq 1.686$
($\delta_{\rm NL,c}\simeq 178$).

In our work we are interested in the moments of turn-around and
collapse.  Here we will follow the formalism as in \cite{pace10}, and
we will use the symbol $\delta_{\rm c}$ for
$\delta_{\rm L,c}\simeq 1.686$ and the symbol $\Delta_{\rm V}$
for $\delta_{\rm NL,c}\simeq 178$. When we refer to the
time(/redshift) of turn-around and collapse, we use
$t_{\rm ta}$(/$z_{\rm ta}$) and $t_{\rm c}$(/$z_{\rm c}$).

We reiterate that $\delta_{\rm c}$ and $\Delta_{\rm V}$ are constant
with redshift in an Einstein - de Sitter (EdS) Universe, while they
evolve with time in a $\Lambda$CDM cosmology, and their evolution
depends on the relative contribution of $\Omega_{\rm \Lambda}(z)$ and
$\Omega_{\rm m}(z)$ to $\Omega_{\rm tot}(z)$. At high redshift (e.g.
$z=5$) when $\Omega_{\rm \Lambda}(z)$ is small, $\delta_{\rm c}$ and
$\Delta_{\rm V}$ are close to their EdS counterparts. As time goes by,
$\Omega_{\rm \Lambda}(z)$ increases and both $\delta_{\rm c}$ and
$\Delta_{\rm V}$ decrease.  This is shown, for instance, in
\cite{pace10}, where they show that $\delta_{\rm c}$ decreases by less
than 1\% from $z=5$ to $z=0$, while in the same timescale
$\Delta_{\rm V}$ decreases from $\sim178$ to $\sim 100$ (see also
\citealp{bryan_norman98}, where they use the symbol $\Delta_{\rm c}$
instead of $\Delta_{\rm V}$). In our work we allow our overdensities
to evolve in the linear regime, so we are interested at the time
when they reach $\delta_{\rm c}$. Given its small evolution with
redshift, we consider it a constant, set as in the EdS universe.

The evolution of a fluctuation is given by its growing mode
$D_{\rm +}(z)$. At a given redshift $z_{\rm 2}$, the overdensity
$\delta_{\rm L}(z_{\rm 2})$ can be computed knowing the overdensity at
another redshift $z_{\rm 1}$ and the value of the growing mode at the
two redshifts, as follows:

\begin{equation} \displaystyle
\delta_{\rm L}(z_{\rm 2}) = \delta_{\rm L}(z_{\rm 1}) \frac{D_{\rm +}(z_{\rm 2})}{D_{\rm +}(z_{\rm 1})}
\label{delta_evol} 
.\end{equation}

\begin{figure} \centering
\includegraphics[width=9.0cm]{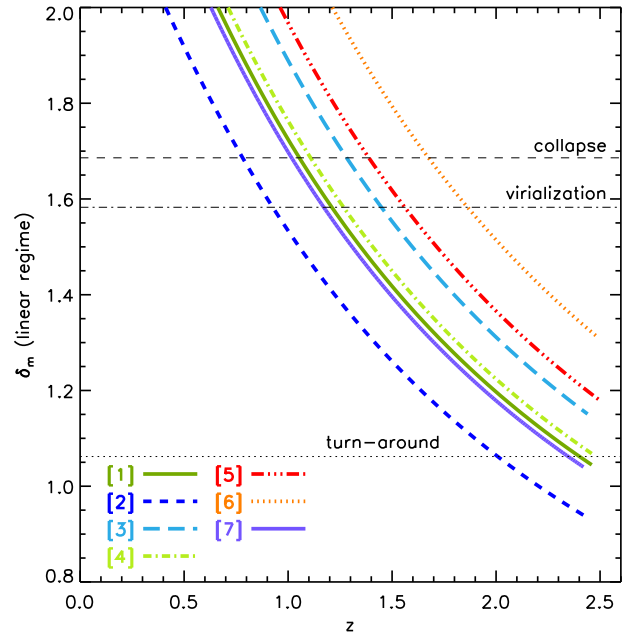}
\caption{Evolution of $\delta_m$ for the seven peaks listed in
  Table~\ref{peaks_evol_tab}, with different line styles as in the
  legend. The evolution is computed in a linear regime for a
  $\Lambda$CDM Universe. For each peak, we start tracking the
  evolution from the redshift of observation (column 2 in
  Table~\ref{peaks_evol_tab}), and we consider as starting
  $\delta_{\rm m}$ the one computed from the corrected
  $\langle \delta_{\rm gal,corr} \rangle$ (column 7 in
  Table~\ref{peaks_elongation_tab}) and transformed into linear
  regime. The horizontal lines represent
  $\delta_{\rm L,ta}\simeq 1.062$, $\delta_{\rm L,vir}\simeq 1.58$ and
  $\delta_{\rm L,c}\simeq 1.686$.  See Sect.~\ref{collapse} for more
  details.}
\label{delta_evol_ps} 
\end{figure}

In a $\Lambda$CDM universe, we define the linear growth factor $g$ as
$g \equiv D_{\rm +}(z)/a$, where $a=(1+z)^{-1}$ is the cosmic scale
factor. By using an approximate expression for $g$ (see
e.g. \citealp{carroll92} and \citealp{hamilton01}), which depends on
$\Omega_{\rm \Lambda}(z)$ and $\Omega_{\rm m}(z)$, we can recover
$D_{\rm +}(z)$ and with equation \ref{delta_evol} derive the time when
our peaks reach $\delta_{\rm L,ta}$ and $\delta_{\rm c}$, starting
from the measured values of $\delta_{\rm L}(z_{\rm obs})$, with
$z_{\rm obs}$ being the redshifts given in
Table~\ref{peaks_tab}. Figure \ref{delta_evol_ps} shows the evolution
of the density contrast of our peaks. In Table~\ref{peaks_evol_tab} we
list the values of $z_{\rm ta}$ and $z_{\rm c}$, together with the
time elapsed from $z_{\rm obs}$ to these two redshifts. As a very
rough comparison, if we considered the entire Hyperion
proto-supercluster with its
$\langle \delta_{\rm gal} \rangle \sim 1.24$ (Sect.~\ref{3D}), and assumed
an elongation equal to the average elongation of the peaks to derive
its $\langle \delta_{\rm gal,corr} \rangle$ and then its
$\langle \delta_{\rm m,corr} \rangle$, the proto-supercluster would
have $\delta_{\rm L} \lesssim 0.8$ at $z=2.46$ (to be compared with
the y-axis of Fig.~\ref{delta_evol_ps}).

We note that the evolutionary status of the peaks depends by
definition on their average density, that is, the higher the density, the
more evolved the overdensity perturbation. The most evolved is peak
[6], which has $\langle \delta_{\rm m,corr} \rangle = 12.66$, almost
twice as large as the second densest peak (peak [5]). According to
the spherical collapse model, peak [6] will be a virialised system by
$z\sim1.7$, that is, in $1.3$ Gyr from the epoch of observation. The least
evolved is peak [2], that will take 0.6 Gyr to reach the turn-around
and then another $\sim3.8$ Gyr to virialise.

This simple exercise, which is based on a strong assumption, shows
that the peaks are possibly at different stages of their evolution,
and will become virialised structures at very different times. In
reality, the peaks' evolution will be more complex, given that they
will possibly accrete mass/subcomponents/galaxies during their
lifetimes, and these results make it desirable to study how we can
combine the density-driven evolution of the individual peaks with
the overall evolution of the Hyperion
proto-supercluster as a whole. Moreover,
by comparing the evolutionary status of each peak with the average
properties of its member galaxies, it will be possible to study the
co-evolution of galaxies and the environment in which they reside.  We
defer these analyses to future works.

\begin{table} 
  \caption{Evolution of the density peaks according to the spherical
    collapse model in linear regime. Columns (1) and (2) are the ID and the
    redshift of the peak, as in Table~\ref{peaks_elongation_tab}. Column (3) is the
    average matter overdensity derived from the average galaxy
    overdensity of column (7) of Table~\ref{peaks_elongation_tab}. Columns (4) and (5)
    are the redshifts when the overdensity reaches the overdensity of
    turn-around and collapse, respectively. Columns (6) and (7) are the
    corresponding time intervals $\Delta t$ since the redshift of observation $z_{\rm obs}$ (column 2) to the redshifts of turn-around and collapse. When
    $z_{\rm ta} < z_{\rm obs}$ the turn-around
    has already been reached before the redshift of observation, and in
    these cases the corresponding $\Delta t $ have not been
    computed. See Sect.~\ref{collapse} for more details.}
\label{peaks_evol_tab} 
\centering 
\begin{tabular}{c c | c c c c c} 
  \hline
  \hline   
  ID & $z$ &  $\langle \delta_{\rm m,corr} \rangle $ & $z_{\rm ta}$  & $z_{\rm c}$   & $\Delta t_{\rm ta}$ &  $\Delta t_{\rm c}$ \\
     &     &      &      &      &  [Gyr] &  [Gyr] \\
 (1) &  (2) &  (3) &  (4) &  (5) &  (6) &  (7) \\
  \hline   
   1 &   2.468  &  4.25  &         2.402  & 1.054 &      0.08 &       3.16 \\
   2 &   2.426  &  2.04  &         2.001  & 0.781 &      0.60 &       4.37 \\
   3 &   2.444  &  6.24  &  $>z_{\rm obs}$  & 1.282 &         - &       2.32 \\
   4 &   2.469  &  4.60  &  $>z_{\rm obs}$  & 1.108 &         - &       2.95 \\
   5 &   2.507  &  6.94  &  $>z_{\rm obs}$  & 1.388 &         - &       2.07 \\
   6 &   2.492  & 12.66  &  $>z_{\rm obs}$  & 1.675 &         - &       1.33 \\
   7 &   2.423  &  4.21  &         2.347  & 1.017 &      0.10 &       3.26 \\
  \hline
  \hline 
\end{tabular} 
\end{table}

\subsection{The proto-supercluster as a whole.}\label{whole_psc}

In the previous section we pretended that the peaks were isolated density
fluctuations and traced their evolution in the absence of
interactions with other components of the proto-supercluster. This is
an oversimplification, because several kinds of interactions are
likely to happen in such a large structure, such as for example accretion of
smaller groups along filaments onto the most dense peaks, as expected
in a $\Lambda$CDM universe.

For instance, for what concerns merger events between proto-clusters,
\cite{lee2016_colossus} examined the merger trees of some of the
density peaks that they identified in realistic mock data sets by
applying the same 3D Ly$\alpha$ forest tomographic mapping that they
applied to the COSMOS field. They found that in the examined mocks,
very few of the proto-structures identified by the tomography at
$z\sim2.4$ and with an elongated shape (such as the `chain' of their
peaks L16a, L16b, and L16c discussed in Sect.~\ref{peak3}) are going to
collapse to one single cluster at z=0.  Similarly, \cite{topping18}
analysed the Small MultiDark Planck Simulation in search for $z\sim3$
massive proto-clusters with a double peak in the galaxy velocity
distribution and with the two peaks separated by about $2000\kms$,
like the one they identified in previous observations
\citep{topping16}. They found that such double-peaked overdensities
are not going to merge into a single cluster at $z=0$.

The structures found by \cite{lee2016_colossus} and \cite{topping16}
are much smaller and with simpler shapes compared to the Hyperion
proto-supercluster, and yet they are unlikely to form a single cluster
at z=0, according to simulations. Interestingly, \cite{topping18} also
found that in their simulation the presence of two massive peaks
separated by $2000\kms$ is a very rare event (one in
$7.4h^3$Gpc$^{-3}$) at $z\sim3$.  These findings indicate that the
evolution of the Hyperion proto-supercluster cannot be simplified as
series of merging events, and that the identification of
massive/complex proto-clusters at high redshift could be useful to
give constraints on dark matter simulations.

Indeed, it would be interesting to know whether or not Hyperion could
be the progenitor of known lower-redshift superclusters. One
difficulty is that there is no unique definition of a supercluster
(but see e.g. \citealp{chon15} for an attempt), and the taxonomy of
known superclusters up to $z\sim1.3$ spans wide ranges of mass (from a
few $10^{14}M_{\odot}$ as in \citealp{swinbank07} to
$>10^{16}M_{\odot}$ as in \citealp{bagchi17}), dimension (a few cMpc
as in \citealp{rosati99_lynx} or $\sim 100$ cMpc as in
\citealp{kim16_sc}), morphology (compact as in \citealp{gilbank08}, or
with multiple overdensities as in \citealp{lubin00,lemaux12}),
and evolutionary status (embedding collapsing cores as in
\citealp{einasto16_SGW} or already virialised clusters as in
\citealp{rumbaugh18}).  This holds also for the well-known
superclusters in the local universe (see e.g.
\citealp{shapley30,shapley34_hercules,delapparent86_greatwall,
  haynes86_pp}), not to mention the category of the so-called Great
Walls, which are sometimes defined as systems of superclusters (like
e.g. the Sloan Great Wall, \citealp{vogeley04, gott05}, and the Boss
Great Wall, \citealp{lietzen16_sc}).

Clearly, Hyperion shares many characteristics with the
above-mentioned superclusters, making it likely that its eventual fate
will be to become a supercluster. A further step would be identifying
which known supercluster is most likely to be similar to the potential
descendant(s) of Hyperion. This would be surely an
important step in understanding how the large-scale structure of the
universe evolves and how it affects galaxy evolution. On the other
hand, it is also interesting to study the likelihood of such
(proto-) superclusters existing in a given cosmological volume, given
their volumes and masses (see e.g. \citealp{sheth11}). For instance,
\cite{lim14} show that the relative abundance of rich superclusters
at a given epoch could be used as a powerful cosmological probe. 

From \cite{lim14} we can qualitatively assess how many superclusters
of the kind that we detect are expected in the volume probed by
VUDS. \cite{lim14} derive the mass function of superclusters, defined
as clusters of clusters according to a Friend of Friend
algorithm. Since the supercluster mass function at $z\sim2.5$ was not
explicitly studied, we adopt here expectations from their study of the
$z=1$ supercluster mass function keeping in mind that this expectation
will be a severe upper limit given that the halo mass function at the
high-mass end decreases by a factor of $\ga100$ from $z=1$ to $z=2.5$
(see, e.g. \citealp{percival05}). With this in mind, we estimate,
using those results of \cite{lim14} that employ a similar cosmology
to the one used in this study, the extreme upper limit to the
number of superclusters with a total mass $>5\times10^{14}$
M$_{\odot}$ expected within the RA-Dec area studied in this paper and
in the redshift range $2<z<4$ to be $\sim$4. We consider this mass
limit, $>5\times10^{14}$ M$_{\odot}$, because it is the sum of the
masses of our peaks, similarly to how they compute the masses of their
superclusters. The extremeness of this upper limit is such that much
more precise comparisons need to be made.  We defer the detailed analysis of number counts and evolution of
proto-superclusters at $z\sim2.5$ in simulated cosmological volumes to a future
work.

\section{Summary and conclusions}\label{summary}

Thanks to the spectroscopic redshifts of VUDS, together with the zCOSMOS-Deep spectroscopic sample, we
unveiled the complex shape of a proto-supercluster at $z\sim2.45$ in
the COSMOS field. We computed the 3D overdensity field over a volume of
$\sim100\times100\times250$ comoving Mpc$^3$ by applying a Voronoi
tessellation technique in overlapping redshift slices. The tracers
catalogue comprises our spectroscopic sample complemented by
photometric redshifts for the galaxies without spectroscopic
redshift. Both spectroscopic and photometric redshifts were treated
statistically, according to their quality flag or their measurement
error, respectively. The main advantage of the Voronoi Tessellation is
that the local density is measured both on an adaptive scale and with
an adaptive filter shape, allowing us to follow the natural
distribution of tracers.  In the explored volume, we identified a
proto-supercluster, dubbed ``Hyperion" for its
  immense size and mass,  extended over a volume of
$\sim60\times60\times150$ comoving Mpc$^3$. We estimated its total
mass to be $\sim 4.8\times 10^{15}{\rm M}_{\odot}$.  Within this
immensely complex structure, we identified seven density peaks in the
range $2.40<z<2.5$, connected by filaments that exceed the average
density of the volume. We analysed the properties of the peaks, as
follows:

\begin{itemize}

\item[-] We estimated the total mass of the individual peaks,
$M_{\rm tot}$, based on their average galaxy density, and found a range of
masses from $\sim 0.1\times 10^{14}{\rm M}_{\odot}$ to
$\sim 2.7\times 10^{14}{\rm M}_{\odot}$. 

\item[-] By assigning spectroscopic members to each peak, we estimated
  the velocity dispersion of the galaxies in the peaks, and then their
  virial mass $M_{\rm vir}$ (under the admittedly strong assumption that they are
  virialised). The agreement between $M_{\rm vir}$ and $M_{\rm tot}$
  is surprisingly good, considering that (almost all) the peaks are
  most probably not yet virialised. 

\item[-] If we assume that the peaks are going to evolve separately,
  without accretion/merger events, the spherical collapse model
  predicts that these peaks have already started or are about to start
  their collapse phase (`turn-around'), and they will all be
  virialised by redshift $z\sim0.8$.

\item[-] We finally performed a careful comparison with the
  literature, given that some smaller components of this proto-supercluster
  had previously been identified in other works using heterogeneous
  galaxy samples (LAEs, 3D Ly$\alpha$ forest tomography, sub-mm
  starbursting galaxies, CO emitting galaxies). In some cases we found
  a one-to-one match between previous findings and our peaks, in other
  cases the match is disputable. We note that a direct comparison is
  often difficult because of the different methods/filters used to
  identify proto-clusters.

\end{itemize}

In summary, with VUDS we obtained, for the first time 
across the central $\sim1$ deg$^2$ of the COSMOS field, a panoramic view
of this large structure that encompasses, connects, and considerably
expands on all previous detections of the various sub-components. The
characteristics of the Hyperion proto-supercluster (its redshift, its richness
over a large volume, the clear detection of its sub-components),
together with the extensive band coverage granted by the COSMOS field,
provide us the unique possibility to study a rich supercluster in
formation 11 billion years ago.

This impressive structure deserves a more detailed analysis. On the
one hand, it would be interesting to compare its mass and volume with
similar findings in simulations, because the relative abundance of
superclusters could be used to probe deviations from the predictions
of the standard $\Lambda$CDM model. On the other hand, it is
crucial to obtain a more complete census of the galaxies residing in
the proto-supercluster and its surroundings. With this new data, it
would be possible to study the co-evolution of galaxies and the
environment in which they reside, at an epoch ($z\sim2-2.5$) when
galaxies are peaking in their star-formation activity.

\begin{acknowledgements} 

  We thank the referee for his/her comments, which allowed us to
  clarify some parts of the paper. This work was supported by funding
  from the European Research Council Advanced Grant
  ERC-2010-AdG-268107-EARLY and by INAF Grants PRIN 2010, PRIN 2012
  and PICS 2013. This work was additionally supported by the National
  Science Foundation under Grant No. 1411943 and NASA Grant Number
  NNX15AK92G. OC acknowledges support from PRIN-INAF 2014 program and
  the Cassini Fellowship program at INAF-OAS. This work is based on
  data products made available at the CESAM data center, Laboratoire
  d'Astrophysique de Marseille.  This work partly uses observations
  obtained with MegaPrime/MegaCam, a joint project of CFHT and
  CEA/DAPNIA, at the Canada-France-Hawaii Telescope (CFHT) which is
  operated by the National Research Council (NRC) of Canada, the
  Institut National des Sciences de l'Univers of the Centre National
  de la Recherche Scientifique (CNRS) of France, and the University of
  Hawaii. This work is based in part on data products produced at
  TERAPIX and the Canadian Astronomy Data Centre as part of the
  Canada--France--Hawaii Telescope Legacy Survey, a collaborative
  project of NRC and CNRS.  This paper is also based in part on data
  products from observations made with ESO Telescopes at the La Silla
  Paranal Observatory under ESO programme ID 179.A-2005 and on data
  products produced by TERAPIX and the Cambridge Astronomy Survey Unit
  on behalf of the UltraVISTA consortium.  OC thanks M.~Roncarelli,
  L.~Moscardini, C.~Fedeli, F.~Marulli, C.~Giocoli, and M.~Baldi for
  useful discussions, and J.R.~Franck and S.S.~McGaugh for their kind
  help in unveiling the details of their work.
\end{acknowledgements}

\bibliographystyle{aa}
\bibliography{biblio}

\begin{thebibliography}{118}
\expandafter\ifx\csname natexlab\endcsname\relax\def\natexlab#1{#1}\fi

\bibitem[{{Allen} {et~al.}(2011){Allen}, {Evrard}, \& {Mantz}}]{allen11}
{Allen}, S.~W., {Evrard}, A.~E., \& {Mantz}, A.~B. 2011, \araa, 49, 409

\bibitem[{{Arnouts} {et~al.}(1999){Arnouts}, {Cristiani}, {Moscardini},
  {Matarrese}, {Lucchin}, {Fontana}, \& {Giallongo}}]{arnouts99}
{Arnouts}, S., {Cristiani}, S., {Moscardini}, L., {et~al.} 1999, \mnras, 310,
  540

\bibitem[{{Bagchi} {et~al.}(2017){Bagchi}, {Sankhyayan}, {Sarkar},
  {Raychaudhury}, {Jacob}, \& {Dabhade}}]{bagchi17}
{Bagchi}, J., {Sankhyayan}, S., {Sarkar}, P., {et~al.} 2017, \apj, 844, 25

\bibitem[{{Beers} {et~al.}(1990){Beers}, {Flynn}, \& {Gebhardt}}]{beers90}
{Beers}, T.~C., {Flynn}, K., \& {Gebhardt}, K. 1990, \aj, 100, 32

\bibitem[{{Bielby} {et~al.}(2013){Bielby}, {Hill}, {Shanks}, {Crighton},
  {Infante}, {Bornancini}, {Francke}, {H{\'e}raudeau}, {Lambas}, {Metcalfe},
  {Minniti}, {Padilla}, {Theuns}, {Tummuangpak}, \& {Weilbacher}}]{bielby13}
{Bielby}, R., {Hill}, M.~D., {Shanks}, T., {et~al.} 2013, \mnras, 430, 425

\bibitem[{{Bryan} \& {Norman}(1998)}]{bryan_norman98}
{Bryan}, G.~L. \& {Norman}, M.~L. 1998, \apj, 495, 80

\bibitem[{{Cai} {et~al.}(2017){Cai}, {Fan}, {Bian}, {Zabludoff}, {Yang},
  {Prochaska}, {McGreer}, {Zheng}, {Kashikawa}, {Wang}, {Frye}, {Green}, \&
  {Jiang}}]{cai17_z23}
{Cai}, Z., {Fan}, X., {Bian}, F., {et~al.} 2017, \apj, 839, 131

\bibitem[{{Cai} {et~al.}(2016){Cai}, {Fan}, {Peirani}, {Bian}, {Frye},
  {McGreer}, {Prochaska}, {Lau}, {Tejos}, {Ho}, \& {Schneider}}]{cai16_method}
{Cai}, Z., {Fan}, X., {Peirani}, S., {et~al.} 2016, \apj, 833, 135

\bibitem[{{Capak} {et~al.}(2011){Capak}, {Riechers}, {Scoville}, {Carilli},
  {Cox}, {Neri}, {Robertson}, {Salvato}, {Schinnerer}, {Yan}, {Wilson}, {Yun},
  {Civano}, {Elvis}, {Karim}, {Mobasher}, \& {Staguhn}}]{capak11}
{Capak}, P.~L., {Riechers}, D., {Scoville}, N.~Z., {et~al.} 2011, \nat, 470,
  233

\bibitem[{{Carlberg} {et~al.}(1997){Carlberg}, {Yee}, {Ellingson}, {Morris},
  {Abraham}, {Gravel}, {Pritchet}, {Smecker-Hane}, {Hartwick}, {Hesser},
  {Hutchings}, \& {Oke}}]{carlberg97}
{Carlberg}, R.~G., {Yee}, H.~K.~C., {Ellingson}, E., {et~al.} 1997, \apjl, 485,
  L13

\bibitem[{{Carroll} {et~al.}(1992){Carroll}, {Press}, \& {Turner}}]{carroll92}
{Carroll}, S.~M., {Press}, W.~H., \& {Turner}, E.~L. 1992, \araa, 30, 499

\bibitem[{{Casey} {et~al.}(2015){Casey}, {Cooray}, {Capak}, {Fu}, {Kovac},
  {Lilly}, {Sanders}, {Scoville}, \& {Treister}}]{casey2015_z247}
{Casey}, C.~M., {Cooray}, A., {Capak}, P., {et~al.} 2015, \apjl, 808, L33

\bibitem[{{Cassata} {et~al.}(2007){Cassata}, {Guzzo}, {Franceschini},
  {Scoville}, {Capak}, {Ellis}, {Koekemoer}, {McCracken}, {Mobasher},
  {Renzini}, {Ricciardelli}, {Scodeggio}, {Taniguchi}, \&
  {Thompson}}]{cassata07_z07}
{Cassata}, P., {Guzzo}, L., {Franceschini}, A., {et~al.} 2007, \apjs, 172, 270

\bibitem[{{Castignani} {et~al.}(2014){Castignani}, {Chiaberge}, {Celotti},
  {Norman}, \& {De Zotti}}]{castignani14}
{Castignani}, G., {Chiaberge}, M., {Celotti}, A., {Norman}, C., \& {De Zotti},
  G. 2014, \apj, 792, 114

\bibitem[{{Chiaberge} {et~al.}(2010){Chiaberge}, {Capetti}, {Macchetto},
  {Rosati}, {Tozzi}, \& {Tremblay}}]{chiaberge10}
{Chiaberge}, M., {Capetti}, A., {Macchetto}, F.~D., {et~al.} 2010, \apjl, 710,
  L107

\bibitem[{{Chiaberge} {et~al.}(2009){Chiaberge}, {Tremblay}, {Capetti},
  {Macchetto}, {Tozzi}, \& {Sparks}}]{chiaberge09}
{Chiaberge}, M., {Tremblay}, G., {Capetti}, A., {et~al.} 2009, \apj, 696, 1103

\bibitem[{{Chiang} {et~al.}(2014){Chiang}, {Overzier}, \&
  {Gebhardt}}]{chiang2014_list}
{Chiang}, Y.-K., {Overzier}, R., \& {Gebhardt}, K. 2014, \apjl, 782, L3

\bibitem[{{Chiang} {et~al.}(2015){Chiang}, {Overzier}, {Gebhardt},
  {Finkelstein}, {Chiang}, {Hill}, {Blanc}, {Drory}, {Chonis}, {Zeimann},
  {Hagen}, {Schneider}, {Jogee}, {Ciardullo}, \& {Gronwall}}]{chiang2015_z244}
{Chiang}, Y.-K., {Overzier}, R.~A., {Gebhardt}, K., {et~al.} 2015, \apj, 808,
  37

\bibitem[{{Chiang} {et~al.}(2017){Chiang}, {Overzier}, {Gebhardt}, \&
  {Henriques}}]{chiang17}
{Chiang}, Y.-K., {Overzier}, R.~A., {Gebhardt}, K., \& {Henriques}, B. 2017,
  \apjl, 844, L23

\bibitem[{{Chon} {et~al.}(2015){Chon}, {B{\"o}hringer}, \& {Zaroubi}}]{chon15}
{Chon}, G., {B{\"o}hringer}, H., \& {Zaroubi}, S. 2015, \aap, 575, L14

\bibitem[{{Cirasuolo} {et~al.}(2010){Cirasuolo}, {McLure}, {Dunlop}, {Almaini},
  {Foucaud}, \& {Simpson}}]{cirasuolo10}
{Cirasuolo}, M., {McLure}, R.~J., {Dunlop}, J.~S., {et~al.} 2010, \mnras, 401,
  1166

\bibitem[{{Coles} \& {Jones}(1991)}]{coles91}
{Coles}, P. \& {Jones}, B. 1991, \mnras, 248, 1

\bibitem[{{Cooper} {et~al.}(2005){Cooper}, {Newman}, {Madgwick}, {Gerke},
  {Yan}, \& {Davis}}]{coooper05}
{Cooper}, M.~C., {Newman}, J.~A., {Madgwick}, D.~S., {et~al.} 2005, \apj, 634,
  833

\bibitem[{{Cucciati} {et~al.}(2010){Cucciati}, {Marinoni}, {Iovino},
  {Bardelli}, {Adami}, {Mazure}, {Scodeggio}, {Maccagni}, {Temporin}, {Zucca},
  {De Lucia}, {Blaizot}, {Garilli}, {Meneux}, {Zamorani}, {Le F{\`e}vre},
  {Cappi}, {Guzzo}, {Bottini}, {Le Brun}, {Tresse}, {Vettolani}, {Zanichelli},
  {Arnouts}, {Bolzonella}, {Charlot}, {Ciliegi}, {Contini}, {Foucaud},
  {Franzetti}, {Gavignaud}, {Ilbert}, {Lamareille}, {McCracken}, {Marano},
  {Merighi}, {Paltani}, {Pell{\`o}}, {Pollo}, {Pozzetti}, {Vergani}, \&
  {P{\'e}rez-Montero}}]{cucciati10}
{Cucciati}, O., {Marinoni}, C., {Iovino}, A., {et~al.} 2010, \aap, 520, A42

\bibitem[{{Cucciati} {et~al.}(2014){Cucciati}, {Zamorani}, {Lemaux},
  {Bardelli}, {Cimatti}, {Le F{\`e}vre}, {Cassata}, {Garilli}, {Le Brun},
  {Maccagni}, {Pentericci}, {Tasca}, {Thomas}, {Vanzella}, {Zucca}, {Amorin},
  {Capak}, {Cassar{\`a}}, {Castellano}, {Cuby}, {de la Torre}, {Durkalec},
  {Fontana}, {Giavalisco}, {Grazian}, {Hathi}, {Ilbert}, {Moreau}, {Paltani},
  {Ribeiro}, {Salvato}, {Schaerer}, {Scodeggio}, {Sommariva}, {Talia},
  {Taniguchi}, {Tresse}, {Vergani}, {Wang}, {Charlot}, {Contini}, {Fotopoulou},
  {L{\'o}pez-Sanjuan}, {Mellier}, \& {Scoville}}]{cucciati2014_z29}
{Cucciati}, O., {Zamorani}, G., {Lemaux}, B.~C., {et~al.} 2014, \aap, 570, A16

\bibitem[{{Darvish} {et~al.}(2015){Darvish}, {Mobasher}, {Sobral}, {Scoville},
  \& {Aragon-Calvo}}]{darvish15}
{Darvish}, B., {Mobasher}, B., {Sobral}, D., {Scoville}, N., \& {Aragon-Calvo},
  M. 2015, \apj, 805, 121

\bibitem[{{de la Torre} {et~al.}(2010){de la Torre}, {Guzzo}, {Kova{\v c}},
  {Porciani}, {Abbas}, {Meneux}, {Carollo}, {Contini}, {Kneib}, {Le F{\`e}vre},
  {Lilly}, {Mainieri}, {Renzini}, {Sanders}, {Scodeggio}, {Scoville},
  {Zamorani}, {Bardelli}, {Bolzonella}, {Bongiorno}, {Caputi}, {Coppa},
  {Cucciati}, {de Ravel}, {Franzetti}, {Garilli}, {Iovino}, {Kampczyk},
  {Knobel}, {Koekemoer}, {Lamareille}, {Le Borgne}, {Le Brun}, {Maier},
  {Mignoli}, {Pell{\'o}}, {Peng}, {Perez-Montero}, {Ricciardelli}, {Silverman},
  {Tanaka}, {Tasca}, {Tresse}, {Vergani}, {Welikala}, {Zucca}, {Bottini},
  {Cappi}, {Cassata}, {Cimatti}, {Fumana}, {Ilbert}, {Leauthaud}, {Maccagni},
  {Marinoni}, {McCracken}, {Memeo}, {Nair}, {Oesch}, {Pozzetti}, {Presotto}, \&
  {Scaramella}}]{delatorre10_clustering}
{de la Torre}, S., {Guzzo}, L., {Kova{\v c}}, K., {et~al.} 2010, \mnras, 409,
  867

\bibitem[{{de Lapparent} {et~al.}(1986){de Lapparent}, {Geller}, \&
  {Huchra}}]{delapparent86_greatwall}
{de Lapparent}, V., {Geller}, M.~J., \& {Huchra}, J.~P. 1986, \apjl, 302, L1

\bibitem[{{Despali} {et~al.}(2013){Despali}, {Tormen}, \& {Sheth}}]{despali13}
{Despali}, G., {Tormen}, G., \& {Sheth}, R.~K. 2013, \mnras, 431, 1143

\bibitem[{{Diener} {et~al.}(2013){Diener}, {Lilly}, {Knobel}, {Zamorani},
  {Lemson}, {Kampczyk}, {Scoville}, {Carollo}, {Contini}, {Kneib}, {Le Fevre},
  {Mainieri}, {Renzini}, {Scodeggio}, {Bardelli}, {Bolzonella}, {Bongiorno},
  {Caputi}, {Cucciati}, {de la Torre}, {de Ravel}, {Franzetti}, {Garilli},
  {Iovino}, {Kova{\v c}}, {Lamareille}, {Le Borgne}, {Le Brun}, {Maier},
  {Mignoli}, {Pello}, {Peng}, {Perez Montero}, {Presotto}, {Silverman},
  {Tanaka}, {Tasca}, {Tresse}, {Vergani}, {Zucca}, {Bordoloi}, {Cappi},
  {Cimatti}, {Coppa}, {Koekemoer}, {L{\'o}pez-Sanjuan}, {McCracken}, {Moresco},
  {Nair}, {Pozzetti}, \& {Welikala}}]{diener2013_list}
{Diener}, C., {Lilly}, S.~J., {Knobel}, C., {et~al.} 2013, \apj, 765, 109

\bibitem[{{Diener} {et~al.}(2015){Diener}, {Lilly}, {Ledoux}, {Zamorani},
  {Bolzonella}, {Murphy}, {Capak}, {Ilbert}, \& {McCracken}}]{diener2015_z245}
{Diener}, C., {Lilly}, S.~J., {Ledoux}, C., {et~al.} 2015, \apj, 802, 31

\bibitem[{{Durkalec} {et~al.}(2015){Durkalec}, {Le F{\`e}vre}, {Pollo}, {de la
  Torre}, {Cassata}, {Garilli}, {Le Brun}, {Lemaux}, {Maccagni}, {Pentericci},
  {Tasca}, {Thomas}, {Vanzella}, {Zamorani}, {Zucca}, {Amor{\'{\i}}n},
  {Bardelli}, {Cassar{\`a}}, {Castellano}, {Cimatti}, {Cucciati}, {Fontana},
  {Giavalisco}, {Grazian}, {Hathi}, {Ilbert}, {Paltani}, {Ribeiro}, {Schaerer},
  {Scodeggio}, {Sommariva}, {Talia}, {Tresse}, {Vergani}, {Capak}, {Charlot},
  {Contini}, {Cuby}, {Dunlop}, {Fotopoulou}, {Koekemoer}, {L{\'o}pez-Sanjuan},
  {Mellier}, {Pforr}, {Salvato}, {Scoville}, {Taniguchi}, \&
  {Wang}}]{durkalec15b}
{Durkalec}, A., {Le F{\`e}vre}, O., {Pollo}, A., {et~al.} 2015, \aap, 583, A128

\bibitem[{{Einasto} {et~al.}(2016){Einasto}, {Lietzen}, {Gramann}, {Tempel},
  {Saar}, {Liivam{\"a}gi}, {Hein{\"a}m{\"a}ki}, {Nurmi}, \&
  {Einasto}}]{einasto16_SGW}
{Einasto}, M., {Lietzen}, H., {Gramann}, M., {et~al.} 2016, \aap, 595, A70

\bibitem[{{Evrard} {et~al.}(2008){Evrard}, {Bialek}, {Busha}, {White}, {Habib},
  {Heitmann}, {Warren}, {Rasia}, {Tormen}, {Moscardini}, {Power}, {Jenkins},
  {Gao}, {Frenk}, {Springel}, {White}, \& {Diemand}}]{evrard08}
{Evrard}, A.~E., {Bialek}, J., {Busha}, M., {et~al.} 2008, \apj, 672, 122

\bibitem[{{Fassbender} {et~al.}(2011){Fassbender}, {Nastasi}, {B{\"o}hringer},
  {{\v S}uhada}, {Santos}, {Rosati}, {Pierini}, {M{\"u}hlegger}, {Quintana},
  {Schwope}, {Lamer}, {de Hoon}, {Kohnert}, {Pratt}, \& {Mohr}}]{fassbender11}
{Fassbender}, R., {Nastasi}, A., {B{\"o}hringer}, H., {et~al.} 2011, \aap, 527,
  L10

\bibitem[{{Foley} {et~al.}(2011){Foley}, {Andersson}, {Bazin}, {de Haan},
  {Ruel}, {Ade}, {Aird}, {Armstrong}, {Ashby}, {Bautz}, {Benson}, {Bleem},
  {Bonamente}, {Brodwin}, {Carlstrom}, {Chang}, {Clocchiatti}, {Crawford},
  {Crites}, {Desai}, {Dobbs}, {Dudley}, {Fazio}, {Forman}, {Garmire}, {George},
  {Gladders}, {Gonzalez}, {Halverson}, {High}, {Holder}, {Holzapfel}, {Hoover},
  {Hrubes}, {Jones}, {Joy}, {Keisler}, {Knox}, {Lee}, {Leitch}, {Lueker},
  {Luong-Van}, {Marrone}, {McMahon}, {Mehl}, {Meyer}, {Mohr}, {Montroy},
  {Murray}, {Padin}, {Plagge}, {Pryke}, {Reichardt}, {Rest}, {Ruhl},
  {Saliwanchik}, {Saro}, {Schaffer}, {Shaw}, {Shirokoff}, {Song}, {Spieler},
  {Stalder}, {Stanford}, {Staniszewski}, {Stark}, {Story}, {Stubbs},
  {Vanderlinde}, {Vieira}, {Vikhlinin}, {Williamson}, \&
  {Zenteno}}]{foley11_SZ}
{Foley}, R.~J., {Andersson}, K., {Bazin}, G., {et~al.} 2011, \apj, 731, 86

\bibitem[{{Franck} \& {McGaugh}(2016)}]{franck16_CCPC}
{Franck}, J.~R. \& {McGaugh}, S.~S. 2016, \apj, 833, 15

\bibitem[{{Fukugita} {et~al.}(1996){Fukugita}, {Ichikawa}, {Gunn}, {Doi},
  {Shimasaku}, \& {Schneider}}]{fukugita96}
{Fukugita}, M., {Ichikawa}, T., {Gunn}, J.~E., {et~al.} 1996, \aj, 111, 1748

\bibitem[{{Galametz} {et~al.}(2012){Galametz}, {Stern}, {De Breuck}, {Hatch},
  {Mayo}, {Miley}, {Rettura}, {Seymour}, {Stanford}, \& {Vernet}}]{galametz12}
{Galametz}, A., {Stern}, D., {De Breuck}, C., {et~al.} 2012, \apj, 749, 169

\bibitem[{{Gerke} {et~al.}(2012){Gerke}, {Newman}, {Davis}, {Coil}, {Cooper},
  {Dutton}, {Faber}, {Guhathakurta}, {Konidaris}, {Koo}, {Lin}, {Noeske},
  {Phillips}, {Rosario}, {Weiner}, {Willmer}, \& {Yan}}]{gerke12}
{Gerke}, B.~F., {Newman}, J.~A., {Davis}, M., {et~al.} 2012, \apj, 751, 50

\bibitem[{{Gilbank} {et~al.}(2008){Gilbank}, {Yee}, {Ellingson}, {Hicks},
  {Gladders}, {Barrientos}, \& {Keeney}}]{gilbank08}
{Gilbank}, D.~G., {Yee}, H.~K.~C., {Ellingson}, E., {et~al.} 2008, \apjl, 677,
  L89

\bibitem[{{Gobat} {et~al.}(2011){Gobat}, {Daddi}, {Onodera}, {Finoguenov},
  {Renzini}, {Arimoto}, {Bouwens}, {Brusa}, {Chary}, {Cimatti}, {Dickinson},
  {Kong}, \& {Mignoli}}]{gobat11}
{Gobat}, R., {Daddi}, E., {Onodera}, M., {et~al.} 2011, \aap, 526, A133

\bibitem[{{Gott} {et~al.}(2005){Gott}, {Juri{\'c}}, {Schlegel}, {Hoyle},
  {Vogeley}, {Tegmark}, {Bahcall}, \& {Brinkmann}}]{gott05}
{Gott}, III, J.~R., {Juri{\'c}}, M., {Schlegel}, D., {et~al.} 2005, \apj, 624,
  463

\bibitem[{{Guzzo} {et~al.}(2007){Guzzo}, {Cassata}, {Finoguenov}, {Massey},
  {Scoville}, {Capak}, {Ellis}, {Mobasher}, {Taniguchi}, {Thompson}, {Ajiki},
  {Aussel}, {B{\"o}hringer}, {Brusa}, {Calzetti}, {Comastri}, {Franceschini},
  {Hasinger}, {Kasliwal}, {Kitzbichler}, {Kneib}, {Koekemoer}, {Leauthaud},
  {McCracken}, {Murayama}, {Nagao}, {Rhodes}, {Sanders}, {Sasaki}, {Shioya},
  {Tasca}, \& {Taylor}}]{guzzo07_z07}
{Guzzo}, L., {Cassata}, P., {Finoguenov}, A., {et~al.} 2007, \apjs, 172, 254

\bibitem[{{Hamilton}(2001)}]{hamilton01}
{Hamilton}, A.~J.~S. 2001, \mnras, 322, 419

\bibitem[{{Hasinger} {et~al.}(2018){Hasinger}, {Capak}, {Salvato}, {Barger},
  {Cowie}, {Faisst}, {Hemmati}, {Kakazu}, {Kartaltepe}, {Masters}, {Mobasher},
  {Nayyeri}, {Sanders}, {Scoville}, {Suh}, {Steinhardt}, \&
  {Yang}}]{hasinger18}
{Hasinger}, G., {Capak}, P., {Salvato}, M., {et~al.} 2018, \apj, 858, 77

\bibitem[{{Haynes} \& {Giovanelli}(1986)}]{haynes86_pp}
{Haynes}, M.~P. \& {Giovanelli}, R. 1986, \apjl, 306, L55

\bibitem[{{Heneka} {et~al.}(2018){Heneka}, {Rapetti}, {Cataneo}, {Mantz},
  {Allen}, \& {von der Linden}}]{heneka18}
{Heneka}, C., {Rapetti}, D., {Cataneo}, M., {et~al.} 2018, \mnras, 473, 3882

\bibitem[{{Higuchi} {et~al.}(2018){Higuchi}, {Ouchi}, {Ono}, {Shibuya},
  {Toshikawa}, {Harikane}, {Kojima}, {Chiang}, {Egami}, {Kashikawa},
  {Overzier}, {Konno}, {Inoue}, {Hasegawa}, {Fujimoto}, {Goto}, {Ishikawa},
  {Ito}, {Komiyama}, \& {Tanaka}}]{higuchi18_silverrush}
{Higuchi}, R., {Ouchi}, M., {Ono}, Y., {et~al.} 2018, ArXiv e-prints
  [\eprint[arXiv]{1801.00531}]

\bibitem[{{Ilbert} {et~al.}(2006){Ilbert}, {Arnouts}, {McCracken},
  {Bolzonella}, {Bertin}, {Le F{\`e}vre}, {Mellier}, {Zamorani}, {Pell{\`o}},
  {Iovino}, {Tresse}, {Le Brun}, {Bottini}, {Garilli}, {Maccagni}, {Picat},
  {Scaramella}, {Scodeggio}, {Vettolani}, {Zanichelli}, {Adami}, {Bardelli},
  {Cappi}, {Charlot}, {Ciliegi}, {Contini}, {Cucciati}, {Foucaud}, {Franzetti},
  {Gavignaud}, {Guzzo}, {Marano}, {Marinoni}, {Mazure}, {Meneux}, {Merighi},
  {Paltani}, {Pollo}, {Pozzetti}, {Radovich}, {Zucca}, {Bondi}, {Bongiorno},
  {Busarello}, {de La Torre}, {Gregorini}, {Lamareille}, {Mathez}, {Merluzzi},
  {Ripepi}, {Rizzo}, \& {Vergani}}]{ilbert2006_pz}
{Ilbert}, O., {Arnouts}, S., {McCracken}, H.~J., {et~al.} 2006, \aap, 457, 841

\bibitem[{{Ilbert} {et~al.}(2013){Ilbert}, {McCracken}, {Le F{\`e}vre},
  {Capak}, {Dunlop}, {Karim}, {Renzini}, {Caputi}, {Boissier}, {Arnouts},
  {Aussel}, {Comparat}, {Guo}, {Hudelot}, {Kartaltepe}, {Kneib}, {Krogager},
  {Le Floc'h}, {Lilly}, {Mellier}, {Milvang-Jensen}, {Moutard}, {Onodera},
  {Richard}, {Salvato}, {Sanders}, {Scoville}, {Silverman}, {Taniguchi},
  {Tasca}, {Thomas}, {Toft}, {Tresse}, {Vergani}, {Wolk}, \&
  {Zirm}}]{ilbert2013}
{Ilbert}, O., {McCracken}, H.~J., {Le F{\`e}vre}, O., {et~al.} 2013, \aap, 556,
  A55

\bibitem[{{Iovino} {et~al.}(2016){Iovino}, {Petropoulou}, {Scodeggio},
  {Bolzonella}, {Zamorani}, {Bardelli}, {Cucciati}, {Pozzetti}, {Tasca},
  {Vergani}, {Zucca}, {Finoguenov}, {Ilbert}, {Tanaka}, {Salvato}, {Kova{\v
  c}}, \& {Cassata}}]{iovino16_wall}
{Iovino}, A., {Petropoulou}, V., {Scodeggio}, M., {et~al.} 2016, \aap, 592, A78

\bibitem[{{Kim} {et~al.}(2016){Kim}, {Im}, {Lee}, {Edge}, {Hyun}, {Kim},
  {Choi}, {Hong}, {Jeon}, {Jun}, {Karouzos}, {Kim}, {Kim}, {Kim}, {Park},
  {Taak}, \& {Yoon}}]{kim16_sc}
{Kim}, J.-W., {Im}, M., {Lee}, S.-K., {et~al.} 2016, \apjl, 821, L10

\bibitem[{{Knobel} {et~al.}(2012){Knobel}, {Lilly}, {Iovino}, {Kova{\v c}},
  {Bschorr}, {Presotto}, {Oesch}, {Kampczyk}, {Carollo}, {Contini}, {Kneib},
  {Le Fevre}, {Mainieri}, {Renzini}, {Scodeggio}, {Zamorani}, {Bardelli},
  {Bolzonella}, {Bongiorno}, {Caputi}, {Cucciati}, {de la Torre}, {de Ravel},
  {Franzetti}, {Garilli}, {Lamareille}, {Le Borgne}, {Le Brun}, {Maier},
  {Mignoli}, {Pello}, {Peng}, {Perez Montero}, {Silverman}, {Tanaka}, {Tasca},
  {Tresse}, {Vergani}, {Zucca}, {Barnes}, {Bordoloi}, {Cappi}, {Cimatti},
  {Coppa}, {Koekemoer}, {L{\'o}pez-Sanjuan}, {McCracken}, {Moresco}, {Nair},
  {Pozzetti}, \& {Welikala}}]{knobel12_groups20k}
{Knobel}, C., {Lilly}, S.~J., {Iovino}, A., {et~al.} 2012, \apj, 753, 121

\bibitem[{{Knobel} {et~al.}(2009){Knobel}, {Lilly}, {Iovino}, {Porciani},
  {Kova{\v c}}, {Cucciati}, {Finoguenov}, {Kitzbichler}, {Carollo}, {Contini},
  {Kneib}, {Le F{\`e}vre}, {Mainieri}, {Renzini}, {Scodeggio}, {Zamorani},
  {Bardelli}, {Bolzonella}, {Bongiorno}, {Caputi}, {Coppa}, {de la Torre}, {de
  Ravel}, {Franzetti}, {Garilli}, {Kampczyk}, {Lamareille}, {Le Borgne}, {Le
  Brun}, {Maier}, {Mignoli}, {Pello}, {Peng}, {Perez Montero}, {Ricciardelli},
  {Silverman}, {Tanaka}, {Tasca}, {Tresse}, {Vergani}, {Zucca}, {Abbas},
  {Bottini}, {Cappi}, {Cassata}, {Cimatti}, {Fumana}, {Guzzo}, {Koekemoer},
  {Leauthaud}, {Maccagni}, {Marinoni}, {McCracken}, {Memeo}, {Meneux}, {Oesch},
  {Pozzetti}, \& {Scaramella}}]{knobel09_groups}
{Knobel}, C., {Lilly}, S.~J., {Iovino}, A., {et~al.} 2009, \apj, 697, 1842

\bibitem[{{Kodama} {et~al.}(2007){Kodama}, {Tanaka}, {Kajisawa}, {Kurk},
  {Venemans}, {De Breuck}, {Vernet}, \& {Lidman}}]{kodama07}
{Kodama}, T., {Tanaka}, I., {Kajisawa}, M., {et~al.} 2007, \mnras, 377, 1717

\bibitem[{{Kova{\v c}} {et~al.}(2010){Kova{\v c}}, {Lilly}, {Cucciati},
  {Porciani}, {Iovino}, {Zamorani}, {Oesch}, {Bolzonella}, {Knobel},
  {Finoguenov}, {Peng}, {Carollo}, {Pozzetti}, {Caputi}, {Silverman}, {Tasca},
  {Scodeggio}, {Vergani}, {Scoville}, {Capak}, {Contini}, {Kneib}, {Le
  F{\`e}vre}, {Mainieri}, {Renzini}, {Bardelli}, {Bongiorno}, {Coppa}, {de la
  Torre}, {de Ravel}, {Franzetti}, {Garilli}, {Guzzo}, {Kampczyk},
  {Lamareille}, {Le Borgne}, {Le Brun}, {Maier}, {Mignoli}, {Pello}, {Perez
  Montero}, {Ricciardelli}, {Tanaka}, {Tresse}, {Zucca}, {Abbas}, {Bottini},
  {Cappi}, {Cassata}, {Cimatti}, {Fumana}, {Koekemoer}, {Maccagni}, {Marinoni},
  {McCracken}, {Memeo}, {Meneux}, \& {Scaramella}}]{kovac10_density}
{Kova{\v c}}, K., {Lilly}, S.~J., {Cucciati}, O., {et~al.} 2010, \apj, 708, 505

\bibitem[{{Kriek} {et~al.}(2015){Kriek}, {Shapley}, {Reddy}, {Siana}, {Coil},
  {Mobasher}, {Freeman}, {de Groot}, {Price}, {Sanders}, {Shivaei}, {Brammer},
  {Momcheva}, {Skelton}, {van Dokkum}, {Whitaker}, {Aird}, {Azadi}, {Kassis},
  {Bullock}, {Conroy}, {Dav{\'e}}, {Kere{\v s}}, \& {Krumholz}}]{kriek15}
{Kriek}, M., {Shapley}, A.~E., {Reddy}, N.~A., {et~al.} 2015, \apjs, 218, 15

\bibitem[{{Laigle} {et~al.}(2016){Laigle}, {McCracken}, {Ilbert}, {Hsieh},
  {Davidzon}, {Capak}, {Hasinger}, {Silverman}, {Pichon}, {Coupon}, {Aussel},
  {Le Borgne}, {Caputi}, {Cassata}, {Chang}, {Civano}, {Dunlop}, {Fynbo},
  {Kartaltepe}, {Koekemoer}, {Le F{\`e}vre}, {Le Floc'h}, {Leauthaud}, {Lilly},
  {Lin}, {Marchesi}, {Milvang-Jensen}, {Salvato}, {Sanders}, {Scoville},
  {Smolcic}, {Stockmann}, {Taniguchi}, {Tasca}, {Toft}, {Vaccari}, \&
  {Zabl}}]{laigle2016}
{Laigle}, C., {McCracken}, H.~J., {Ilbert}, O., {et~al.} 2016, \apjs, 224, 24

\bibitem[{{Le F{\`e}vre} {et~al.}(2013){Le F{\`e}vre}, {Cassata}, {Cucciati},
  {Garilli}, {Ilbert}, {Le Brun}, {Maccagni}, {Moreau}, {Scodeggio}, {Tresse},
  {Zamorani}, {Adami}, {Arnouts}, {Bardelli}, {Bolzonella}, {Bondi},
  {Bongiorno}, {Bottini}, {Cappi}, {Charlot}, {Ciliegi}, {Contini}, {de la
  Torre}, {Foucaud}, {Franzetti}, {Gavignaud}, {Guzzo}, {Iovino}, {Lemaux},
  {L{\'o}pez-Sanjuan}, {McCracken}, {Marano}, {Marinoni}, {Mazure}, {Mellier},
  {Merighi}, {Merluzzi}, {Paltani}, {Pell{\`o}}, {Pollo}, {Pozzetti},
  {Scaramella}, {Tasca}, {Vergani}, {Vettolani}, {Zanichelli}, \&
  {Zucca}}]{lefevre2013a}
{Le F{\`e}vre}, O., {Cassata}, P., {Cucciati}, O., {et~al.} 2013, \aap, 559,
  A14

\bibitem[{{Le F{\`e}vre} {et~al.}(2003){Le F{\`e}vre}, {Saisse}, {Mancini},
  {Brau-Nogue}, {Caputi}, {Castinel}, {D'Odorico}, {Garilli}, {Kissler-Patig},
  {Lucuix}, {Mancini}, {Pauget}, {Sciarretta}, {Scodeggio}, {Tresse}, \&
  {Vettolani}}]{lefevre2003}
{Le F{\`e}vre}, O., {Saisse}, M., {Mancini}, D., {et~al.} 2003, in Instrument
  Design and Performance for Optical/Infrared Ground-based Telescopes. Edited
  by Iye, Masanori; Moorwood, Alan F. M. Proceedings of the SPIE, Volume 4841,
  pp. 1670-1681 (2003)., 1670--1681

\bibitem[{{Le F{\`e}vre} {et~al.}(2015){Le F{\`e}vre}, {Tasca}, {Cassata},
  {Garilli}, {Le Brun}, {Maccagni}, {Pentericci}, {Thomas}, {Vanzella},
  {Zamorani}, {Zucca}, {Amorin}, {Bardelli}, {Capak}, {Cassar{\`a}},
  {Castellano}, {Cimatti}, {Cuby}, {Cucciati}, {de la Torre}, {Durkalec},
  {Fontana}, {Giavalisco}, {Grazian}, {Hathi}, {Ilbert}, {Lemaux}, {Moreau},
  {Paltani}, {Ribeiro}, {Salvato}, {Schaerer}, {Scodeggio}, {Sommariva},
  {Talia}, {Taniguchi}, {Tresse}, {Vergani}, {Wang}, {Charlot}, {Contini},
  {Fotopoulou}, {L{\'o}pez-Sanjuan}, {Mellier}, \&
  {Scoville}}]{lefevre2015_vuds}
{Le F{\`e}vre}, O., {Tasca}, L.~A.~M., {Cassata}, P., {et~al.} 2015, \aap, 576,
  A79

\bibitem[{{Lee} {et~al.}(2016){Lee}, {Hennawi}, {White}, {Prochaska},
  {Font-Ribera}, {Schlegel}, {Rich}, {Suzuki}, {Stark}, {Le F{\`e}vre},
  {Nugent}, {Salvato}, \& {Zamorani}}]{lee2016_colossus}
{Lee}, K.-G., {Hennawi}, J.~F., {White}, M., {et~al.} 2016, \apj, 817, 160

\bibitem[{{Lee} {et~al.}(2018){Lee}, {Krolewski}, {White}, {Schlegel},
  {Nugent}, {Hennawi}, {M{\"u}ller}, {Pan}, {Prochaska}, {Font-Ribera},
  {Suzuki}, {Glazebrook}, {Kacprzak}, {Kartaltepe}, {Koekemoer}, {Le
  F{\`e}vre}, {Lemaux}, {Maier}, {Nanayakkara}, {Rich}, {Sanders}, {Salvato},
  {Tasca}, \& {Tran}}]{lee2017_clamato}
{Lee}, K.-G., {Krolewski}, A., {White}, M., {et~al.} 2018, \apjs, 237, 31

\bibitem[{{Lee} {et~al.}(2014){Lee}, {Dey}, {Hong}, {Reddy}, {Wilson},
  {Jannuzi}, {Inami}, \& {Gonzalez}}]{lee14_NB}
{Lee}, K.-S., {Dey}, A., {Hong}, S., {et~al.} 2014, \apj, 796, 126

\bibitem[{{Lemaux} {et~al.}(2014){Lemaux}, {Cucciati}, {Tasca}, {Le F{\`e}vre},
  {Zamorani}, {Cassata}, {Garilli}, {Le Brun}, {Maccagni}, {Pentericci},
  {Thomas}, {Vanzella}, {Zucca}, {Amor{\'{\i}}n}, {Bardelli}, {Capak},
  {Cassar{\`a}}, {Castellano}, {Cimatti}, {Cuby}, {de la Torre}, {Durkalec},
  {Fontana}, {Giavalisco}, {Grazian}, {Hathi}, {Ilbert}, {Moreau}, {Paltani},
  {Ribeiro}, {Salvato}, {Schaerer}, {Scodeggio}, {Sommariva}, {Talia},
  {Taniguchi}, {Tresse}, {Vergani}, {Wang}, {Charlot}, {Contini}, {Fotopoulou},
  {Gal}, {Kocevski}, {L{\'o}pez-Sanjuan}, {Lubin}, {Mellier}, {Sadibekova}, \&
  {Scoville}}]{lemaux2014_z33}
{Lemaux}, B.~C., {Cucciati}, O., {Tasca}, L.~A.~M., {et~al.} 2014, \aap, 572,
  A41

\bibitem[{{Lemaux} {et~al.}(2012){Lemaux}, {Gal}, {Lubin}, {Kocevski},
  {Fassnacht}, {McGrath}, {Squires}, {Surace}, \& {Lacy}}]{lemaux12}
{Lemaux}, B.~C., {Gal}, R.~R., {Lubin}, L.~M., {et~al.} 2012, \apj, 745, 106

\bibitem[{{Lemaux} {et~al.}(2018){Lemaux}, {Le F{\`e}vre}, {Cucciati},
  {Ribeiro}, {Tasca}, {Zamorani}, {Ilbert}, {Thomas}, {Bardelli}, {Cassata},
  {Hathi}, {Pforr}, {Smol{\v c}i{\'c}}, {Delvecchio}, {Novak}, {Berta},
  {McCracken}, {Koekemoer}, {Amor{\'{\i}}n}, {Garilli}, {Maccagni}, {Schaerer},
  \& {Zucca}}]{lemaux2018_z45}
{Lemaux}, B.~C., {Le F{\`e}vre}, O., {Cucciati}, O., {et~al.} 2018, \aap, 615,
  A77

\bibitem[{{Lemaux} {et~al.}(2009){Lemaux}, {Lubin}, {Sawicki}, {Martin},
  {Lagattuta}, {Gal}, {Kocevski}, {Fassnacht}, \& {Squires}}]{lemaux09}
{Lemaux}, B.~C., {Lubin}, L.~M., {Sawicki}, M., {et~al.} 2009, \apj, 700, 20

\bibitem[{{Lietzen} {et~al.}(2016){Lietzen}, {Tempel}, {Liivam{\"a}gi},
  {Montero-Dorta}, {Einasto}, {Streblyanska}, {Maraston},
  {Rubi{\~n}o-Mart{\'{\i}}n}, \& {Saar}}]{lietzen16_sc}
{Lietzen}, H., {Tempel}, E., {Liivam{\"a}gi}, L.~J., {et~al.} 2016, \aap, 588,
  L4

\bibitem[{{Lilly} {et~al.}(2009){Lilly}, {Le Brun}, {Maier}, {Mainieri},
  {Mignoli}, {Scodeggio}, {Zamorani}, {Carollo}, {Contini}, {Kneib}, {Le
  F{\`e}vre}, {Renzini}, {Bardelli}, {Bolzonella}, {Bongiorno}, {Caputi},
  {Coppa}, {Cucciati}, {de la Torre}, {de Ravel}, {Franzetti}, {Garilli},
  {Iovino}, {Kampczyk}, {Kovac}, {Knobel}, {Lamareille}, {Le Borgne}, {Pello},
  {Peng}, {P{\'e}rez-Montero}, {Ricciardelli}, {Silverman}, {Tanaka}, {Tasca},
  {Tresse}, {Vergani}, {Zucca}, {Ilbert}, {Salvato}, {Oesch}, {Abbas},
  {Bottini}, {Capak}, {Cappi}, {Cassata}, {Cimatti}, {Elvis}, {Fumana},
  {Guzzo}, {Hasinger}, {Koekemoer}, {Leauthaud}, {Maccagni}, {Marinoni},
  {McCracken}, {Memeo}, {Meneux}, {Porciani}, {Pozzetti}, {Sanders},
  {Scaramella}, {Scarlata}, {Scoville}, {Shopbell}, \& {Taniguchi}}]{lilly2009}
{Lilly}, S.~J., {Le Brun}, V., {Maier}, C., {et~al.} 2009, \apjs, 184, 218

\bibitem[{{Lilly} {et~al.}(2007){Lilly}, {Le F{\`e}vre}, {Renzini}, {Zamorani},
  {Scodeggio}, {Contini}, {Carollo}, {Hasinger}, {Kneib}, {Iovino}, {Le Brun},
  {Maier}, {Mainieri}, {Mignoli}, {Silverman}, {Tasca}, {Bolzonella},
  {Bongiorno}, {Bottini}, {Capak}, {Caputi}, {Cimatti}, {Cucciati}, {Daddi},
  {Feldmann}, {Franzetti}, {Garilli}, {Guzzo}, {Ilbert}, {Kampczyk}, {Kovac},
  {Lamareille}, {Leauthaud}, {Le Borgne}, {McCracken}, {Marinoni}, {Pello},
  {Ricciardelli}, {Scarlata}, {Vergani}, {Sanders}, {Schinnerer}, {Scoville},
  {Taniguchi}, {Arnouts}, {Aussel}, {Bardelli}, {Brusa}, {Cappi}, {Ciliegi},
  {Finoguenov}, {Foucaud}, {Franceschini}, {Halliday}, {Impey}, {Knobel},
  {Koekemoer}, {Kurk}, {Maccagni}, {Maddox}, {Marano}, {Marconi}, {Meneux},
  {Mobasher}, {Moreau}, {Peacock}, {Porciani}, {Pozzetti}, {Scaramella},
  {Schiminovich}, {Shopbell}, {Smail}, {Thompson}, {Tresse}, {Vettolani},
  {Zanichelli}, \& {Zucca}}]{lilly2007}
{Lilly}, S.~J., {Le F{\`e}vre}, O., {Renzini}, A., {et~al.} 2007, \apjs, 172,
  70

\bibitem[{{Lim} \& {Lee}(2014)}]{lim14}
{Lim}, S. \& {Lee}, J. 2014, \apj, 783, 39

\bibitem[{{Lovell} {et~al.}(2018){Lovell}, {Thomas}, \& {Wilkins}}]{lovell18}
{Lovell}, C.~C., {Thomas}, P.~A., \& {Wilkins}, S.~M. 2018, \mnras, 474, 4612

\bibitem[{{Lubin} {et~al.}(2000){Lubin}, {Brunner}, {Metzger}, {Postman}, \&
  {Oke}}]{lubin00}
{Lubin}, L.~M., {Brunner}, R., {Metzger}, M.~R., {Postman}, M., \& {Oke}, J.~B.
  2000, \apjl, 531, L5

\bibitem[{{Marinoni} {et~al.}(2002){Marinoni}, {Davis}, {Newman}, \&
  {Coil}}]{marinoni02}
{Marinoni}, C., {Davis}, M., {Newman}, J.~A., \& {Coil}, A.~L. 2002, \apj, 580,
  122

\bibitem[{{Massey} {et~al.}(2007){Massey}, {Rhodes}, {Ellis}, {Scoville},
  {Leauthaud}, {Finoguenov}, {Capak}, {Bacon}, {Aussel}, {Kneib}, {Koekemoer},
  {McCracken}, {Mobasher}, {Pires}, {Refregier}, {Sasaki}, {Starck},
  {Taniguchi}, {Taylor}, \& {Taylor}}]{massey07}
{Massey}, R., {Rhodes}, J., {Ellis}, R., {et~al.} 2007, \nat, 445, 286

\bibitem[{{McCracken} {et~al.}(2012){McCracken}, {Milvang-Jensen}, {Dunlop},
  {Franx}, {Fynbo}, {Le F{\`e}vre}, {Holt}, {Caputi}, {Goranova}, {Buitrago},
  {Emerson}, {Freudling}, {Hudelot}, {L{\'o}pez-Sanjuan}, {Magnard}, {Mellier},
  {M{\o}ller}, {Nilsson}, {Sutherland}, {Tasca}, \& {Zabl}}]{mcCracken12}
{McCracken}, H.~J., {Milvang-Jensen}, B., {Dunlop}, J., {et~al.} 2012, \aap,
  544, A156

\bibitem[{{Miley} \& {De Breuck}(2008)}]{miley08}
{Miley}, G. \& {De Breuck}, C. 2008, \aapr, 15, 67

\bibitem[{{Muldrew} {et~al.}(2018){Muldrew}, {Hatch}, \& {Cooke}}]{muldrew18}
{Muldrew}, S.~I., {Hatch}, N.~A., \& {Cooke}, E.~A. 2018, \mnras, 473, 2335

\bibitem[{{Munari} {et~al.}(2013){Munari}, {Biviano}, {Borgani}, {Murante}, \&
  {Fabjan}}]{munari13}
{Munari}, E., {Biviano}, A., {Borgani}, S., {Murante}, G., \& {Fabjan}, D.
  2013, \mnras, 430, 2638

\bibitem[{{Oke}(1974)}]{oke74}
{Oke}, J.~B. 1974, \apjs, 27, 21

\bibitem[{{Ouchi} {et~al.}(2005){Ouchi}, {Shimasaku}, {Akiyama}, {Sekiguchi},
  {Furusawa}, {Okamura}, {Kashikawa}, {Iye}, {Kodama}, {Saito}, {Sasaki},
  {Simpson}, {Takata}, {Yamada}, {Yamanoi}, {Yoshida}, \& {Yoshida}}]{ouchi05}
{Ouchi}, M., {Shimasaku}, K., {Akiyama}, M., {et~al.} 2005, \apjl, 620, L1

\bibitem[{{Pace} {et~al.}(2010){Pace}, {Waizmann}, \& {Bartelmann}}]{pace10}
{Pace}, F., {Waizmann}, J.-C., \& {Bartelmann}, M. 2010, \mnras, 406, 1865

\bibitem[{Padmanabhan(1993)}]{padmanabhan}
Padmanabhan, T. 1993, Structure Formation in the Universe (Cambridge University
  Press)

\bibitem[{{Pentericci} {et~al.}(2000){Pentericci}, {Kurk}, {R{\"o}ttgering},
  {Miley}, {van Breugel}, {Carilli}, {Ford}, {Heckman}, {McCarthy}, \&
  {Moorwood}}]{pentericci00}
{Pentericci}, L., {Kurk}, J.~D., {R{\"o}ttgering}, H.~J.~A., {et~al.} 2000,
  \aap, 361, L25

\bibitem[{{Percival}(2005)}]{percival05}
{Percival}, W.~J. 2005, \aap, 443, 819

\bibitem[{{Perna} {et~al.}(2015){Perna}, {Brusa}, {Salvato}, {Cresci},
  {Lanzuisi}, {Berta}, {Delvecchio}, {Fiore}, {Lutz}, {Le Floc'h}, {Mainieri},
  \& {Riguccini}}]{perna15}
{Perna}, M., {Brusa}, M., {Salvato}, M., {et~al.} 2015, \aap, 583, A72

\bibitem[{{Peter} {et~al.}(2007){Peter}, {Shapley}, {Law}, {Steidel}, {Erb},
  {Reddy}, \& {Pettini}}]{peter07}
{Peter}, A.~H.~G., {Shapley}, A.~E., {Law}, D.~R., {et~al.} 2007, \apj, 668, 23

\bibitem[{{Roncarelli} {et~al.}(2015){Roncarelli}, {Carbone}, \&
  {Moscardini}}]{roncarelli15}
{Roncarelli}, M., {Carbone}, C., \& {Moscardini}, L. 2015, \mnras, 447, 1761

\bibitem[{{Rosati} {et~al.}(1999){Rosati}, {Stanford}, {Eisenhardt}, {Elston},
  {Spinrad}, {Stern}, \& {Dey}}]{rosati99_lynx}
{Rosati}, P., {Stanford}, S.~A., {Eisenhardt}, P.~R., {et~al.} 1999, \aj, 118,
  76

\bibitem[{{Rumbaugh} {et~al.}(2018){Rumbaugh}, {Lemaux}, {Tomczak}, {Shen},
  {Pelliccia}, {Lubin}, {Kocevski}, {Wu}, {Gal}, {Mei}, {Fassnacht}, \&
  {Squires}}]{rumbaugh18}
{Rumbaugh}, N., {Lemaux}, B.~C., {Tomczak}, A.~R., {et~al.} 2018, \mnras, 478,
  1403

\bibitem[{{Salimbeni} {et~al.}(2009){Salimbeni}, {Castellano}, {Pentericci},
  {Trevese}, {Fiore}, {Grazian}, {Fontana}, {Giallongo}, {Boutsia},
  {Cristiani}, {de Santis}, {Gallozzi}, {Menci}, {Nonino}, {Paris}, {Santini},
  \& {Vanzella}}]{salimbeni09}
{Salimbeni}, S., {Castellano}, M., {Pentericci}, L., {et~al.} 2009, \aap, 501,
  865

\bibitem[{{Schmidt} {et~al.}(2009){Schmidt}, {Vikhlinin}, \& {Hu}}]{schmidt09}
{Schmidt}, F., {Vikhlinin}, A., \& {Hu}, W. 2009, \prd, 80, 083505

\bibitem[{{Scoville} {et~al.}(2013){Scoville}, {Arnouts}, {Aussel}, {Benson},
  {Bongiorno}, {Bundy}, {Calvo}, {Capak}, {Carollo}, {Civano}, {Dunlop},
  {Elvis}, {Faisst}, {Finoguenov}, {Fu}, {Giavalisco}, {Guo}, {Ilbert},
  {Iovino}, {Kajisawa}, {Kartaltepe}, {Leauthaud}, {Le F{\`e}vre}, {LeFloch},
  {Lilly}, {Liu}, {Manohar}, {Massey}, {Masters}, {McCracken}, {Mobasher},
  {Peng}, {Renzini}, {Rhodes}, {Salvato}, {Sanders}, {Sarvestani}, {Scarlata},
  {Schinnerer}, {Sheth}, {Shopbell}, {Smol{\v c}i{\'c}}, {Taniguchi}, {Taylor},
  {White}, \& {Yan}}]{scoville13_env}
{Scoville}, N., {Arnouts}, S., {Aussel}, H., {et~al.} 2013, \apjs, 206, 3

\bibitem[{{Scoville} {et~al.}(2007){Scoville}, {Aussel}, {Benson}, {Blain},
  {Calzetti}, {Capak}, {Ellis}, {El-Zant}, {Finoguenov}, {Giavalisco}, {Guzzo},
  {Hasinger}, {Koda}, {Le F{\`e}vre}, {Massey}, {McCracken}, {Mobasher},
  {Renzini}, {Rhodes}, {Salvato}, {Sanders}, {Sasaki}, {Schinnerer}, {Sheth},
  {Shopbell}, {Taniguchi}, {Taylor}, \& {Thompson}}]{scoville07_env}
{Scoville}, N., {Aussel}, H., {Benson}, A., {et~al.} 2007, \apjs, 172, 150

\bibitem[{{Sereno} \& {Ettori}(2015)}]{sereno15}
{Sereno}, M. \& {Ettori}, S. 2015, \mnras, 450, 3675

\bibitem[{{Shapley}(1934)}]{shapley34_hercules}
{Shapley}, H. 1934, \mnras, 94, 791

\bibitem[{{Shapley} \& {Ames}(1930)}]{shapley30}
{Shapley}, H. \& {Ames}, A. 1930, Harvard College Observatory Bulletin, 880, 1

\bibitem[{{Sheth} \& {Diaferio}(2011)}]{sheth11}
{Sheth}, R.~K. \& {Diaferio}, A. 2011, \mnras, 417, 2938

\bibitem[{{Smol{\v c}i{\'c}} {et~al.}(2017){Smol{\v c}i{\'c}}, {Miettinen},
  {Tomi{\v c}i{\'c}}, {Zamorani}, {Finoguenov}, {Lemaux}, {Aravena}, {Capak},
  {Chiang}, {Civano}, {Delvecchio}, {Ilbert}, {Jurlin}, {Karim}, {Laigle}, {Le
  F{\`e}vre}, {Marchesi}, {McCracken}, {Riechers}, {Salvato}, {Schinnerer},
  {Tasca}, \& {Toft}}]{smolcic17}
{Smol{\v c}i{\'c}}, V., {Miettinen}, O., {Tomi{\v c}i{\'c}}, N., {et~al.} 2017,
  \aap, 597, A4

\bibitem[{{Spitler} {et~al.}(2012){Spitler}, {Labb{\'e}}, {Glazebrook},
  {Persson}, {Monson}, {Papovich}, {Tran}, {Poole}, {Quadri}, {van Dokkum},
  {Kelson}, {Kacprzak}, {McCarthy}, {Murphy}, {Straatman}, \&
  {Tilvi}}]{spitler12}
{Spitler}, L.~R., {Labb{\'e}}, I., {Glazebrook}, K., {et~al.} 2012, \apjl, 748,
  L21

\bibitem[{{Steidel} {et~al.}(1998){Steidel}, {Adelberger}, {Dickinson},
  {Giavalisco}, {Pettini}, \& {Kellogg}}]{steidel98}
{Steidel}, C.~C., {Adelberger}, K.~L., {Dickinson}, M., {et~al.} 1998, \apj,
  492, 428

\bibitem[{{Steidel} {et~al.}(2005){Steidel}, {Adelberger}, {Shapley}, {Erb},
  {Reddy}, \& {Pettini}}]{steidel05}
{Steidel}, C.~C., {Adelberger}, K.~L., {Shapley}, A.~E., {et~al.} 2005, \apj,
  626, 44

\bibitem[{{Steidel} {et~al.}(2000){Steidel}, {Adelberger}, {Shapley},
  {Pettini}, {Dickinson}, \& {Giavalisco}}]{steidel00}
{Steidel}, C.~C., {Adelberger}, K.~L., {Shapley}, A.~E., {et~al.} 2000, \apj,
  532, 170

\bibitem[{{Strazzullo} {et~al.}(2013){Strazzullo}, {Gobat}, {Daddi}, {Onodera},
  {Carollo}, {Dickinson}, {Renzini}, {Arimoto}, {Cimatti}, {Finoguenov}, \&
  {Chary}}]{strazzullo13}
{Strazzullo}, V., {Gobat}, R., {Daddi}, E., {et~al.} 2013, \apj, 772, 118

\bibitem[{{Swinbank} {et~al.}(2007){Swinbank}, {Edge}, {Smail}, {Stott},
  {Bremer}, {Sato}, {Van Breukelen}, {Jarvis}, {Waddington}, {Clewley},
  {Bergeron}, {Cotter}, {Dye}, {Geach}, {Gonzalez-Solares}, {Hirst}, {Ivison},
  {Rawlings}, {Simpson}, {Smith}, {Verma}, \& {Yamada}}]{swinbank07}
{Swinbank}, A.~M., {Edge}, A.~C., {Smail}, I., {et~al.} 2007, \mnras, 379, 1343

\bibitem[{{Tanaka} {et~al.}(2010){Tanaka}, {De Breuck}, {Venemans}, \&
  {Kurk}}]{tanaka10}
{Tanaka}, M., {De Breuck}, C., {Venemans}, B., \& {Kurk}, J. 2010, \aap, 518,
  A18

\bibitem[{{Taniguchi} {et~al.}(2015){Taniguchi}, {Kajisawa}, {Kobayashi},
  {Shioya}, {Nagao}, {Capak}, {Aussel}, {Ichikawa}, {Murayama}, {Scoville},
  {Ilbert}, {Salvato}, {Sanders}, {Mobasher}, {Miyazaki}, {Komiyama}, {Le
  F{\`e}vre}, {Tasca}, {Lilly}, {Carollo}, {Renzini}, {Rich}, {Schinnerer},
  {Kaifu}, {Karoji}, {Arimoto}, {Okamura}, {Ohta}, {Shimasaku}, \&
  {Hayashino}}]{taniguchi2015}
{Taniguchi}, Y., {Kajisawa}, M., {Kobayashi}, M.~A.~R., {et~al.} 2015, \pasj,
  67, 104

\bibitem[{{Taniguchi} {et~al.}(2007){Taniguchi}, {Scoville}, {Murayama},
  {Sanders}, {Mobasher}, {Aussel}, {Capak}, {Ajiki}, {Miyazaki}, {Komiyama},
  {Shioya}, {Nagao}, {Sasaki}, {Koda}, {Carilli}, {Giavalisco}, {Guzzo},
  {Hasinger}, {Impey}, {LeFevre}, {Lilly}, {Renzini}, {Rich}, {Schinnerer},
  {Shopbell}, {Kaifu}, {Karoji}, {Arimoto}, {Okamura}, \&
  {Ohta}}]{taniguchi2007}
{Taniguchi}, Y., {Scoville}, N., {Murayama}, T., {et~al.} 2007, \apjs, 172, 9

\bibitem[{{Tomczak} {et~al.}(2017){Tomczak}, {Lemaux}, {Lubin}, {Gal}, {Wu},
  {Holden}, {Kocevski}, {Mei}, {Pelliccia}, {Rumbaugh}, \& {Shen}}]{tomczak17}
{Tomczak}, A.~R., {Lemaux}, B.~C., {Lubin}, L.~M., {et~al.} 2017, \mnras, 472,
  3512

\bibitem[{{Topping} {et~al.}(2016){Topping}, {Shapley}, \&
  {Steidel}}]{topping16}
{Topping}, M.~W., {Shapley}, A.~E., \& {Steidel}, C.~C. 2016, \apjl, 824, L11

\bibitem[{{Topping} {et~al.}(2018){Topping}, {Shapley}, {Steidel}, {Naoz}, \&
  {Primack}}]{topping18}
{Topping}, M.~W., {Shapley}, A.~E., {Steidel}, C.~C., {Naoz}, S., \& {Primack},
  J.~R. 2018, \apj, 852, 134

\bibitem[{{Toshikawa} {et~al.}(2018){Toshikawa}, {Uchiyama}, {Kashikawa},
  {Ouchi}, {Overzier}, {Ono}, {Harikane}, {Ishikawa}, {Kodama}, {Matsuda},
  {Lin}, {Onoue}, {Tanaka}, {Nagao}, {Akiyama}, {Komiyama}, {Goto}, \&
  {Lee}}]{toshikawa18_goldrush}
{Toshikawa}, J., {Uchiyama}, H., {Kashikawa}, N., {et~al.} 2018, \pasj, 70, S12

\bibitem[{{Trump} {et~al.}(2009){Trump}, {Impey}, {Elvis}, {McCarthy},
  {Huchra}, {Brusa}, {Salvato}, {Capak}, {Cappelluti}, {Civano}, {Comastri},
  {Gabor}, {Hao}, {Hasinger}, {Jahnke}, {Kelly}, {Lilly}, {Schinnerer},
  {Scoville}, \& {Smol{\v c}i{\'c}}}]{trump09}
{Trump}, J.~R., {Impey}, C.~D., {Elvis}, M., {et~al.} 2009, \apj, 696, 1195

\bibitem[{{Venemans} {et~al.}(2002){Venemans}, {Kurk}, {Miley},
  {R{\"o}ttgering}, {van Breugel}, {Carilli}, {De Breuck}, {Ford}, {Heckman},
  {McCarthy}, \& {Pentericci}}]{venemans02}
{Venemans}, B.~P., {Kurk}, J.~D., {Miley}, G.~K., {et~al.} 2002, \apjl, 569,
  L11

\bibitem[{{Vogeley} {et~al.}(2004){Vogeley}, {Hoyle}, {Rojas}, \&
  {Goldberg}}]{vogeley04}
{Vogeley}, M.~S., {Hoyle}, F., {Rojas}, R.~R., \& {Goldberg}, D.~M. 2004, in
  IAU Colloq. 195: Outskirts of Galaxy Clusters: Intense Life in the Suburbs,
  ed. A.~{Diaferio}, 5--11

\bibitem[{{Wang} {et~al.}(2016){Wang}, {Elbaz}, {Daddi}, {Finoguenov}, {Liu},
  {Schreiber}, {Mart{\'{\i}}n}, {Strazzullo}, {Valentino}, {van der Burg},
  {Zanella}, {Ciesla}, {Gobat}, {Le Brun}, {Pannella}, {Sargent}, {Shu}, {Tan},
  {Cappelluti}, \& {Li}}]{wang2016_z250}
{Wang}, T., {Elbaz}, D., {Daddi}, E., {et~al.} 2016, \apj, 828, 56

\end{thebibliography}

\appendix


\section{Stability of the peaks properties}\label{app_stability}

We investigated the extent to which the choice of a 5$\sigma_{\delta}$ threshold
affects some of the properties of the identified peaks. Namely, we
varied the overdensity threshold from 4.5$\sigma_{\delta}$ to
5.5$\sigma_{\delta}$, and verified the variation of $M_{\rm tot}$
(Table~\ref{peaks_tab}), velocity dispersion
(Table~\ref{peaks_tab_veldisp}) and elongation
(Table~\ref{peaks_elongation_tab}) as a function of the used threshold.

\subsection{Total mass}\label{app_Mtot_sigma}

\begin{figure} \centering
\includegraphics[width=9.0cm]{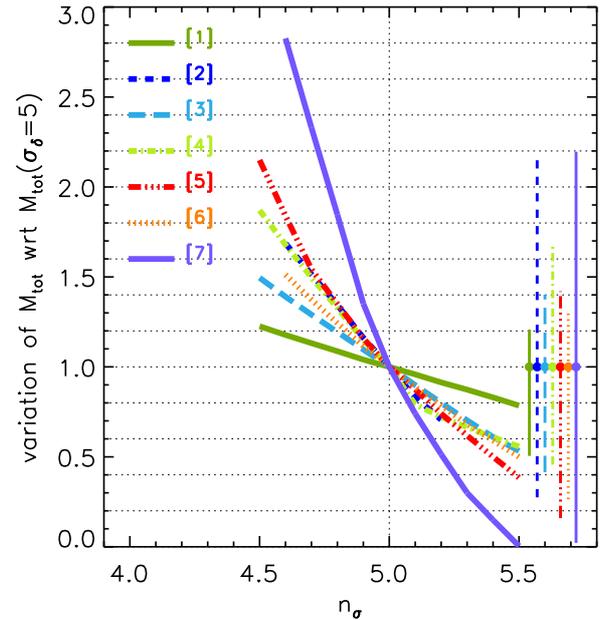}
\caption{ Fractional variation of the total mass $M_{\rm tot}$
  (Table~\ref{peaks_tab}) for the seven peaks as a function of the
  overdensity threshold adopted to identify them, expressed in terms
  of the corresponding multiples $n_{\sigma}$ of
  $\sigma_{\delta}$. The reference total mass value is the one at the
  5$\sigma_{\delta}$ threshold. The different lines correspond to the
  different peaks as in the legend. The filled symbols on the right,
  with their error bars, correspond to the fractional variation of
  $M_{\rm tot}$ calculate at $5\sigma_{\delta}$ resulting from the
  uncertainties on the density reconstruction quoted in
  Table~\ref{peaks_tab}. The position of the error bars on the x-axis
  is arbitrary. In all cases, these errors are much larger than the
  uncertainty resulting from slightly modulating the overdensity
  threshold employed.}
\label{Mtot_vs_sigma} 
\end{figure}

Figure \ref{Mtot_vs_sigma} shows the fractional variation of $M_{\rm tot}$ 
(Table~\ref{peaks_tab}) as a function of the adopted threshold,
which is expressed in terms of the corresponding multiple of
$\sigma_{\delta}$. Five peaks out of seven show roughly the same
variation, while peak [1] has a much smaller variation and peak [7] a
much steeper one. This might imply that the (baryonic) matter
distribution within peak [7] is less peaked toward the centre with
respect to the other peaks, while the matter distribution within peak
[1] is more peaked. 

Given that we are probing very dense peaks (they are about to
collapse, see Sect.~\ref{discussion}), we expect the total mass
enclosed above a given overdensity threshold to have large variations
if we vary the overdensity threshold by much. If instead we focus on a
small $n_{\sigma}$ range around our nominal value of $n_{\sigma}=5$,
for instance the interval $5\pm0.2$, we see that the variation of the
total mass is much smaller than the uncertainty on the
total mass quoted in Table~\ref{peaks_tab}, which was computed by
using the density maps obtained with $\delta_{\rm gal,16}$ and
$\delta_{\rm gal,84}$ (see Sect.~\ref{method}).

This means that, although the total mass of our peaks depends on the
chosen overdensity threshold, because of the very nature of the mass
distribution in these peaks, at the chosen threshold the uncertainty
is dominated by the uncertainty on the reconstruction of the density
field and not by our precise definition of `overdensity peak'.

\subsection{Velocity dispersion}\label{app_vdisp_sigma}

Similarly to the variation of the total mass, we verified how the
velocity dispersion $\sigma_{\rm v}$ varies as a function of the
adopted overdensity threshold, for the seven identified peaks. For
each threshold, the velocity dispersion and its error are computed as
described in Sect.~\ref{vel_disp}, and only when we could use at least
three spectroscopic galaxies. For all the peaks, $\sigma_{\rm v}$ is
relatively stable in the entire range of the explored overdensity
thresholds, and its small variations (due to the increasing or
decreasing number of spectroscopic members) are always much smaller
than the uncertainties computed on the velocity dispersion itself, at
fixed $n_{\sigma}$. For this reason we consider the virial masses
quoted in Table~\ref{peaks_tab_veldisp} to be independent from small
variations of the overdensity threshold.

We remind the reader that for the computation of the velocity
dispersion we used a more relaxed definition of galaxy membership
within each peak so as to increase the number of the available galaxies
(see Sect.~\ref{vel_disp}). Even with this broader definition, for
peak [7] we had only two galaxies available if we used $n_{\sigma}=5$ to
define the peak, while their number increased to four by using
$n_{\sigma}=4.9$. For this reason, we decided that the most reliable
value of $\sigma_{\rm v}$ for peak [7] is the one computed using
$n_{\sigma}=4.9$, and we quote this $\sigma_{\rm v}$ in
Table~\ref{peaks_tab_veldisp}.

\subsection{Elongation}\label{app_elongation}

Here we approximately estimate how the elongation depends on the
typical dimension of our density peaks. Our estimation is based on the
following simplistic assumptions: 1) the intrinsic shape of a
proto-cluster is a sphere with radius $r_{\rm int}$, and its measured
dimensions on the $x-$ and $y-$axis ($r_x$ and $r_y$) correspond to
the intrinsic dimension $r_{int}$, i.e. $r_x=r_y=r_{\rm int}$, and 2) the
measured dimension on the $z-$axis ($r_z$) corresponds to $r_{int}$
plus a constant factor $\Delta r$, which is the result of the complex
interaction among the several factors that might cause the elongation
(the depth of the redshift slices, the photometric redshift error
etc), i.e. $r_z=r_{int}+\Delta r$. From these assumptions it follows:

\begin{equation} \displaystyle
\frac{r_z}{r_{xy}} = 1 + \frac{\Delta r}{r_{int}},
\label{elongation_eq} 
\end{equation}

\noindent where $r_{xy}$ is the average between $r_x$ and $r_y$, and
in our example we have $r_{xy}=r_x=r_y=r_{\rm int}$. If we
substitute $r_x$, $r_y$ and $r_z$ with $R_{e,x}$, $R_{e,y}$ and
$R_{e,z}$ as defined in Sect.~\ref{3D_peaks}, from
Eq.~\ref{elongation_eq} follows:

\begin{equation} \displaystyle
E_{\rm z/xy} = 1 + \frac{\Delta r}{R_{e,xy}},
\label{elongation_eq2} 
\end{equation}

\noindent with $E_{\rm z/xy}$ and $R_{e,xy}$ as defined in
Sect.~\ref{3D_peaks}. This means that the measured elongation depends
on the circularised 2D effective radius as $y=1+A/x$.

\begin{figure} \centering
\includegraphics[width=9.0cm]{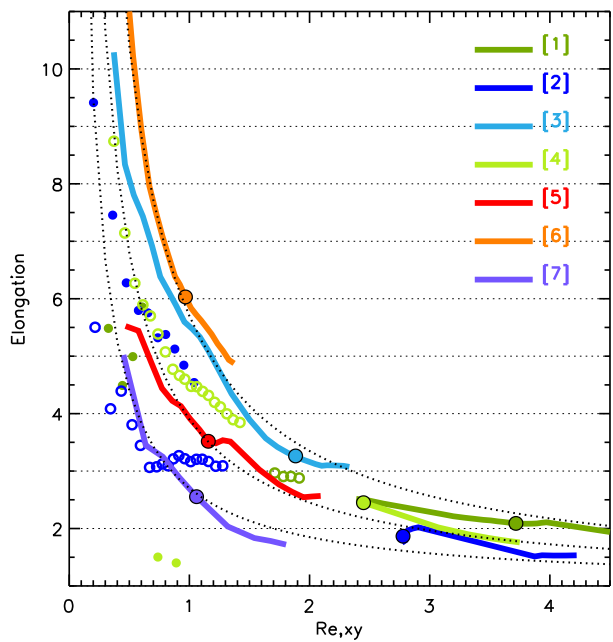}
\caption{Elongation $E_{\rm z/xy}$ as a function of $R_{e,xy}$. The
  different colours refer to the different peaks as in the
  legend. $E_{\rm z/xy}$ and $R_{e,xy}$ are measured by fixing
  different thresholds (number of $\sigma_{\rm \delta}$ above the mean density) to
  define the peaks themselves, ranging from 4.1 to 7 $\sigma_{\rm \delta}$.
  $E_{\rm z/xy}$ and $R_{e,xy}$ measured at the $5\sigma_{\rm \delta}$ threshold
  are highlighted with a filled circle, and are the same quoted in
  Table~\ref{peaks_elongation_tab}.  The peaks [1] and [2] are
  split into two smaller peaks when $\delta_{\rm gal}$ is above
  5.5$\sigma_{\rm \delta}$ and 5.7$\sigma_{\rm \delta}$ above the mean density, respectively:
  this is shown in the plot by splitting the curve of the two peaks
  into two series of circles (filled and empty). The three dotted
  lines corresponds to the curves $y=1+A/x$, with $A=4.3,2.9,1.5$ from
  top to bottom. The values of A are chosen to make the curves
  overlap with some of the data, to guide the eye.}
\label{elongation_fig} 
\end{figure}

To verify this dependence, we measured $E_{\rm z/xy}$ and $R_{e,xy}$
for our seven peaks for different thresholds, expressed in terms of
the multiples $n_{\sigma}$ of $\sigma_{\rm \delta}$. In this case, we
made the threshold vary from 4.1 to 7 $\sigma_{\rm \delta}$, because
the two peaks [1] and [4] merge in one huge structure if we use a
threshold $<4.1\sigma_{\delta}$. We notice that peak [5] disappears
for $\sigma_{\rm \delta}>5.8$ above the mean density, and peak [7] for
$\sigma_{\rm \delta}>5.4$. The peaks [1], [2] and [4] are split into
two smaller peaks when $\delta_{\rm gal}$ is above
6.5$\sigma_{\rm \delta}$, 5.2$\sigma_{\rm \delta}$ and
5.1$\sigma_{\rm \delta}$ above the mean density, respectively. Figure
\ref{elongation_fig} shows how $E_{\rm z/xy}$ varies as a function of
$R_{e,xy}$. The three curves with equation $y=1+A/x$ are shown to
guide the eye, with $A$ tuned by eye to match the normalisation of
some of the observed trends. It is evident that the foreseen
dependence of $E_{\rm z/xy}$ on $R_{e,xy}$ is confirmed. In the
Figure, $A$ increases by a factor of $\sim3$ from the lowest curve
(corresponding e.g. to peak [7]) to the highest one (matching e.g.
peak [6]). The specific value of $A$ is likely due to a complex
combination of peculiar velocities, spectral sampling, reconstruction
methods (e.g. slice size relative to the true l.o.s. extent), and
photometric redshift errors. It is beyond the scope of this paper to
precisely quantify the contribution of each for each individual peak.
Nevertheless, although in some cases $E_{\rm z/xy}$ quickly vary for
small changes of $R_{e,xy}$ (i.e. small changes in the threshold),
this plot confirms that its measured values are reasonably consistent
with our expectations.

\section{Details on individual peaks}\label{app_peaks}

We show here the projections on the RA-Dec and $z$-Dec planes of the
four most massive peaks (``Theia'', ``Eos'', ``Helios'', and
``Selene''), to highlight their complex shape. The remaining peaks
have very regular shapes on the RA-Dec and $z$-Dec planes, so we do
not show them here. The projections that we show include the peak
isodensity contours in the 3D cube and the position of the
spectroscopic member galaxies. The $z$-Dec projection is associated to
the velocity distribution of the spectroscopic members.

\begin{figure} \centering
\includegraphics[width=9.0cm]{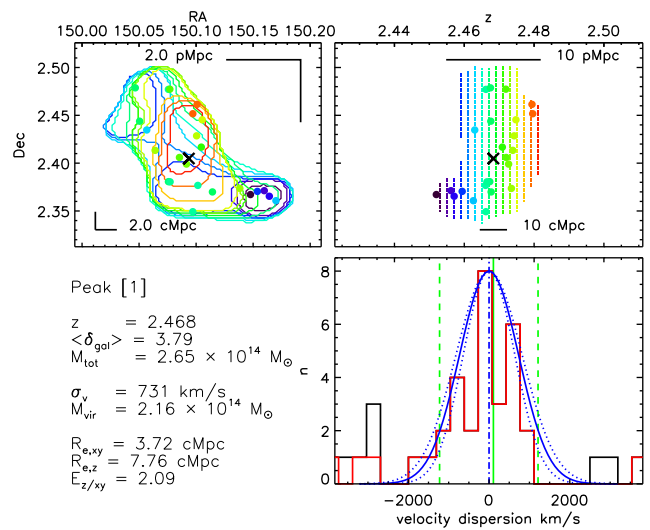}
\caption{For peak [1], ``Theia'', the {\it top-left} panel show the projection on
  the RA-Dec plane of the $5\sigma_{\rm \delta}$ contours which
  identify the peak in the 3D overdensity cube; the different colours
  indicate the different redshift slices (from blue to red, they go
  from the lowest to the highest redshift). Filled circles are the
  spectroscopic galaxies which are members of the peak (flag=X2/X2.5,
  X3, X4, X9), with the same colour code as the the contours. The
  black cross is the RA-Dec barycenter of the peak. In the top-right
  and bottom-left corners we show the scale in pMpc and cMpc,
  respectively, for both RA and Dec. {\it Top-right.}  Projections on
  the $z$-Dec plane of the same contours shown in the top-left panel,
  with the same colour code. The filled circles and the black cross
  are as in the top-left panel. On the top and on the bottom of the
  panel we show the scale in pMpc and cMpc, respectively. {\it
    Bottom-right}. The black histogram represents the velocity
  distribution of the spectroscopic galaxies which fall in the same
  RA-Dec region as the proto-cluster. The red histogram includes only
  VUDS and zCOSMOS galaxies with reliable quality flag, and flags
  X1/X1.5 for galaxies within the peak volume (see
  Sect.~\ref{vel_disp} for details). The vertical solid green line
  indicates the barycenter along the l.o.s (the top x-axis is the same
  as the one in the top-right panel), and the two dashed vertical
  lines the maximum extent of the peak. The dotted-dashed blue
  vertical line is the $z_{\rm BI}$ of Table~\ref{peaks_tab_veldisp},
  around which we center the Gaussian (blue solid curve) with the same
  $\sigma_v$ as in Table~\ref{peaks_tab_veldisp}. The two dotted blue
  curves are the uncertainties on the Gaussian due to the
  uncertainties on $\sigma_v$. In the {\it bottom-left} corner of the
  figure we summarise some of the peak properties, which are all
  already mentioned in the Tables or in the text. }
\label{peak1_fig} 
\end{figure}

\begin{figure} \centering
\includegraphics[width=9.0cm]{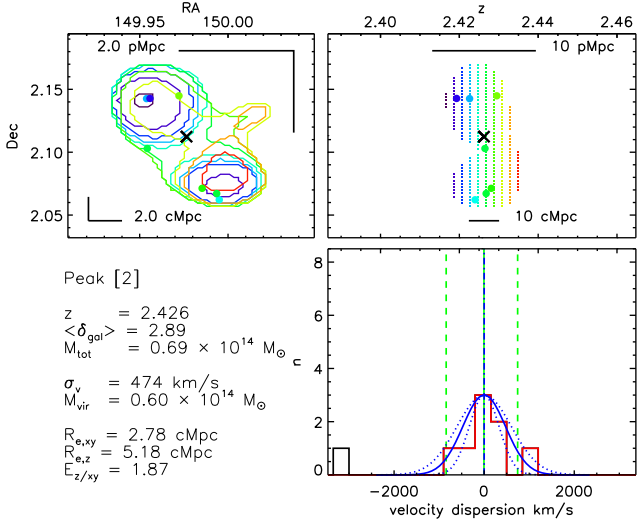}
\caption{As in Fig.\ref{peak1_fig}, but for Peak [2], ``Eos''.}
\label{peak2_fig} 
\end{figure}

\begin{figure} \centering
\includegraphics[width=9.0cm]{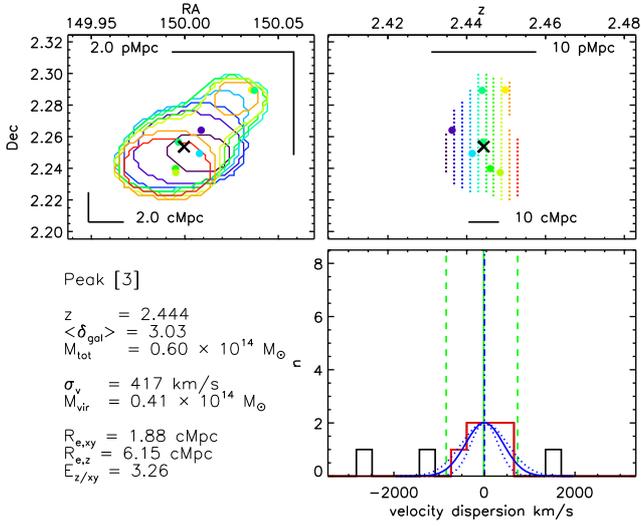}
\caption{As in Fig.\ref{peak1_fig}, but for Peak [3], ``Helios''. }
\label{peak3_fig} 
\end{figure}

\begin{figure} \centering
\includegraphics[width=9.0cm]{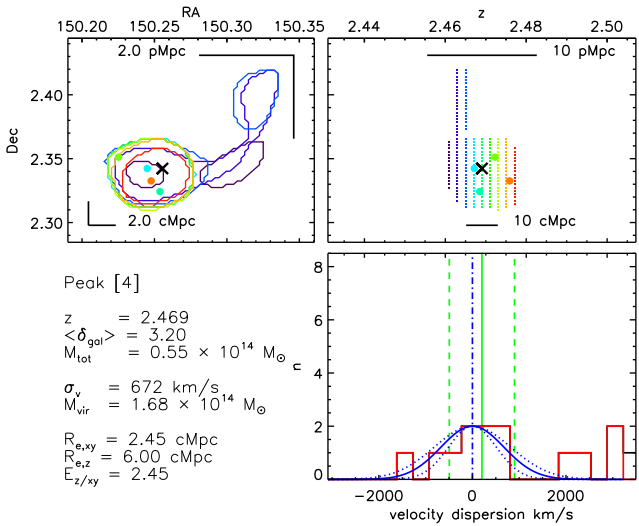}
\caption{As in Fig.\ref{peak1_fig}, but for Peak [4], ``Selene''.}
\label{peak4_fig} 
\end{figure}

\end{document}